\documentclass{aa}  
\usepackage[varg]{txfonts}
\usepackage{graphicx}
\graphicspath{{Figures/}}
\usepackage{txfonts}
\usepackage{float}
\usepackage{hyperref}
\usepackage{multirow}
\usepackage{color}
\usepackage{csquotes}
\usepackage{caption}
\newcommand{\cla}[1]{{#1}}

\begin{document}

\title{Determination of small-scale magnetic fields on Sun-like stars in the near-infrared using CRIRES$^+$}
\subtitle{}
   \author{A. Hahlin\inst{1}, O. Kochukhov\inst{1}, A. D. Rains\inst{1}, A. Lavail\inst{1,2}, A. Hatzes\inst{3}, N. Piskunov\inst{1}, A. Reiners\inst{4}, U. Seemann\inst{5,4}, L. Boldt-Christmas\inst{1}, E. W. Guenther\inst{3}, U. Heiter\inst{1}, L. Nortmann\inst{4},  F. Yan\inst{6}, D. Shulyak\inst{7},  J. V. Smoker\inst{8,9}, F. Rodler\inst{8}, P. Bristow\inst{5}, R. J. Dorn\inst{5}, Y. Jung\inst{5}, T. Marquart\inst{1}, E. Stempels\inst{1}}
   \institute{
       Department of Physics and Astronomy, Uppsala University, Box 516, SE-751 20 Uppsala, Sweden \\\email{axel.hahlin@physics.uu.se}
       \and 
       Institut de Recherche en Astrophysique et Plan\'etologie, Universit\'e de Toulouse, CNRS, IRAP/UMR 5277, 14 avenue Edouard Belin, F-31400, Toulouse, France
       \and
       Thüringer Landessternwarte Tautenburg, Sternwarte 5, Tautenburg, 07778, Germany
       \and 
       Institut für Astrophysik und Geophysik, Georg-August-Universität, Friedrich-Hund-Platz 1, 37077 Göttingen, Germany
       \and 
       European Southern Observatory, Karl-Schwarzschild-Str. 2, 85748 Garching, Germany
       \and  
       Department of Astronomy, University of Science and Technology of China, Hefei 230026, China 
       \and 
       Instituto de Astrof\'{\i}sica de Andaluc\'{\i}a - CSIC, c/ Glorieta de la Astronom\'{\i}a s/n, 18008 Granada, Spain 
       \and 
       European Southern Observatory, Alonso de Cordova 3107, Vitacura, Santiago, Chile
       \and 
       UK Astronomy Technology Centre, Royal Observatory, Blackford Hill, Edinburgh EH9 3HJ, UK}

    \date{received: 2023-03-03, accepted: 2023-05-09}
    
\abstract
{}%Context(optional)
{We aim to characterise the small-scale magnetic fields for a sample of 16 Sun-like stars and investigate the capabilities of the newly upgraded near-infrared (NIR) instrument CRIRES$^+$ at the Very Large Telescope (VLT) in the context of small-scale magnetic field studies. Our targets also had their magnetic fields studied in the optical, which allows us to compare magnetic field properties at different spatial scales on the stellar surface and to contrast small-scale magnetic field measurements at different wavelengths.}%Aims
{We analyse the Zeeman broadening signature for six magnetically sensitive and insensitive \ion{Fe}{I} lines in the H-band to measure small-scale magnetic fields on the stellar surface. We use polarised radiative transfer modelling and non-local thermodynamic equilibrium departure coefficients in combination with Markov Chain Monte Carlo sampling to determine magnetic field characteristics together with non-magnetic stellar parameters. We use two different approaches to describe small-scale magnetic fields. The first is a two-component model with a single magnetic region and a free magnetic field strength. The second model contains multiple magnetic components with fixed magnetic field strengths.}%Methods 
{We find average magnetic field strengths ranging from $\sim 0.4$\,kG down to $<0.1$\,kG. The results align closely with other results from high resolution NIR spectrographs such as SPIRou. It appears that the typical magnetic field strength in the magnetic region is slightly stronger than 1.3\,kG and that for most stars in our sample this strength is between 1 and 2\,kG. We also find that the small-scale fields correlate with the large-scale fields and that the small-scale fields are at least 10 times stronger than the large-scale fields inferred with Zeeman Doppler imaging. The two- and multi-component models produce systematically different results as the strong fields from the multi-component model increase the obtained mean magnetic field strength. When comparing our results with the optical measurements of small-scale fields we find a systematic offset of 2--3 times stronger fields in the optical. This discrepancy can not be explained by uncertainties in stellar parameters. Care should therefore be taken when comparing results obtained at different wavelengths until a clear cause can be established.}%Results
{} %Conclusions(optional)


\authorrunning{A. Hahlin et al.}
\maketitle
%
%-------------------------------------------------------------------
\section{Introduction}

The study of stellar magnetic fields involves \cla{the} analysis of magnetic structures at multiple spatial scales, with one of the most powerful magnetic diagnostic techniques known as Zeeman Doppler Imaging \citep[ZDI,][]{donati:2009} that utilises time series of circularly polarised spectra, ZDI is however only sensitive to the large-scale magnetic field structures. Other methods that rely on unpolarised spectra, such as Zeeman broadening or intensification, are sensitive to smaller spatial scales. Comparing the two scales, 
one finds that the large-scale field strengths are only about 10\,\% of the 
total field strength on the surface of stars
\citep[e.g.][]{reiners:2012,see:2019,kochukhov:2020a}, which suggests that the latter is dominated by the small-scale component. Due to this large difference, magnetic analysis based on only a single spatial scale might miss important information causing inaccurate conclusions to be drawn. Large-scale fields are generally understood to reach out beyond the stellar surface and into the surrounding environment, thereby governing processes such as star-planet interactions \citep[e.g.][]{vidotto:2018,alvaradogomez:2022} and the rotational spin-down of stars due to magnetised stellar wind \citep[e.g.][]{kawaler:1988,barnes:2003}.
On the other hand, when studying stellar evolution, predictions of surface magnetic field strengths required to sufficiently inflate the star made by \cite{feiden:2013} have been found to better align with small-scale field measurements \citep{kochukhov:2019,hahlin:2021} showing that the stronger small-scale fields are more relevant for stellar structure and evolution. 
While the different spatial scales dominate different processes they are intimately connected, one such example is from \cite{reiners:2022} that found a relationship between the small-scale magnetic fields and rotation rate. This shows that as the star slows down due to the large-scale field interacting with the stellar wind, the generation of magnetic fields on all spatial scales are affected. 
As the two spatial scales play a role in different processes in and around stars it is therefore important to be mindful of all spatial scales in order to fully describe the magnetic properties of the star.

In the past, most cool-star magnetic field studies have focused on singular spatial scales and the primary reason for this is due to the fact that most ZDI results have been obtained with optical spectrographs. The magnetic splitting caused by the Zeeman effect is strongly wavelength dependent, which makes studies of Zeeman broadening using optical spectra quite challenging for most active stars. While Zeeman intensification \citep[e.g.][]{basri:1992,kochukhov:2020a} has been used successfully to study magnetic fields in the optical, the method introduces a large number of systematics as the change in equivalent width is degenerate with several other stellar parameters. For this reason it is beneficial to study active stars in the near-infrared (NIR) as that increases the broadening signal significantly owing to the $\lambda^2$ dependence of the Zeeman effect. Recently, several high-resolution spectrographs covering the NIR wavelengths, such as SPIRou \citep{donati:2020} and CRIRES$^+$ \citep{dorn:2023}, have begun operation. These spectrographs also have the capability to perform polarimetric observations and are therefore well suited for the investigation of stellar magnetic fields. With the upgraded CRIRES$^+$ being the most recently commissioned and the highest resolution NIR spectrometer, it offers an excellent opportunity to reach new levels of precision in detecting and measuring small-scale fields on moderately active cool stars. This study is the first to test the capabilities of CRIRES$^+$ for stellar magnetic field diagnostics.

In this paper, we aim to describe the analysis of small-scale magnetic fields for a sample of 16 Sun-like stars. These stars have had their magnetic fields investigated in the past, which allows comparison with previous results. For most stars included in our sample, large-scale fields were determined with ZDI. Some also had their small-scale fields measured with the Zeeman intensification using optical spectra \citep{kochukhov:2020a} and, more recently, with NIR SPIRou data \citep{petit:2021}. The goal of this study is to investigate if the small-scale magnetic fields measured using CRIRES$^+$ spectra agree with those measured in the optical as well as seeing if the strength of the large- and small-scale fields behave in a similar way.

The rest of this paper is organised as follows. In Sect. \ref{sec:Obs} the observations and targets \cla{analysed} in this study are introduced. This section also describes the data reduction process that was performed to obtain the final spectra. Sect.\ref{sec:modeling} discusses the impact of departures from local thermodynamic equilibrium (LTE) for Sun-like stars in the H-band as well as the selection of suitable lines for a magnetic field investigation. The magnetic field inference is presented in Sect. \ref{sec:inference} and the results are discussed in Sect. \ref{sec:discussion}, including assessment of various systematic effects and biases that could potentially impact magnetic field diagnostics.

\section{Observations}
\label{sec:Obs}
%Observations that were carried out
Observational data analysed in this study were taken throughout the ESO's observing period 108 between October 2021 and March 2022. The observations were carried out using CRIRES$^+$,
the recently upgraded NIR spectrograph at the 8-m UT3 Very Large Telescope. CRIRES$^+$ is a cross-dispersed echelle spectrometer and spectropolarimeter operating in conjunction with the MACAO adaptive optics system. The spectrometer is able to reach a spectral resolving power of 100\,000 and obtain observations in the Y, J, H, K, L, and M bands using a mosaic of three 2K$\times$2K detectors. CRIRES$^+$ is also capable of polarimetric observations in the Y, J, H, and K bands.

For the magnetic field investigation of Sun-like stars, we focused on unpolarised spectra in the H-band which is known to have several magnetically sensitive spectral lines that have been used in previous studies on both the Sun and other stars \citep[e.g.][]{valenti:1995,lavail:2017,petit:2021}. Observations with CRIRES$^+$ typically have gaps in the spectral coverage due to both the separation between each detector but also between each echelle order. 
We opted to use the H1567
setting of CRIRES$^+$ 
as it contains lines that are both magnetically sensitive and insensitive spectral lines (see Sect.~\ref{subsec:lines}).

\subsection{Target stars}
The list of targets selected for this study can be seen in Table \ref{tab:targets}. These are Sun-like stars with known magnetic activity and large-scale fields that were studied in the past (see Table~\ref{tab:targets} for obtained values and references). 
We focus here on Sun-like stars of spectral types 
spanning from late-F to early-K
with a variety of ages and activity levels. Several of the stars in this sample are also known or candidate exoplanet hosts, HD 3651 \citep[e.g.][]{fischer:2003}, HD 22049 \citep[e.g][]{hatzes:2000}, HD 73256 \citep[e.g.][]{udry:2003}, HD 75732 \citep[e.g.][]{butler:1997}, HD 102195 \citep[e.g.][]{melo:2007}, HD 130322 \citep[e.g.][]{udry:2000},  HD 179949 \citep[e.g.][]{tinney:2001}). The characterisation of stellar magnetic fields is particularly interesting and consequential in these cases as magnetic activity shapes the space environments of exoplanet systems and causes additional noise (activity jitter) that can hide or distort planetary signals during both transit and radial velocity measurements \citep[see e.g.][respectively]{salz:2018, yu:2017}.

\subsection{Data reduction}

Most of the observations were obtained in the adaptive optics mode using 0.2\arcsec\ slit. In all cases, nodding was employed to allow accurate background removal. The observing log is summarised in Table~\ref{tab:obslog}, which gives the observing date, the total exposure time for each setting, and the resulting signal-to-noise ratio (S/N) per pixel. 

The data were reduced using the CRIRES pipeline\footnote{\url{http://www.eso.org/sci/software/pipelines/cr2res/cr2res-pipe-recipes.html}}. 
The reduction process included the standard steps of obtaining master dark and flat field images; deriving order positions and wavelength calibration with U-Ne and Fabry-Perot etalon exposures; and optimal extraction of the stellar spectra at the two nodding positions. 
The resulting spectra were then co-added and divided by the blaze function (derived from the master flat field) for each wavelength setting. Then, line-free regions in the observed spectra were fitted with a linear function to normalise to the continuum.

In the NIR, there is significant absorption by Earth's atmosphere resulting in numerous telluric lines. These tellurics were removed using {\tt Molecfit} tool \citep{smette:2015}. 
{\tt Molecfit} uses recorded weather data from the location of the telescope to model a synthetic spectra of Earth's atmosphere with prominent molecules such as H$_2$O, CO$_2$, etc. by fitting to an observed spectrum. Molecfit has been shown by \cite{smette:2015} to be able to fit unsaturated telluric absorption features to within 2\% of the continuum.
{\tt Molecfit} only models the telluric contributions to the spectra, which means that it might misinterpret stellar spectral lines as atmospheric absorption. This issue manifests in two different ways. First, a stellar spectral line that is blended with a telluric line will cause {\tt Molecfit} to interpret the reduced flux as a stronger telluric contribution to the spectra, effectively resulting in an overestimation of the telluric contamination. Secondly, stellar lines that are not blended with any tellurics will instead be interpreted as a reduced continuum flux as {\tt Molecfit} has no information about the presence of stellar features. 
To mitigate these issues we produced a mask using a synthetic spectrum generated with \texttt{SYNMAST} \citep{kochukhov:2007,Kochukhov:2010} using line data from VALD \citep{ryabchikova:2015} and \cla{plane-parallel} {\tt MARCS} model atmospheres \citep{gustafsson:2008}. 
By identifying regions with stellar absorption we can remove as many stellar features as possible from the telluric removal procedure by masking out any wavelengths with significant absorption in the synthetic stellar spectra shifted to the stellar rest frame. To minimise the risk of continuum mismatch and improve repeatability, we apply our own linear continuum normalisation separately to each order instead of relying on {\tt Molecfit}'s built-in continuum fitting functionality.

Once the spectra were prepared, the telluric removal was performed using an iterative approach where the first iteration generated a first guess for the telluric model. Using the initial telluric model, the continuum normalisation could be improved by finding the regions in the observed spectra affected by neither telluric nor stellar absorption. This was particularly beneficial in the regions heavily contaminated by tellurics that also contain the most information about the telluric influence on the observed spectrum. With this improved continuum normalisation, the second iteration produces the telluric-corrected spectrum that will be used for the magnetic field inference in Sect.~\ref{sec:inference}. To mitigate any non-linear deviations from the blaze function that are not modelled by the linear adjustment of the continuum we further refined the continuum of the orders containing magnetically interesting lines (see Sect.~\ref{subsec:lines}) using a third-order polynomial.

\begin{table*}
  \centering
  \caption{Stellar parameters for the stars employed in this study, including previously obtained average large-scale magnetic field strength from Stokes $V$ if available.}
  \label{tab:targets}
  \begin{tabular}{lcccccc}
  \hline\hline
  Star & T$_{\mathrm{eff}}$\,(K) & $\log g$ (cms$^{-2}$) &$\varepsilon_{\mathrm{Fe}}$& $v\sin i$\,(kms$^{-1}$) & $\langle B\rangle_{V}$\,(G) & References \\
  \hline
  HD 1835 & 5837 & 4.47 & -4.29 & 6.3& 19 & 1, 2\\
  HD 3651 & 5284 & 4.53 & -4.38 & 1.1 & 3.58 & 3, 4\\
  HD 9986 & 5805 & 4.45 & -4.46 & 2.6 & 0.605 & 1, 4 \\
  HD 10476 & 5181 & 4.54 & -4.58 & 0.1 & 3.3 & 1, 5 \\
  HD 20630 & 5742 & 4.49 & -4.42 & 4.7 & 26.3 & 1, 4 \\ 
  HD 22049 & 5146 & 4.57 & -4.57 & 2.4 & 10 -- 20 & 1, 6, 7 \\
  HD 59967 & $5848$ & $4.54$ & -4.57 & 4.0 & -- & 8, 9 \\ 
  HD 73256 & $5532$ & $4.49$ & -4.28 & 3.2 & 2.7 & 3, 10 \\ 
  HD 73350 & 5802 & 4.48 & -4.42 & 3.2 & 11 & 1, 4 \\
  HD 75732 & 5235 & 4.45 & -4.23 & 2.5 & 3.4 & 1, 11\\
  HD 76151 & 5790 & 4.55 & -4.43 & 0.0 & 2.99 & 1, 4 \\
  HD 102195 & 5330 & 4.37 & -4.44 & 2.9 & 10.7 & 10,12 \\
  HD 130322 & 5308 & 4.41 & -4.53 & 1.6 & 2.34 & 1,10 \\
  HD 131156 A & 5570 & 4.65 & -4.58 & 4.9 & 22.2 -- 61.8 & 1, 4 \\
  HD 179949 & 6168 & 4.34 & -4.40 & 7.0 & 2.6 -- 3.7 & 10\\
  HD 206860 & 5974 & 4.47 & -4.56 & 10.1& 11 -- 24 & 1, 2 \\

  \hline
  \end{tabular}
  \tablefoot{1: \cite{valenti:2005}, 2: \cite{rosen:2016}, 3: \cite{bonfanti:2015}, 4: \cite{see:2019}, 5: \cite{marsden:2014},  6: \cite{jeffers:2014}, 7: \cite{petit:2021}, 8: \cite{yana-galarza:2019}, 9: \cite{lorenzooliveira:2019}, 10: \cite{fares:2013}, 11: \cite{folsom:2020}, 12: \cite{ge:2006}. \cla{Fe abundance is defined as $\log(N_{\mathrm{Fe}}/N_{\mathrm{total}})$}.}
\end{table*}

\section{NLTE effects and line selection}
\label{sec:modeling}

Investigations of small-scale magnetic fields usually
focus on a small selection of lines that have suitable magnetic sensitivities. In order to obtain reliable estimates of the magnetic field, accurate line parameters are required. This section goes through the process of selecting suitable lines as well as determining more accurate parameters for them. Another aspect that is explored is the impact of non-local thermodynamic equilibrium (NLTE) in the NIR. While investigations of M dwarfs have revealed that \ion{Fe}{I} lines in the H-band are not significantly affected by NLTE effects \citep[e.g.][]{olander:2021}, \cite{lind:2012} showed that the impact of NLTE for \ion{Fe}{i} increases with temperature. It is therefore worthwhile to assess what the impact of NLTE is for the temperature range of our stellar sample, especially since magnetic field studies often assumes LTE. 

\subsection{Impact of NLTE}
\begin{figure*}
\centering
\includegraphics[width=0.49\linewidth]{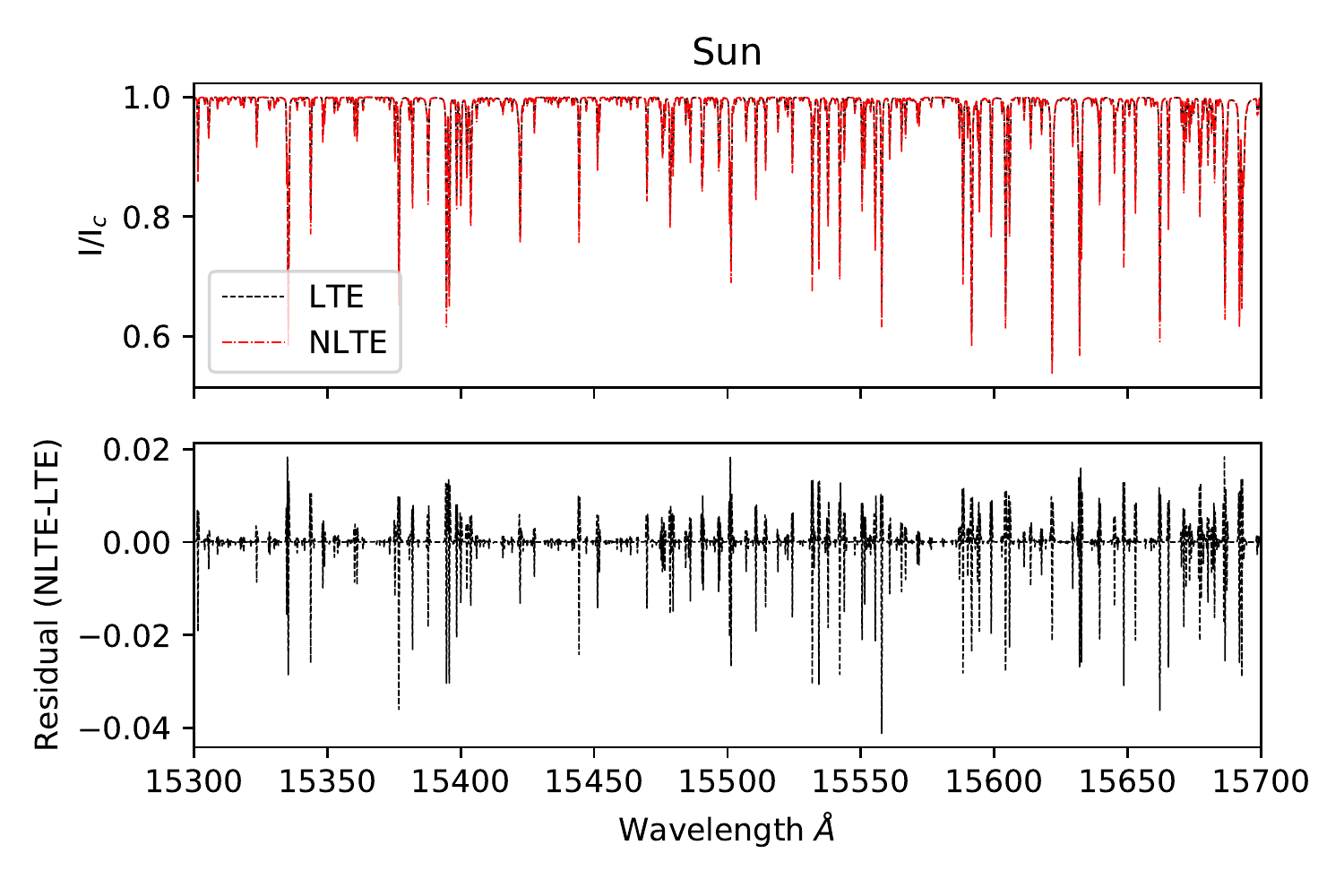}
\includegraphics[width=0.49\linewidth]{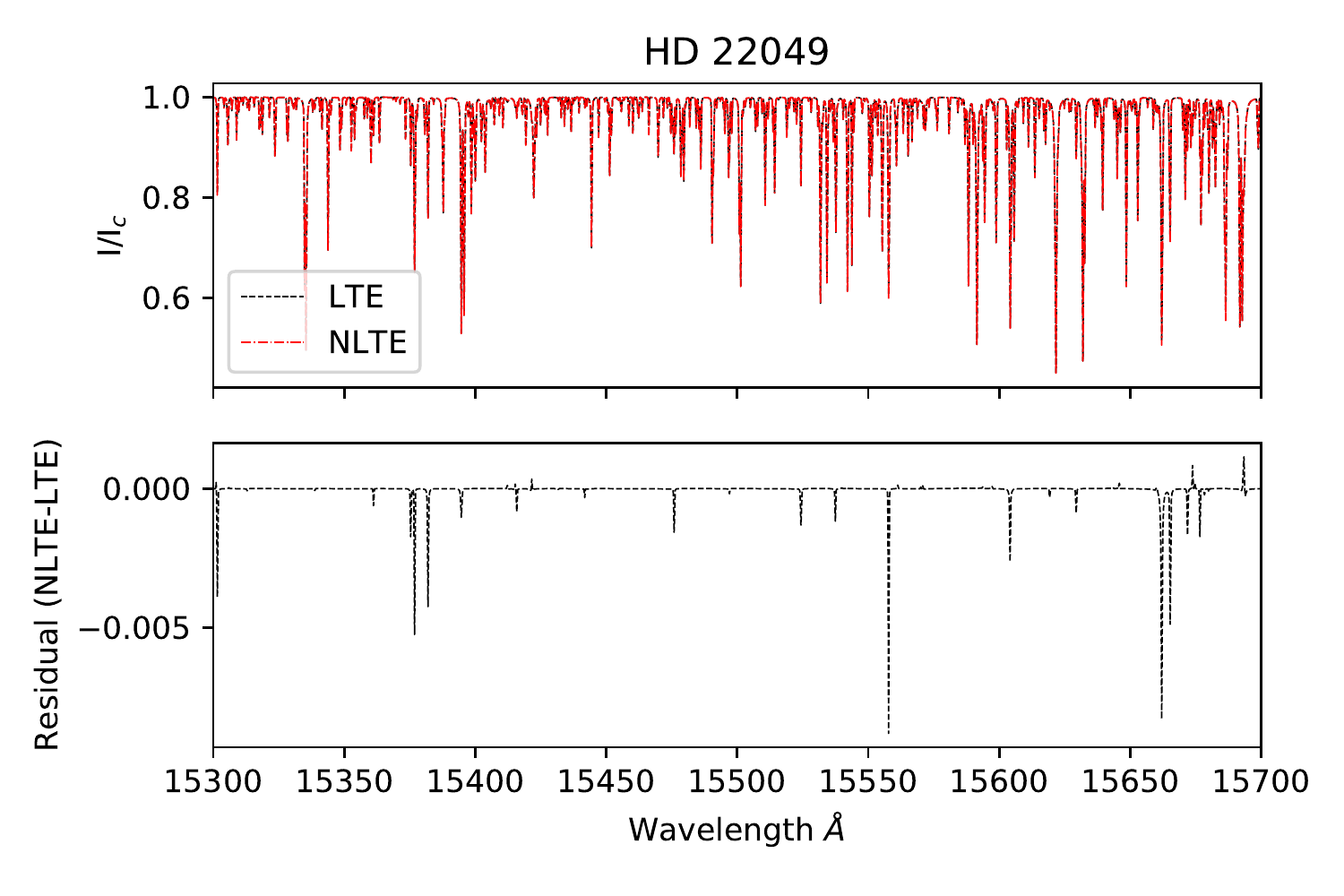}
\caption{Investigation of the influence of NLTE effects in spectra of stars with different stellar parameters in the H-band. \textit{Left:} synthetic spectra for solar parameters; \textit{right:} calculations for parameters of HD 22049, one of the coolest stars in our sample. \textit{Top} panels show synthetic spectra for LTE (solid black line) and NLTE (dashed red line) while the \textit{bottom} panels show the difference between the NLTE and LTE synthetic spectra.}
\label{fig:nlte-vs-lte}
\end{figure*}

In order to investigate possible effects of NLTE we calculated synthetic spectra of the H-band containing lines that have previously been used for the study of small-scale magnetic fields. We choose stellar parameters corresponding to the Sun and HD 22049 ($\epsilon$ Eri) with effective temperatures of 5780 and 5146~K \citep{valenti:2005} respectively and use Spectroscopy Made Easy \citep[{\tt SME},][]{valenti:1996,Piskunov:2017} to generate synthetic spectra, we adopt solar metallicities from \cite{asplund:2009}. This spectrum synthesis code allows one to take into account NLTE effects by using pre-calculated tables of NLTE departure coefficients from the work of \cite{Amarsi:2022}. The two chosen stars are close to the upper and lower boundaries of the temperature range of stars in our study and thus give a good indication of the range of sensitivity to NLTE effects in our stellar sample. We calculated spectra both with and without NLTE departure coefficients and compared the resulting synthetic spectra. Both calculations correspond to stellar flux spectra with \cla{a macroturbulent velocity of 3\,kms$^{-1}$ and a $v\sin i$ of 1.9\,kms$^{-1}$}. This comparison can be seen in Fig. \ref{fig:nlte-vs-lte}. For HD 22049, the effect of NLTE on the synthetic spectra is small, with the difference between two calculations exceeding $\sim 0.5$ \% of the continuum in only a few lines. For the solar spectrum the situation is quite different. We find that many lines show NLTE effects at the level of a few \% of the continuum. Furthermore, NLTE also changes line shape: the line cores in the LTE spectrum are underestimated while the wings appear deeper compared to the NLTE spectrum. This change in line shape could possibly impact the Zeeman broadening analysis as that effect also causes a change in line shape. For this reason we elect to use NLTE departure coefficients for \ion{Fe}{I} lines in the remaining analysis in this work. The polarised radiative transfer code \texttt{SYNMAST} 
that is used for the magnetic spectrum synthesis in this work is capable of using pre-calculated departure coefficients to account for NLTE effects similar to their treatment by {\tt SME}.

\subsection{Magnetically sensitive lines}
\label{subsec:lines}

\begin{table}
  \centering
  \caption{\ion{Fe}{i} lines selected for magnetic inference.}
  \label{tab:lines}
  \resizebox{\linewidth}{!}{
  \begin{tabular}{cccccc}
  \hline\hline
  \multirow{2}{*}{Wavelength (\AA)}  & \multirow{2}{*}{$g_{\mathrm{eff}}$} & \multicolumn{2}{c}{$\log{gf}$} & \multicolumn{2}{c}{vdw} \\
   & & This work & VALD & This work & VALD \\
  \hline
  15343.788 & 2.63 & $-0.662$ & $-0.582$ & $-7.054$ & $-7.520$\\%Other stars, e.g. eps eri
  15381.960 & 0.00 & $-0.590$ & $-0.460$ & $-6.909$ & $-7.430$\\%Other stars, e.g. eps eri
  15534.245 & 1.95 & $-0.369$ & $-0.382$ & $-7.061$ & $-7.510$\\%Used in solar studies
  15542.079 & 1.52 & $-0.577$ & $-0.377$ & $-7.062$ & $-7.450$\\%Used in solar studies
  15648.510 & 3.00 & $-0.702$ & $-0.599$ & $-7.143$ & $-7.490$\\%Other stars, e.g. eps eri
  15652.871 & 1.50 & $-0.080$ & $-0.161$ & $-6.996$ & $-7.320$\\%Used in solar studies
  \hline
  \end{tabular}
  }
  \tablefoot{From left to right: central wavelength of line, effective Land\'e factor, $\log{gf}$ values both corrected and original VALD, and van der Waals broadening parameters given as the logarithm of the broadening width per unit pertuber number density at 10\,000~K in units of rad s$^{-1}$ cm$^{-3}$.
  }
\end{table}

We aim to select a group of lines that are suitable for the determination of magnetic fields for Sun-like stars in the H-band. Magnetic field investigations for stars in the H-band have been performed previously \citep[e.g.][]{petit:2021,lavail:2017} and give insights into magnetically sensitive and insensitive lines that are particularly useful for such an analysis. Another source for suitable lines can be found in the studies of magnetic fields on the surface of the Sun. \cite{smitha:2017} highlight two NIR lines that are commonly used to study the solar magnetic field and also suggest a new pair of \ion{Fe}{I} lines that could be a useful magnetic field indicator. Our final line selection can be seen in Table~\ref{tab:lines} and includes 6 lines with effective Land\'e factors ranging from 0 to 3. As the effective Land\'e factor is a measure of the magnetic sensitivity of the line, a large range of values is useful to disentangle the magnetic broadening from any non-magnetic sources of broadening. 
Lines with an effective Lande factor of 0, such as the \ion{Fe}{I} 15381.96~\AA\ line, are particularly useful as they provide a reference for non-magnetic parameters. 
 
Before these lines can be used to study magnetic fields for our sample of stars, the key line parameters must be verified. Both our own and previous work \citep[e.g.][]{petit:2021} finds that the $\log{gf}$ values of NIR lines provided by VALD are often inaccurate due to the fact that no laboratory measurements are available for many of these lines. 
The line parameters of the selected lines for this study are instead reliant on theoretical calculations by \cite{kurucz:2014}. While such calculations are useful to gain a large amount of line data, they still introduce errors that can cause mismatch with observations, particularly when only a few lines are used. 
In order to get reliable results, other studies corrected these values by fitting synthetic spectra to observations of the Sun or other benchmark stars to obtain astrophysical $\log{gf}$ values \citep[see e.g.][]{valenti:1995,petit:2021}. In addition, the van der Waals broadening parameter is also often adjusted empirically. For example, when \cite{petit:2021} made corrections to a large number of $\log{gf}$ values, they found some synthetic lines to be discrepant with their observations unless the van der Waals broadening parameter is fitted as well. Another approach is to use a more sophisticated theoretical calculation of van der Waals broadening. To this end, \cite{quinteronoda:2021} performed calculations on some of the lines suitable for our work using the code from \citet{barklem:1998}\footnote{\url{https://github.com/barklem/abo-cross}}. We completed these calculations for the remaining lines using the same code. It appears that these theoretical van der Waals damping coefficients give consistently stronger broadening compared to the values provided by VALD for all selected \ion{Fe}{i} lines.

\cla{The inclusion of NLTE in this study has some impact on the obtained $\log gf$-values. Comparing NLTE and LTE fitting results we find that the difference in $\log gf$ is $\sim 0.01$ dex. In addition, this change is not uniform across the lines. This means that the lines employed here have a variable sensitivity to NLTE effects, which possibly could cause issues in accurately determining magnetic field strengths. This effect is, however, smaller than the impact of updating the van der Waals broadening coefficients, where the difference between $\log gf$-values obtained using the VALD and our updated van der Waals broadening coefficients could reach $\sim 0.1$ dex.}

In order to find the $\log{gf}$ values we use the intensity atlas of the quiet Sun NIR spectrum observed by \citet{livingston:1991}\footnote{\url{https://nso.edu/data/historical-archive/\#ftp}}. We used the program {\tt BinMag6} \citep{kochukhov:2018a}\footnote{\url{https://www.astro.uu.se/~oleg/binmag.html}} that interfaces with \texttt{SYNMAST} to find the best fitting $\log{gf}$ value for each line using $\chi^2$-minimisation. We also allowed the radial velocity for each line to be adjusted individually during this process. We use {\tt MARCS} model atmospheres, VALD linelists, and NLTE departure coefficients, both with the original van der Waals coefficients as well as those calculated by \cite{quinteronoda:2021} and by us. We found that using our updated van der Waals broadening parameters significantly improves the fit, particularly in the line wings. Consequently, we elected to use these new values. This is important because much of the information on magnetic broadening will reside in the wings, possibly resulting in spurious results if the van der Waals broadening parameters are not accurate enough. The resulting oscillator strength and van der Waals damping parameter values are summarised in Table~\ref{tab:lines}, as well as the original values from VALD.

The lines that we use have been shown by \cite{Smoker:2022} to overlap with Diffuse Interstellar Bands (DIBs) that can cause changes in the estimated equivalent width. All of our stars are significantly closer than 100~pc so we do not expect any significant contribution from DIBs to the stellar spectra. We also do not find any broadband features similar to that of DIBs in our observations that could affect our spectral lines from Table \ref{tab:lines}.

\section{Magnetic field inference}
\label{sec:inference}
To describe magnetic field strength distribution on the surfaces of moderately active Sun-like stars, a two-component model is often used \citep[e.g.][]{valenti:1995, kochukhov:2020a}. This model consists of one spectral component corresponding to a magnetic field with strength $B$ that covers a fraction of the surface $f_B$. The other component has no magnetic field and covers the rest of the stellar surface. Both $B$ and $f_B$ are free parameters in this model. The synthetic model spectrum is obtained by combining the magnetic and non-magnetic continuum-normalised component spectra according to
\begin{equation}
  \label{eq:combination}
  S = S_{B}f_{B}+S_{0}(1-f_{B}),
\end{equation}
where $S_B$ and $S_0$ represent the continuum normalised synthetic spectra, $S\equiv S^{(l)}/S^{(c)}$, calculated with and without magnetic fields respectively. 

Another common approach is to use a multi-component model with fixed magnetic field strengths but several magnetic filling factors \citep[e.g.][]{yang:2011}. In this case, the model synthetic spectrum is given by
\begin{equation}
    \label{eq:multicomp}
    S = \sum_{i}S_{i}f_{i}.
\end{equation}
Here $i$ is the index spanning different magnetic field components, including a zero-strength component. In this model, all filling factors $f_i$ except the non-magnetic $f_0$ is a free parameter. In principle, there is no limit to the number of magnetic components. Studies of very active, slowly rotating stars for which Zeeman broadening signatures are very prominent often used more than 5 components \citep[e.g.][]{lavail:2019,shulyak:2019}. For stars with weaker magnetic broadening, it is often sufficient to use 3--4 components. Using 3 components is roughly similar to the two-component approach in Eq. (\ref{eq:combination}) as this uses the same number of free magnetic field parameters for the inference. 
Differences are however likely to appear if there is a significant spread of magnetic field strengths on the stellar surface as the two component model will have difficulties in reproducing the broadening pattern of a complex field distribution. Rotation is likely to reduce the difference between the two approaches as the magnetic broadening becomes smeared out and less distinct.

Similar to the determination of astrophysical $\log{gf}$ values, we allowed for shifts of 2~km\,s$^{-1}$ in the radial velocity for individual lines in order to mitigate any uncertainties in the line positions or wavelength calibration errors. Other non-magnetic parameters in the model included the \ion{Fe}{i} abundance and macroturbulent broadening $v_{\mathrm{mac}}$. The projected rotational velocity, $v\sin{i}$, was kept fixed according to the literature values in Table~\ref{tab:targets}. For the microturbulent broadening, $v_\mathrm{mic}$, we assumed a fixed value of 1\,km\,s$^{-1}$ for all stars in our sample. This is close to $v_{\mathrm{mic}}$ values of Sun-like stars obtained by \cite{jofre:2015}.

In contrast to similar Zeeman broadening studies of cooler stars \citep[e.g.][]{kochukhov:2019,shulyak:2019}
we elected to not apply continuum scaling to individual lines in the process of finding best fit solutions. This continuum scaling approach was meant to compensate for weak blends of molecular lines missing from theoretical line lists. But our spectra in the region around 15500~\AA\ are relatively free from molecular absorption, so this problem is absent for our hotter stars compared to M dwarfs.

The spectral model grids were generated with {\tt SYNMAST} using {\tt MARCS} model atmospheres and VALD line lists modified by the procedure outlined in Sect.~\ref{subsec:lines}. \cla{The rotational and radial-tangential macroturbulent broadening was implemented in the same way as in the {\tt SME} code as described by \cite{valenti:1996}, by applying convolution with appropriate kernels to the synthetic spectra generated at 7 different limb angles.} When using \texttt{SYNMAST} we assumed a uniform radial magnetic field configuration. This assumption is commonly used when modelling magnetic field effects in the unpolarised stellar spectra. Its accuracy has been tested by \citet{kochukhov:2021}, who showed that no significant deviations between different assumptions of uniform field orientation can be detected for disk-integrated stellar spectra. \cla{The synthetic spectra were then interpolated in $T_{\mathrm{eff}}$ and $\log g$ to the values in Table~\ref{tab:targets}} 

When generating the synthetic grid for the magnetic field descriptions presented in Eq. (\ref{eq:combination}) and (\ref{eq:multicomp}) different considerations need to be taken into account when selecting the magnetic field grid step. For the two-component model, the goal is to represent the magnetic field as a single value. For this reason the grid step needs to be sufficiently small as to allow the interpolation to produce a profile as close as possible to a profile generated from a single field strength. We elect to use a grid step of 0.1\,kG for the two-component spectral grid, the same step size that was used by \cite{kochukhov:2020a}. 
In the case of a multi-component model we aim to represent the magnetic field as several distinct regions on the stellar surface. 
If the step size is too small, the inference will have difficulty distinguishing them, resulting in a significant degeneracy between the filling factors of different magnetic components.
If, instead, the step size is too large then the synthetic profile will contain distinct distortions that originate from Zeeman components. In order to mitigate this, an intermediate step size should be selected. The typical value for previous multi-component studies has been 2\,kG steps \citep[e.g.][]{shulyak:2019,hahlin:2022}. These studies are, however, using data obtained at shorter wavelengths (and lower spectral resolutions) than the H-band, reducing the sensitivity to the Zeeman splitting. Many targets in these studies are also more rapidly rotating compared to our sample, which smears out the Zeeman components due to Doppler broadening. For our targets, the Zeeman splitting is often the dominant contributor to the line broadening, particularly in the wings of magnetically sensitive lines such as the \ion{Fe}{I} 15648.51~\AA\ line, meaning that a step size of 2\,kG produces distinct bumps in the spectral profile. 
%For this reason we elect to use a smaller grid step of 1\,kG for the multi-component modelling.
To \cla{address} this issue, \cite{reiners:2022} elected to apply a threshold of 5~km\,s$^{-1}$ for when to shift from a step size of 1 to 2~kG, while we do have 3 stars where this criterion is satisfied we still elect to apply a 1~kG step size for all stars to better quantify the systematic differences between the two- and multi-component model for different stellar parameters.

We carried out Markov Chain Monte-Carlo (MCMC) sampling with the modified {\tt SoBAT} library \citep{anfinogentov:2021}\footnote{\url{https://github.com/Sergey-Anfinogentov/SoBAT}} to determine optimal magnetic field parameters. We used flat priors between 0 and 1 for filling factors and implemented an additional constraint on their sum such that
\begin{equation*}
    \sum_{i}f_{i}=1
\end{equation*}
in order to ensure a physically realistic solution. 
We utilised {\tt SoBAT}'s functionality of including observational error as a free parameter. This allows us to account for potential systematics that often dominate high-S/N studies which makes accurate uncertainties difficult to determine from just observational errors.

We let our sampler walk until the effective sample size \citep[ESS, ][]{sharma:2017} has reached 1000. This was done by allowing the sampler to take 10\,000 steps before calculating the ESS from the auto-correlation time and comparing it to the threshold. For the multi-component model we selected the number of filling factors with the help of the Bayesian information criterion \citep[BIC,][]{sharma:2017}
\begin{equation}
    \label{eq:BIC}
    \mathrm{BIC}=-2\ln p(Y|\hat{\theta})+d\ln{n}.
\end{equation}
BIC weights the quality of the fit $\ln p(Y|\hat{\theta})$ given observations $Y$ and parameters $\hat{\theta}$ with the number of parameters $d$ scaled by the number of data points $n$. This places a limit on when adding additional free magnetic parameters can no longer be justified as no significant improvement of the fit is obtained.

The obtained median magnetic field parameters as well as their 68\,\% confidence regions are reported in Table~\ref{tab:results}. The median average magnetic field strengths $\langle B\rangle_{I}$ in Table \ref{tab:results} were determined by calculating the average magnetic field strength of every point in the sampling from Eq.~(\ref{eq:combination}) and (\ref{eq:multicomp}), not by using the median magnetic field parameters presented in Table~\ref{tab:results}. For this reason, the product of the median $B$ and $f_B$
might deviate slightly from the reported $\langle B\rangle_{I}$. 
Also included are the nuisance parameters of iron abundance $\varepsilon_{\mathrm{Fe}}$ and $v_{\mathrm{mac}}$. We can see that the $v_{\mathrm{mac}}$ obtained from the two-component model is systematically larger compared to the multi-component result, this is likely due to the stronger magnetic fields reducing the need for non-magnetic broadening. Our abundances agree well between the two approaches even if there is a small shift towards lower abundances for the multi-component model. The obtained abundances also agree reasonably well with those reported from other works (see Table~\ref{tab:targets} and references therein), this shows that accounting for magnetic fields can produce reasonable abundance estimate even from magnetically sensitive lines. Still\cla{,} care should be taken as only six lines were used, meaning that the results are subjected to small number statistics. The radial velocity scatter for each line is not included but does not exceed 0.5 km\,s$^{-1}$ which corresponds to a wavelength shift of $\sim0.03$ \AA. 

\cla{Comparison between running the inference with and without fixed observational errors reveals that the median values of fitted parameters do not change significantly but their uncertainties decrease by about 10--15\% when using prescribed observational errors. Therefore, we do not expect our approach to treating observational uncertainties to create any systematic differences in the analysis. At the same time, it provides somewhat higher, more conservative errors, likely including additional sources of errors beyond the random observational noise.}

\begin{figure*}
  \centering
  \includegraphics[width=\textwidth]{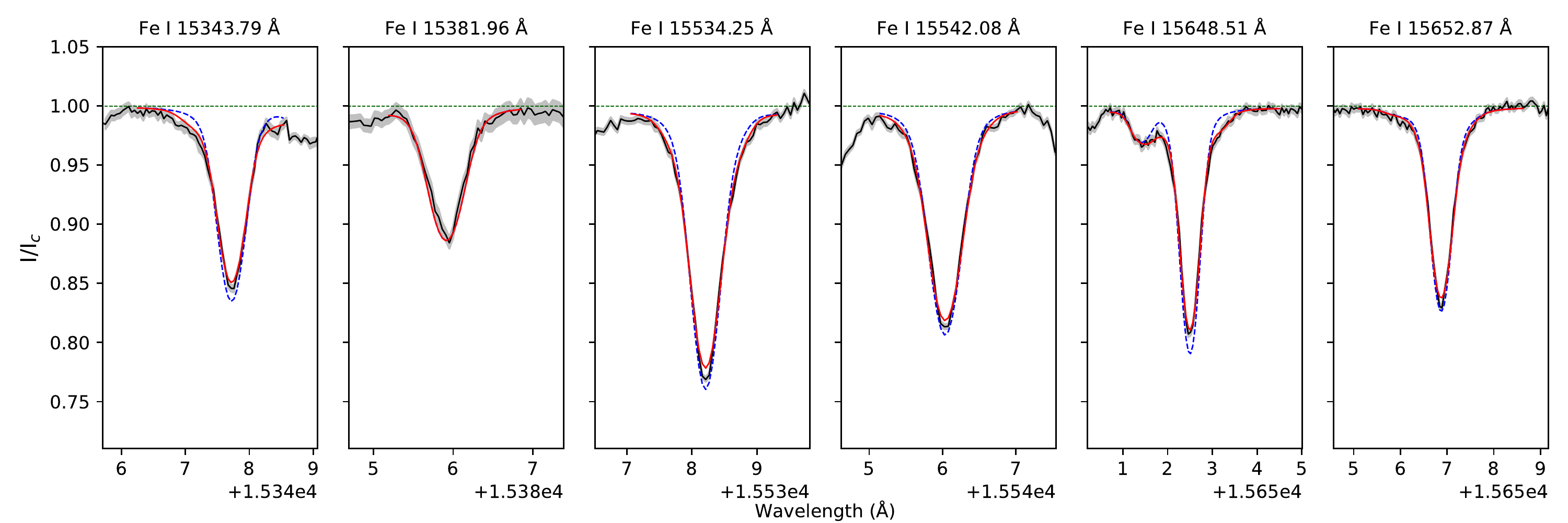}
  \caption{\cla{Two-component model} fit to the spectrum of HD\,20630 showing the observed spectra in black, the best fit synthetic model in red and the synthetic model with the same stellar parameters but without magnetic field in dashed blue. The continuum is also shown as the yellow dotted line.
  }
  \label{fig:mcmc_fit}
\end{figure*}

\begin{figure}
  \centering
  \includegraphics[width=\linewidth]{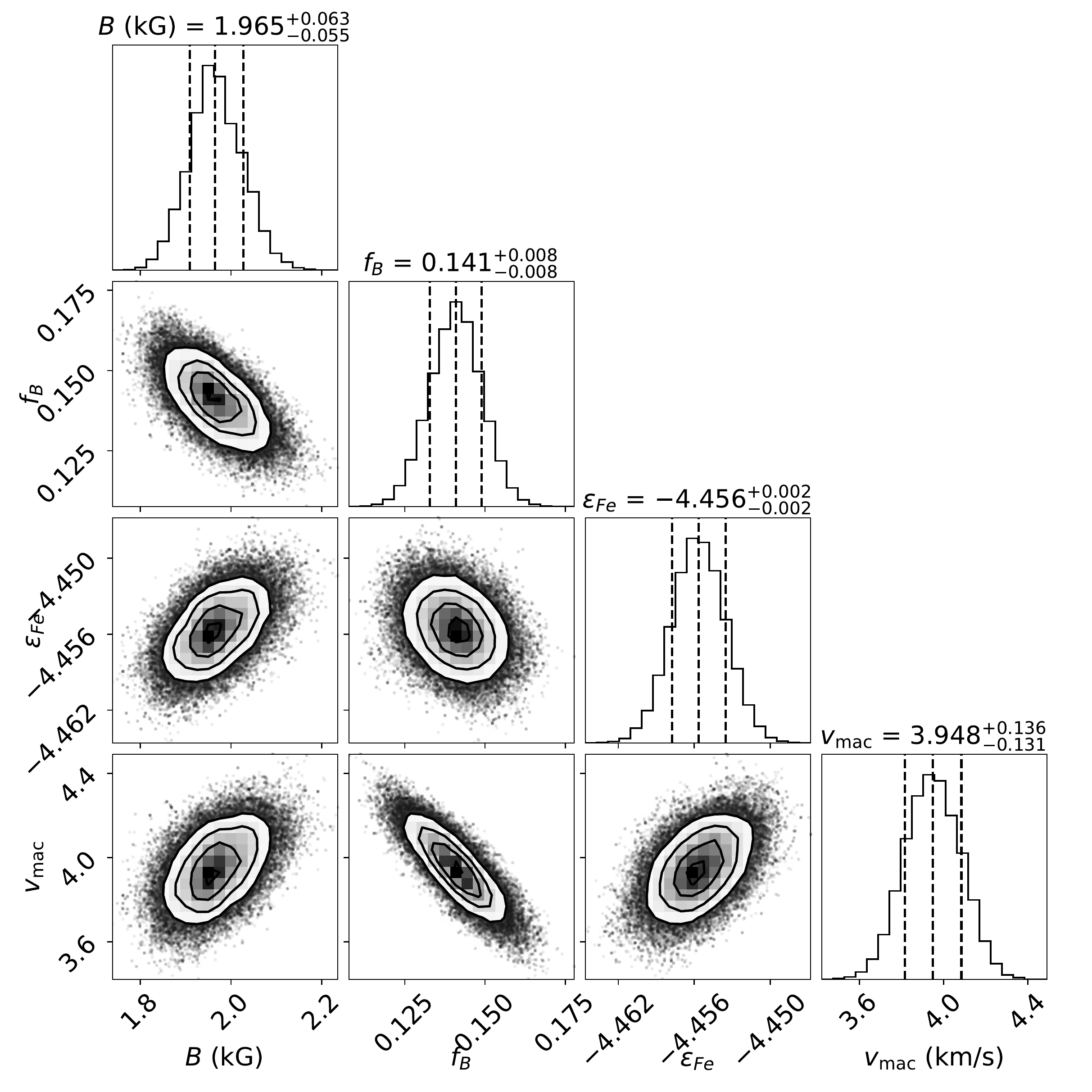}
  \includegraphics[width=\linewidth]{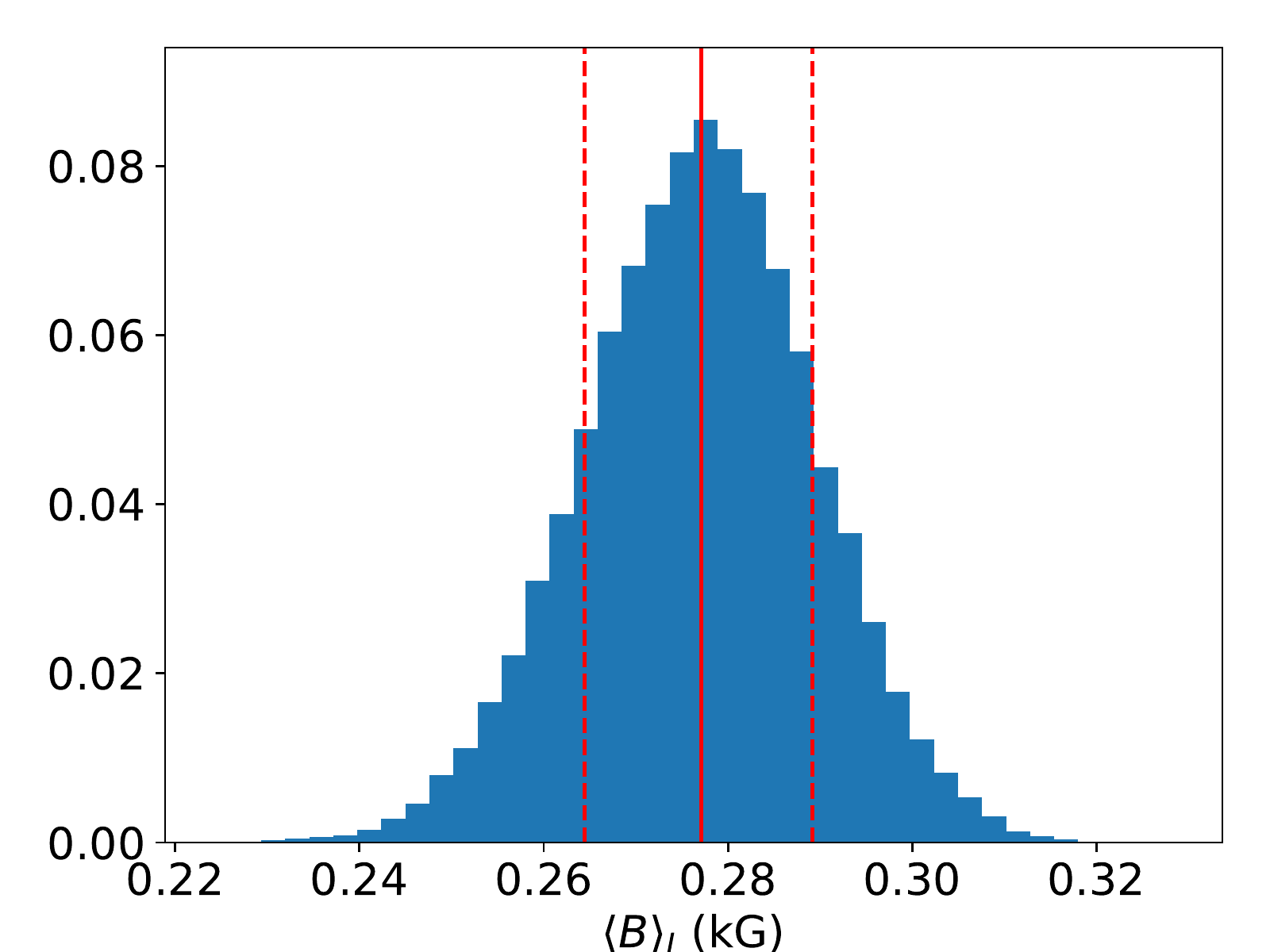}
  \caption{\textit{Top}. Posterior distributions of the parameters in the sampling of HD 20630 using the two-component model. \textit{Bottom}. Posterior distribution of the average magnetic field strength $\langle B\rangle_{I}$ given by $B f_{B}$.}
  \label{fig:mcmc_corner}
\end{figure}

An example of the fit to CRIRES$^+$ spectrum is presented for HD~20630 in Fig.~\ref{fig:mcmc_fit}. The posterior distributions of the parameters corresponding to this inference are illustrated in Fig. \ref{fig:mcmc_corner}. This figure also shows the posterior distribution for the average magnetic field strength.
Similar illustrations of the spectral fits to observations of other stars together with the corresponding posterior distributions can be found in Appendix~\ref{sec:complete_fits}.

\begin{table*}
  \centering
  \caption{Results from the magnetic field inference including both two- and multi-component models.} %Change format to () instead of +/- to reduce size. 
  \label{tab:results}
  \resizebox{\textwidth}{!}{
  \begin{tabular}{lc|ccccc|cccccc}
  \hline\hline
   &  & \multicolumn{5}{c|}{Two-component model} &\multicolumn{6}{c}{Multi-component model} \\
  Star & Night & $\langle B\rangle_{I}$ (kG) & $B$ (kG) & $f_{B}$ & $\varepsilon_{\mathrm{Fe}}$ & $v_{\mathrm{mac}}$ (km\,s$^{-1}$) & $\langle B\rangle_{I}$ (kG) & $f_{1}$ & $f_{2}$ & $f_{3}$ &$\varepsilon_{\mathrm{Fe}}$ & $v_{\mathrm{mac}}$ (km\,s$^{-1}$) \\
   \hline
  HD\,1835 & 21-10-30 &$0.279(17)$ & $1.517(61)$ & $0.184(13)$ & $-4.350(2)$ & $4.28(20)$ & $0.330(20)$ & $0.170(20)$ & $0.080(10)$ & -- & $-4.381(2)$ & $4.06(20)$ \\ %Value is closer to that obtained by Saar 1987
  HD\,3651 & 21-10-30 & $0.092(06)$ & $1.547(70)$ & $0.059(05)$ & $-4.413(1)$ & $1.47(8)$ & $0.121(07)$ & $0.059(08)$ & $0.031(03)$ & -- & $-4.419(2)$ & $1.15(12)$\\
  HD\,9986 & 21-10-30 & $0.043\substack{+0.050\\-0.020}$ & $0.267\substack{+1.590\\-0.017}$ & $0.15\substack{+0.65\\-0.14}$ & $-4.479(3)$ & $2.84(12)$ & $0.049(09)$ & $0.014(09)$ & $0.017(05)$ & -- & $-4.477(2)$ & $0.83(12)$\\
  HD\,10476 & 21-10-30 & $0.065(10)$ & $0.870(86)$ & $0.075(15)$ & $-4.569(2)$ & $1.85(12)$ & $0.060(09)$ & $0.060(09)$ & -- & -- & $-4.573(2)$ & $1.90(10)$\\
  HD\,20630 & 22-02-07 &$0.277(13)$ & $1.965\substack{+0.063\\-0.055}$ & $0.141(08)$ & $-4.460(3)$ & $3.95(13)$ &$0.382(20)$ & $0.103(18)$ & $0.087(11)$ & $0.035(07)$ & $-4.463(3)$ & $3.35(18)$\\ %Quite close to estimates by Saar & Balinuas 1992
  HD\,22049 & 22-02-09 & $0.209(07)$ & $1.646(34)$ & $0.127(05)$ & $-4.569(1)$ & $2.07(9)$ & $0.316(11)$ & $0.124(09)$ & $0.060(05)$ & $0.024(04)$ & $-4.572(2)$ & $1.12(20)$\\
  HD\,59967 & 21-10-30 &$0.250(15)$ & $1.712(59)$ & $0.146(10)$ & $-4.536(2)$ & $3.72(15)$ & $0.306(18)$ & $0.102(18)$ & $0.102(08)$ & -- & $-4.542(2)$ & $3.40(18)$\\
  HD\,59967 & 22-02-09 & $0.375(13)$ & $1.940(42)$ & $0.193(08)$ & $-4.524(2)$ & $3.47(14)$ & $0.498(22)$ & $0.115(19)$ & $0.132(11)$ & $0.040(07)$ & $-4.530(3)$ &$2.65(25)$\\
  HD\,59967 & 22-03-23 &$0.316(14)$ & $1.949(57)$ & $0.162(09)$ & $-4.534(3)$ & $3.50(15)$ &$0.431(24)$ & $0.108(20)$ & $0.109(11)$ & $0.035(06)$ & $-4.540(3)$ & $2.76(25)$\\
  HD\,59967 & 22-03-24 &$0.339(14)$ & $2.145(59)$ & $0.158(08)$ & $-4.517(3)$ & $3.87(13)$ & $0.474(23)$& $0.101(22)$ & $0.100(11)$ & $0.057(07)$ & $-4.524(3)$ & $3.00(22)$\\
  HD\,59967 & 22-03-30 &$0.336(14)$ & $1.969\substack{+0.055\\-0.047}$ & $0.171(09)$ & $-4.521(2)$ & $3.79(14)$ & $0.440(23)$ & $0.096(21)$ & $0.122(11)$ & $0.033(08)$ & $-4.528(3)$ & $3.15(22)$ \\
  HD\,73256 & 21-10-08 &$0.279(11)$ & $1.611(51)$ & $0.174(09)$ & $-4.290(3)$ & $3.79(11)$ & $0.359(15)$ & $0.153(15)$ & $0.103(07)$ & -- & $-4.300(2)$ & $3.33(15)$\\
  HD\,73256 & 21-10-30 &$0.255(12)$ & $1.556(49)$ & $0.164(10)$ & $-4.299(3)$ & $3.31(13)$ & $0.322(15)$ & $0.144(15)$ & $0.089(07)$ & -- & $-4.308(3)$ & $2.90(14)$\\
  HD\,73256 & 22-02-07 &$0.264(09)$ & $1.513(31)$ & $0.174(07)$ & $-4.310(2)$ & $2.81(10)$ & $0.336(12)$ & $0.162(11)$ & $0.087(05)$ & -- & $-4.318(2)$ & $2.29(14)$\\
  HD\,73256 & 22-03-30 &$0.262(10)$ & $1.544(30)$ & $0.170(08)$ & $-4.316(2)$ & $2.76(11)$ & $0.344(13)$ & $0.160(11)$ & $0.092(05)$ & -- & $-4.326(2)$ & $2.08(17)$\\
  HD\,73350 & 22-02-09 &$0.157(12)$ & $1.585(77)$ & $0.099(10)$ & $-4.432(2)$ & $3.66(11)$ & $0.210(15)$ & $0.090(14)$ & $0.060(06)$ & -- & $-4.438(2)$ & $3.39(13)$\\
  HD\,73350 & 22-03-10 &$0.152(12)$ & $1.359(55)$ & $0.112(11)$ & $-4.455(2)$ & $3.51(11)$ & $0.190(16)$ & $0.112(14)$ & $0.039(06)$ & -- & $-4.459(2)$ & $3.37(13)$\\
  HD\,73350 & 22-03-30 &$0.137(14)$ & $1.412(73)$ & $0.097(11)$ & $-4.452(2)$ & $3.68(11)$ & $0.176(16)$ & $0.094(15)$ & $0.041(06)$ & -- & $-4.457(2)$ & $3.53(14)$\\
  HD\,75732 & 22-03-23 &$0.052(08)$ & $0.927(91)$ & $0.056(11)$ & $-4.331(2)$ & $1.56(10)$ & $0.050(07)$ & $0.050(07)$ & -- & -- & $-4.338(2)$ & $1.57(10)$\\
  HD\,76151 & 22-02-07 &$0.111(10)$ & $1.160(66)$ & $0.096(11)$ & $-4.451(2)$ & $2.76(9)$ & $0.135(12)$ & $0.101(12)$ & $0.017(05)$ & -- & $-4.456(2)$ & $2.65(10)$\\
  HD\,76151 & 22-03-30 &$0.096(10)$ & $1.133(66)$ & $0.084(11)$ & $-4.460(2)$ & $2.67(9)$ & $0.115(12)$ & $0.087(11)$ & $0.014(05)$ & -- & $-4.464(2)$ & $2.58(10)$\\
  HD\,102195 & 22-03-11 &$0.226(08)$ & $1.562\substack{+0.043\\-0.033}$ & $0.145(07)$ & $-4.416(2)$ & $2.33(10)$ & $0.335(13)$ & $0.143(10)$ & $0.061(07)$ & $0.023(05)$ & $-4.421(2)$ & $1.56(16)$\\
  HD\,130322 & 22-03-23 &$0.081(10)$ & $1.27\substack{+0.12\\-0.10}$ & $0.064(11)$ & $-4.522(2)$ & $2.38(10)$ & $0.107(12)$ & $0.064(11)$ & $0.021(05)$ & -- & $-4.526(2)$ & $2.26(10)$\\
  HD\,131156A & 22-03-23 &$0.299(13)$ & $1.692(46)$ & $0.177(10)$ & $-4.582(2)$ & $2.87(17)$ & $0.361(13)$ & $0.126(18)$ & $0.118(06)$ & -- & $-4.590(2)$ & $2.28(25)$\\
  HD\,179949 & 21-10-08 &$0.188(21)$ & $1.89(14)$ & $0.100(13)$ & $-4.380(3)$ & $3.72(24)$ & $0.209(23)$ & $0.049(28)$ & $0.080(13)$ & -- & $-4.388(3)$ & $3.59(25)$\\
  HD\,206860 & 21-10-28 &$0.270(29)$ & $1.88(24)$ & $0.144(29)$ & $-4.527(4)$ & $3.95(14)$ & $0.302(35)$ & $0.102(60)$ & $0.099(27)$ & -- & $-4.535(4)$ & $4.58(35)$\\
  \hline
  \end{tabular}
  }
  \tablefoot{From left to right: HD number, date at the start of the observing night (yy-mm-dd), the average field strength, field strength of the magnetic component of the synthetic model, the filling factor, the Fe abundance in \cla{$\log(N_{\mathrm{Fe}}/N_{\mathrm{total}})$ units, and macroturbulent velocity}. For multi-component models the columns give the average field strength and filling factor values for the 1, 2, and 3\,kG magnetic components. Filling factors marked with \enquote{--} did not significantly improve the fit according to the BIC given by Eq.~(\ref{eq:BIC}). \cla{Uncertainties derived from asymmetric posterior distributions (see Appendix~\ref{sec:complete_fits}) are indicated with $\pm$; in other cases error in the last  significant digits is provided in brackets.}}
\end{table*}

\begin{figure}
    \centering
    \includegraphics[width=\linewidth]{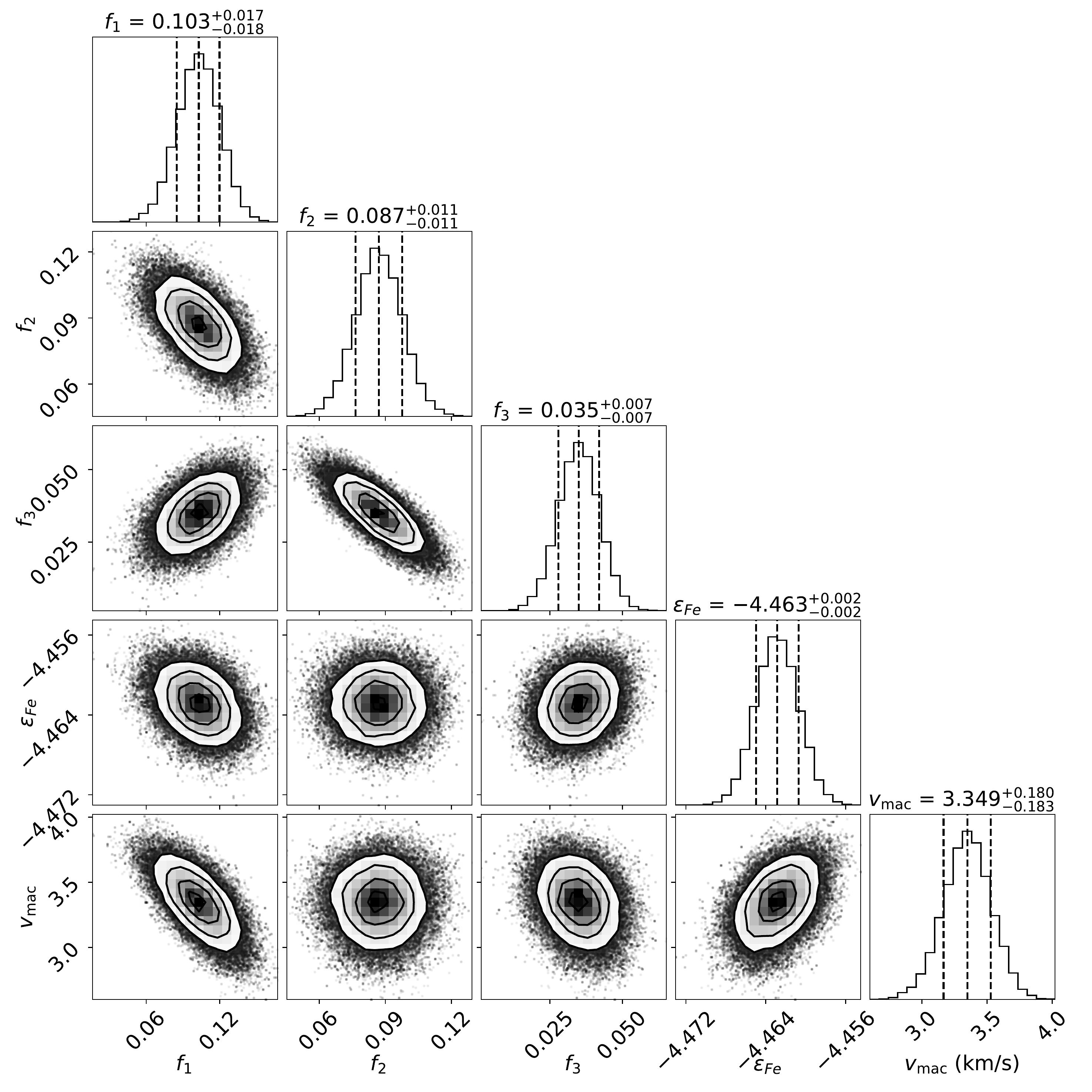}
    \includegraphics[width=\linewidth]{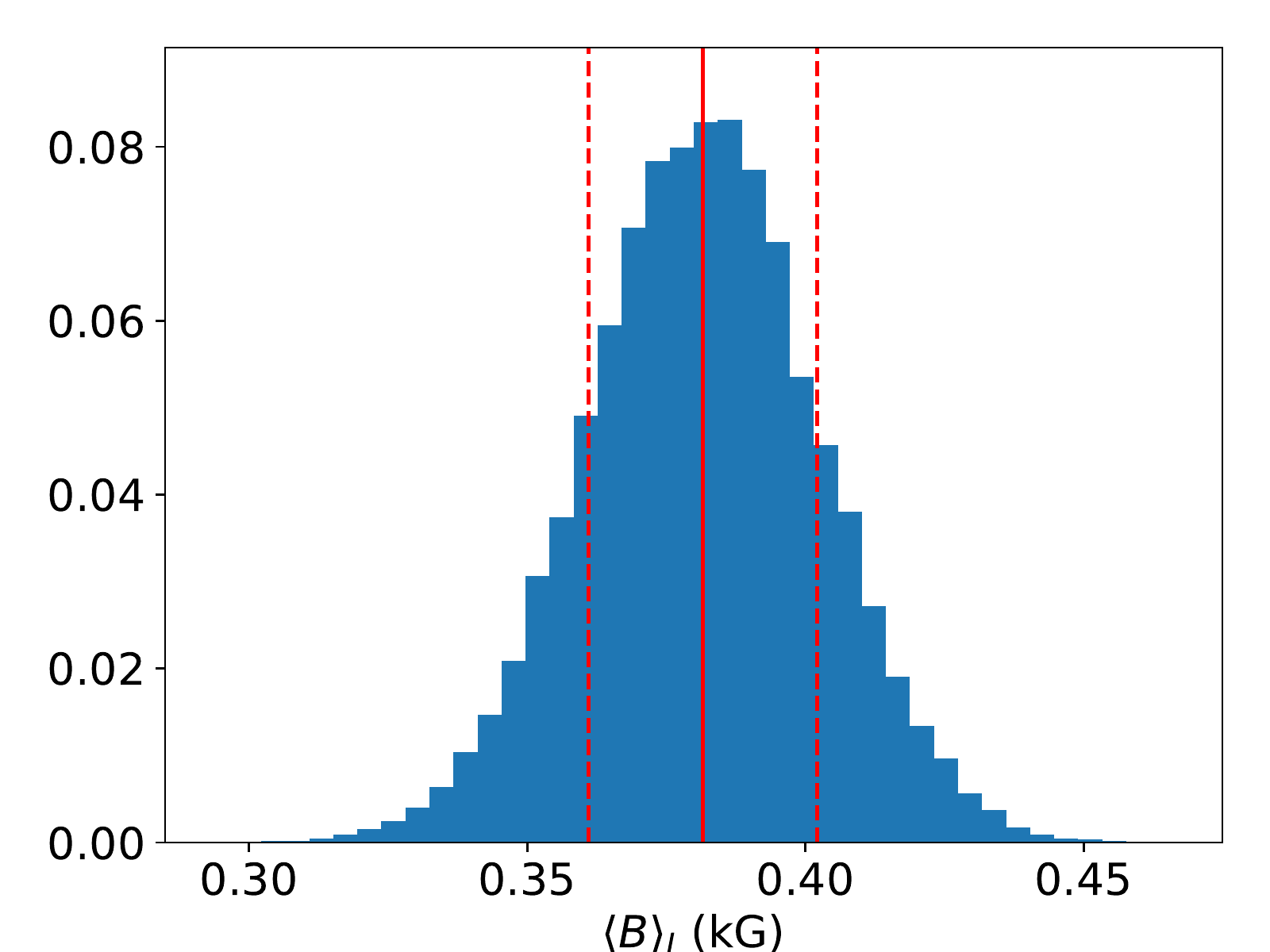}
    \caption{\cla{Same as Fig~\ref{fig:mcmc_corner} but for the multi-component model applied to HD\,20630.}}
    \label{fig:mcmc_corner_multi}
\end{figure}

\section{Discussion}
\label{sec:discussion}
As magnetic fields of the stars in this study were repeatedly studied in the past, we are able to compare the capabilities of CRIRES$^+$ with the measurements by other instruments. In addition, evaluating the ability of polarimetric large-scale magnetic field studies to recover the total stellar magnetic field strength for these stars is also interesting. Since many previous studies have been performed in the optical, we can also compare results of the magnetic measurements at the same spatial scale but at different wavelengths. This could give an indication of systematic differences obtained using spectral lines at different wavelengths and using different analysis methods.

\subsection{Sensitivity to Zeeman broadening}
\label{sec:ZB-sens}
The results reported in Table \ref{tab:results} can give some indication to the requirement for a detectable Zeeman broadening signature. We can do this by considering the stars with the weakest magnetic field signal detected in this study. The weakest field value obtained is for HD 9986. This target does, however, have a posterior distribution that does not show a strong preference to a particular field strength (posterior distributions are available in the Appendix, Fig. \ref{fig:HD9986_obs}). In fact, this posterior shows multiple peaks which indicates that different lines prefer different magnetic field strength. This means that the magnetic field is poorly constrained for this star and assumptions on both line and stellar atmosphere parameters could have a large effect on the obtained magnetic field value. The other stars with particularly weak fields are HD 10476 and HD 75732. In both cases there is a clear peak in the field strength distribution at 0.87 and 0.93\,kG respectively. This is an indication of the broadening sensitivity that is obtained using CRIRES$^+$: it should be possible to detect average magnetic field strengths of $\gtrsim 50$\,G provided that the strength of the magnetic field regions approaches 1\,kG. The magnetic regions only need to cover $\lesssim 10$\% of the stellar surface to become detectable.

As long as the strength of the magnetic field is sufficient, the covering fraction our method is sensitive to can be rather small, as average field strengths of about 
100~G
should be reliably detectable. Larger S/N should be able to increase the sensitivity, either by performing longer observations or by combining multiple observations as was done in e.g. \cite{petit:2021} and \cite{kochukhov:2020a}. The second method synergises well with ZDI studies as such techniques already require an extended sequence of observations that can then be combined in order to study the average properties of the small-scale field at a significantly improved S/N.

\subsubsection{Temperature sensitivity}
\label{subsec:Teff}
\begin{table*}
  \centering
  \caption{Changes of the mean magnetic field and Fe abundance inferred in the two-component modelling caused by temperature variations.}
  \label{tab:deltaT}
  \begin{tabular}{lrrrr|rrr}
  \hline\hline
  \multirow{2}{*}{Star} & \cla{\multirow{2}{*}{$T_\mathrm{eff}$ (K)}} & \multicolumn{3}{c}{Change in $T_{\mathrm{eff}}$}  & \multicolumn{3}{c}{Magnetic spots} \\
   & & $\Delta T_{\mathrm{eff}}$ (K) & $\Delta \langle B\rangle_{I}$ (G) & $\Delta \varepsilon_{\mathrm{Fe}}$ & $\Delta T_{\rm spot}$ (K) & $\Delta \langle B\rangle_{I}$ (G) & $\Delta \varepsilon_{\mathrm{Fe}}$\\
  \hline
  \multirow{2}{*}{HD 1835} & \cla{\multirow{2}{*}{5837}} & $+50$ & $-0.29$ & $0.023$ & $+200$ & $8.52$ & $0.016$\\
   & & $-50$ & $0.77$ & $-0.023$ & $-200$ & $-11.56$ & $-0.019$ \\
   \multirow{2}{*}{HD 9986} & \cla{\multirow{2}{*}{5805}} & $+50$ & $20.53$ & $0.021$ & $+200$ & $-7.64$ & $0.006$ \\
   & & $-50$ & $-7.42$ & $-0.021$ & $-200$ & $41.15$ & $0.041$ \\
  \multirow{2}{*}{HD 22049} & \cla{\multirow{2}{*}{5146}} & $+50$ & $8.46$ & $0.015$ & $+200$ & $-1.76$ & $0.008$\\
   & & $-50$ & $-7.66$ & $-0.014$ & $-200$ & $-1.02$ & $-0.006$  \\
  \multirow{2}{*}{HD 75732} & \cla{\multirow{2}{*}{5235}} & $+50$ & $0.50$ & $0.016$ & $+200$ & $-24.60$ & $0.070$ \\
  & & $-50$ & $-1.13$ & $-0.016$ & $-200$ & $-13.90$ & $0.000$  \\
  \hline
  \end{tabular}
  \tablefoot{Left half of the table corresponds to a change in $T_{\mathrm{eff}}$ by $\pm50$\,K. Right half shows the impact of the assumption that the magnetically active regions correlate with temperature inhomogeneities on the stellar surface by considering a different temperature in the magnetic and non-magnetic areas. 
  }
\end{table*}

In this study we assume a fixed stellar temperature and do not allow any variations of $T_{\mathrm{eff}}$ during the inference. This is commonly done in magnetic field studies due to the limited number of studied lines making it difficult to fit multiple parameters that affect the line depth. Any systematic variation in depth is already accounted for by the element abundance parameter. However, there can be second-order effects such as differential line strength changes of lines with different temperature sensitivity. Therefore, it is still worthwhile to assess how a change in temperature modifies our results. For this test we selected a subsample of stars spanning our temperature range as well as measured magnetic field strength, HD 1835, HD 9986, HD 22049, and HD 75732. We then modified adopted effective temperature of these stars by $\pm50$\,K, which is around the level of commonly reported observational errors of $T_{\mathrm{eff}}$ \citep[e.g.][]{valenti:2005}. 

The difference in magnetic field measurements $\langle B\rangle_{I}$, as well as the change in Fe abundance caused by shifting $T_{\mathrm{eff}}$ by 50\,K, can be seen in Table \ref{tab:deltaT}. It appears that there is some difference in sensitivity of the magnetic inference results to temperature errors depending on the temperatures of the star. The magnetic field of HD 22049, which is the coolest star in our sample, has a stronger response than the field of one of the hottest; HD 1835. This could be either due to the fact that the studied lines become more temperature sensitive at lower temperatures or due to the increased strength of spectral lines at lower temperatures causing increased blends in the wings of the studied lines which impacts interpretation of the Zeeman broadening signal. 
This is likely a fairly sharp threshold as HD 75732, slightly hotter than HD 22049, also shows a weak response to variations in $T_{\mathrm{eff}}$.
The weakly active star HD 9986 shows a larger change for $\langle B\rangle_{I}$ compared to the other stars, this illustrate larger uncertainties of the magnetic inference for these objects when the magnetic signal is very weak.
Even if there is a large variation in sensitivity to $T_{\mathrm{eff}}$, the differences does not exceed the 68~\% credence regions shown in Table~\ref{tab:results} in all cases but HD 22049 which have differences slightly above this level.
Overall, this analysis shows that temperature errors can modify the measured mean field strength by up to 50\% for weakly active stars with $\langle B\rangle_{I}<100$~G but by just a few \% for more active stars with several hundred G mean fields.

Another form of potential temperature sensitivity comes from the possibility that the magnetic field is correlated with temperature inhomogeneities on the stellar surface.
While particularly strong magnetic fields are believed to coincide with cool starspots, it is also possible that the weaker background fields affect the photospheric temperature as well, albeit with a smaller effect.
This could result in an over- or underestimation of the covering fraction of the stellar surface as hotter regions will contribute more to the overall flux. In order to test this we used the same approach, except that when combining spectra corresponding to different temperatures, we cannot use continuum normalised spectra as done in Eq. (\ref{eq:combination}). Instead, we combine the synthetic line and continuum flux spectra from the magnetic and non-magnetic regions to obtain multi-component normalised spectrum,
\begin{equation}
\label{eq:Flux}
  S = \frac{S^{(l)}_{B}f_{B} + S^{(l)}_{0}(1-f_{B})}{S^{(c)}_{B}f_{B} + S^{(c)}_{0}(1-f_{B})}.
\end{equation}
$S^{(l)}$ represents the synthetic flux profiles and $S^{(c)}$ represents the corresponding continuum level. In this case we test the outcome of a magnetic region temperature difference of $\pm200$ K for the same stars as studied in the test of uniform temperature variation.
The result can be seen in the rightmost columns of Table \ref{tab:deltaT}. The magnetic field variation appears to be stronger when introducing temperature inhomogeneities rather than shifting $T_{\mathrm{eff}}$, with the only exception being HD 22049 where the magnetic variation seems to be on a lower level. HD 9986 has the strongest change in magnetic field strength of 41.15~G, which corresponds to a field strength about a factor of 2 compared to its measured strength. While still within the uncertainties of the measurement, it further shows the sensitivity of the magnetic field measurement to assumed parameters of the weakly active stars.

The abundance changes are more varied in the latter experiment than in the uniform $T_{\mathrm{eff}}$ variation case, but the abundance response does comparatively appear to be systematically weaker for the spot model. It also seems that for a colder star, e.g. HD 75732, the response is stronger for increasing temperature while for a hotter star, e.g. HD 9986, the response is strongest towards decreasing temperature. An explanation for this could be that weak blends in the wings of the lines start to become more prominent \cla{at temperatures below 5250\,K. Shifting} the spot temperature towards this threshold could play a role in this change. For the more active stars in our test sample, the more rapidly rotating star, HD 1835, shows a similar abundance shift as in the $T_{\mathrm{eff}}$ case while the more slowly rotating HD 22049 exhibits a smaller change in abundance. When comparing the changes in abundance to the uncertainties of our \cla{inference} (see \cla{Table~\ref{tab:results}}) the abundance changes significantly exceed the uncertainties. This is due to the fact that temperature and abundance both affect the depth of lines, changing one will inevitably force the other to change as well. 
In any case, since our work only considers 6 lines -- fewer than is typical for stellar abundance analysis -- we treat abundance as a nuisance parameter rather than using it for meaningful insight into stellar chemistry.

While a uniform $T_{\mathrm{eff}}$ change and temperature inhomogeneities do introduce variations that are comparable or lower to the 
uncertainty of the magnetic parameters in the inference
for the stars in our sample, they do not hugely change the magnetic properties derived for the more active stars. When considering the impact on the continuum flux for different temperatures, we see that a change of 200\,K modifies the 
continuum level by about 4\,\% in the H-band. 
\cla{This means that inhomogeneities in temperature, where the continuum level will affect the relative contribution of different regions, should not play a significant role in measuring strong magnetic field signals.}
%For this reason it makes sense that a strong magnetic field signal would be only marginally affected by a temperature change. 
More care should be taken for the weakly active part of our sample. For example, HD 75732 and HD 9986 show $\langle B\rangle_{I}$ changes of above 20\,G, which corresponds to 50\% or more of their average magnetic field strengths. For our weakly active targets this indicates that the derived mean magnetic field estimates are likely affected by a significant amount of systematics, such as uncertainties in temperature and other parameters.

\subsubsection{Impact of different modelling approaches}

\label{subsec:other_assumtions}

Both the large- and small-scale magnetic field of HD 22049 
was recently investigated by \cite{petit:2021} using SPIRou NIR spectra. While using a similar set of \ion{Fe}{i} lines as in our study, they made different assumptions in the modelling of the magnetic fields in stellar spectra. They did not consider potential NLTE effects and used the standard van der Waals broadening coefficients from VALD for the lines used for their magnetic study. Their astrophysical $\log{gf}$-values also differed from ours. Even with these differences in assumptions, the obtained magnetic field strength for HD 22049 in both studies are consistent with each other \citep[$209\pm7$ and $237\pm36$~G from this work and][respectively]{petit:2021}. While we showed that NLTE effects would play a minimal role in stars with temperatures similar to HD 22049, the similarity between our results and the work by \cite{petit:2021} also shows that the Zeeman broadening signal is not particularly sensitive to the other assumptions in the spectroscopic modelling, including the selection of \ion{Fe}{i} lines used for the inference. 

\cla{This is further supported by performing the magnetic field inference assuming LTE, which yields no significant change in the results for HD 22049. The difference for stars with similar parameters to the Sun is also negligible, although this is likely because correcting $\log gf$-values under LTE or NLTE to solar observations cancels out most effects at these temperatures. For the hottest star in our sample, HD 179949, the difference between LTE and NLTE results appears larger at $\sim$\,15~G, although still within reported uncertainties. This difference is higher than the variation caused by changing the effective temperature for the more active stars in our sample. The difference is also larger than the uncertainties of many of our other results. 
This suggests that NLTE could become significant for more slowly rotating late-F stars. We also note that the discrepancy in measured magnetic field strengths increases if $\log gf$-values determined under NLTE is used for LTE spectrum synthesis and vice versa. In this case, the results for HD 179949 are deviating significantly, which indicates that while $\log gf$-corrections can somewhat mimic NLTE effects, consistent treatment of NLTE is important.}

\begin{figure}
    \centering
    \includegraphics[width=\linewidth]{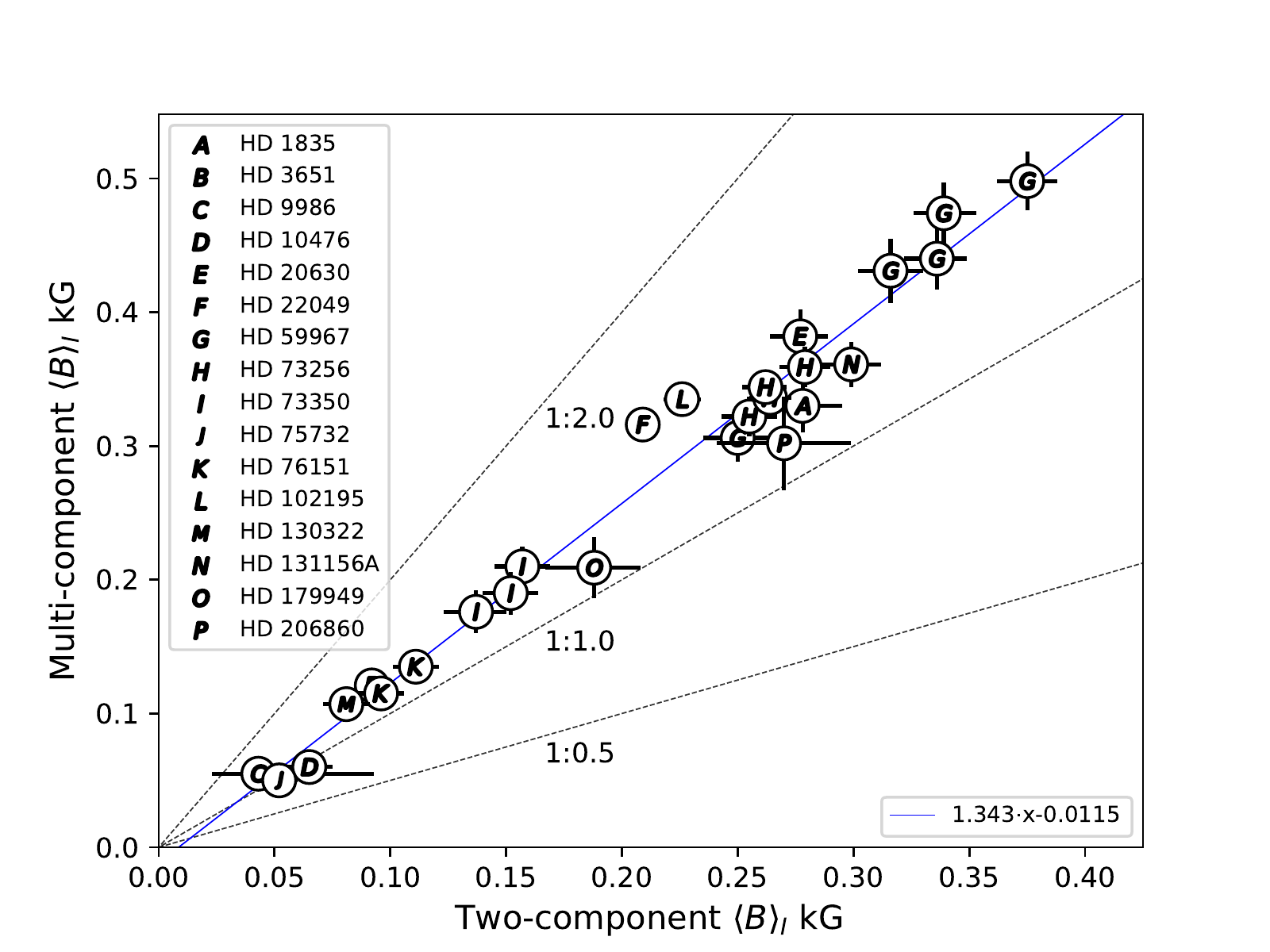}
    \caption{Comparison between the magnetic field values obtained using different model prescriptions of Eq. (\ref{eq:combination}) and (\ref{eq:multicomp}). The solid blue line represents a linear fit to the data. }
    \label{fig:model_comp}
\end{figure}

Another aspect that affects the magnetic inference results is the parameterisation of the magnetic field strength distribution adopted in the model. Since we have used both the two- and multi-component descriptions to determine mean magnetic fields, comparing the different approaches is possible. The relationship between the two models can be seen in Fig.~\ref{fig:model_comp}. We find a linear trend with a slope of $\sim$1.34 indicating a systematic difference between the two model approaches. The multi-component approach gives a systematically higher magnetic field strength compared to the two-component model. It also appears that including more magnetic components increases this discrepancy, e.g. HD\,22049 and HD\,102195 that have moderate magnetic field strengths but benefit from adding a third component both lie above the trend in Fig.~\ref{fig:model_comp} (marked as \textbf{\textit{F}} and \textbf{\textit{L}}). It also appears that the more rapidly rotating stars in our sample, such as HD 179949 and HD 206860 (\textbf{\textit{O}} and \textbf{\textit{P}}) have smaller deviations between the two models. This indicates that the difficulty of describing the magnetic field as a single value becomes less acute for more rapidly rotating stars as the Zeeman broadening signal becomes 
hidden in the rotational Doppler broadening.

The resulting change of $\langle B\rangle_{I}$ exceeds both the uncertainties as well as any temperature variations discussed in Sect. \ref{subsec:Teff} for all but the most weakly active stars in our sample. This highlights the importance of selecting a magnetic field parameterisation that can describe the entire range of the magnetic field strength on the stellar surface. At the same time, one has to be careful when including strong fields that do not significantly improve the model fit since a small fraction of strong fields can cause a major contribution to the average magnetic field strength. 

\subsection{Magnetic fields at different spatial scales}
\label{sec:scale-comp}

\begin{figure}
\centering
\includegraphics[width=\linewidth]{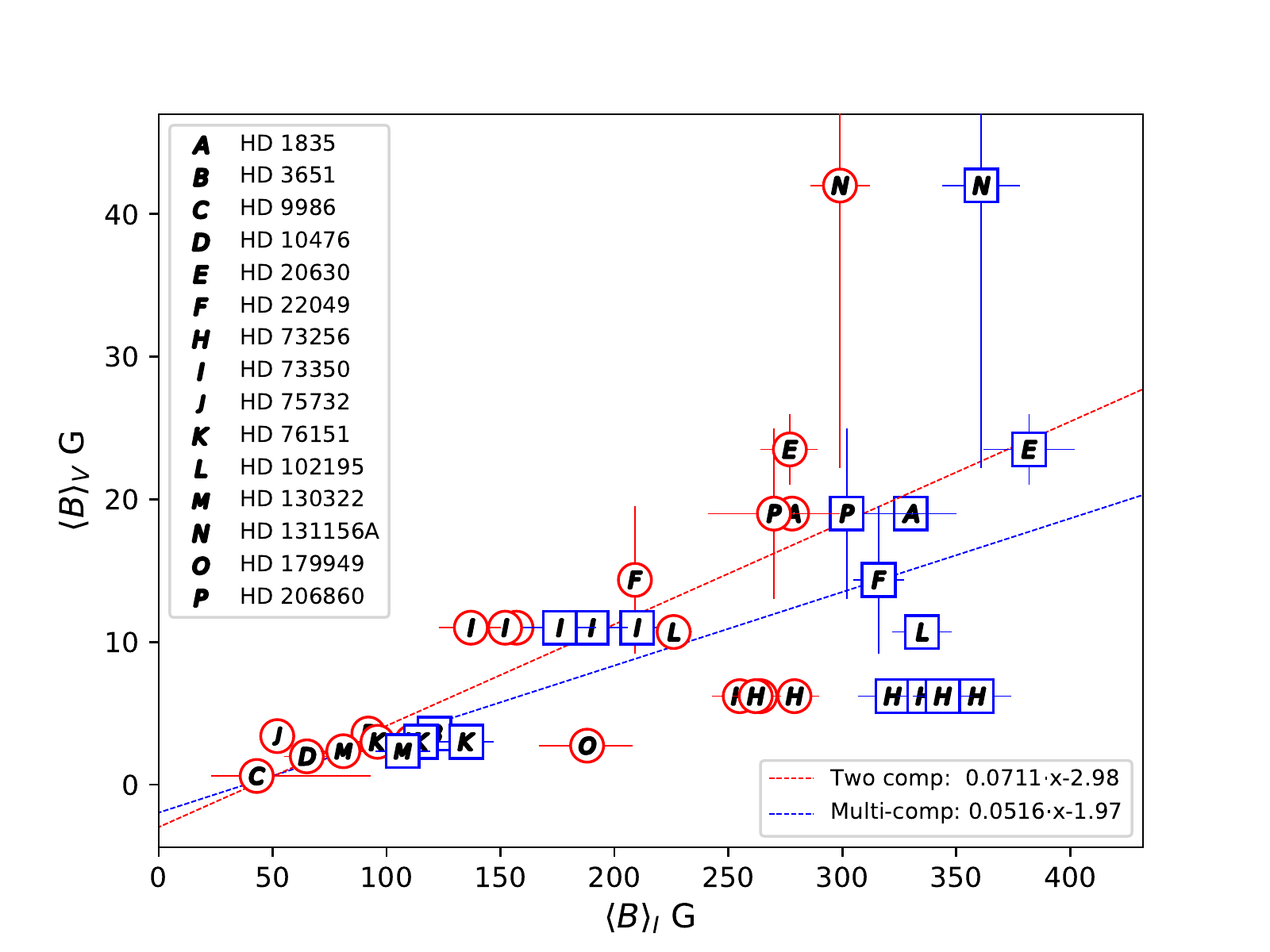}
\caption{Mean strengths of the large- and small-scale fields from Table \ref{tab:targets} and \ref{tab:results} respectively. Red circles represents the two-component results. Blue squares represent multi-component results that are significantly different compared to their two-component counterpart. Dashed lines show linear fits, red for the two-component model and blue for the multi-component model. Uncertainties in $\langle B\rangle_V$ represent scatter in the ZDI results corresponding to different observing epochs. Stellar labels are the same as in Fig.~\ref{fig:model_comp}}
\label{fig:small-vs-large}
\end{figure}

The large-scale magnetic fields reconstructed from polarimetry and the small-scale fields measured from unpolarised spectra are known to be vastly different in strength \citep[see e.g.][]{see:2019,kochukhov:2020a}. For Sun-like stars in particular, the average large-scale field strengths recovered by ZDI are generally only about 10\% or less of the small-scale magnetic field strength. We find this to hold for the stars in our sample when comparing our NIR magnetic measurements with the optical ZDI studies (see Table~\ref{tab:targets} for references). This comparison is shown in Fig.~\ref{fig:small-vs-large}.
Despite different field strengths measured by the two techniques, there seems to be a loose correlation between our results and mean ZDI global field strength. This suggests a certain degree of relative consistency between the magnetic fields measured at different spatial scales: a star with a stronger global field is more likely to exhibit a stronger small-scale field as well. While we observe a significant spread in the relationship between small- and large-scale fields, the results support the discovery from \cite{vidotto:2014} that found similar trends between both small- and large-scale fields with the rotation period and Rossby number, indicating a coupling between the two spatial scales.

As shown in Sect.~\ref{subsec:other_assumtions}, the multi-component model produces stronger fields for all but the most weakly active stars in our sample. As a result, the trend between small- and large-scale magnetic fields becomes less steep for the multi-component approach, as can be seen in  Fig.~\ref{fig:small-vs-large}.

\subsection{Small-scale magnetic fields at different wavelengths}
\label{sec:wl-comp}

\begin{figure}
\centering
\includegraphics[width=\linewidth]{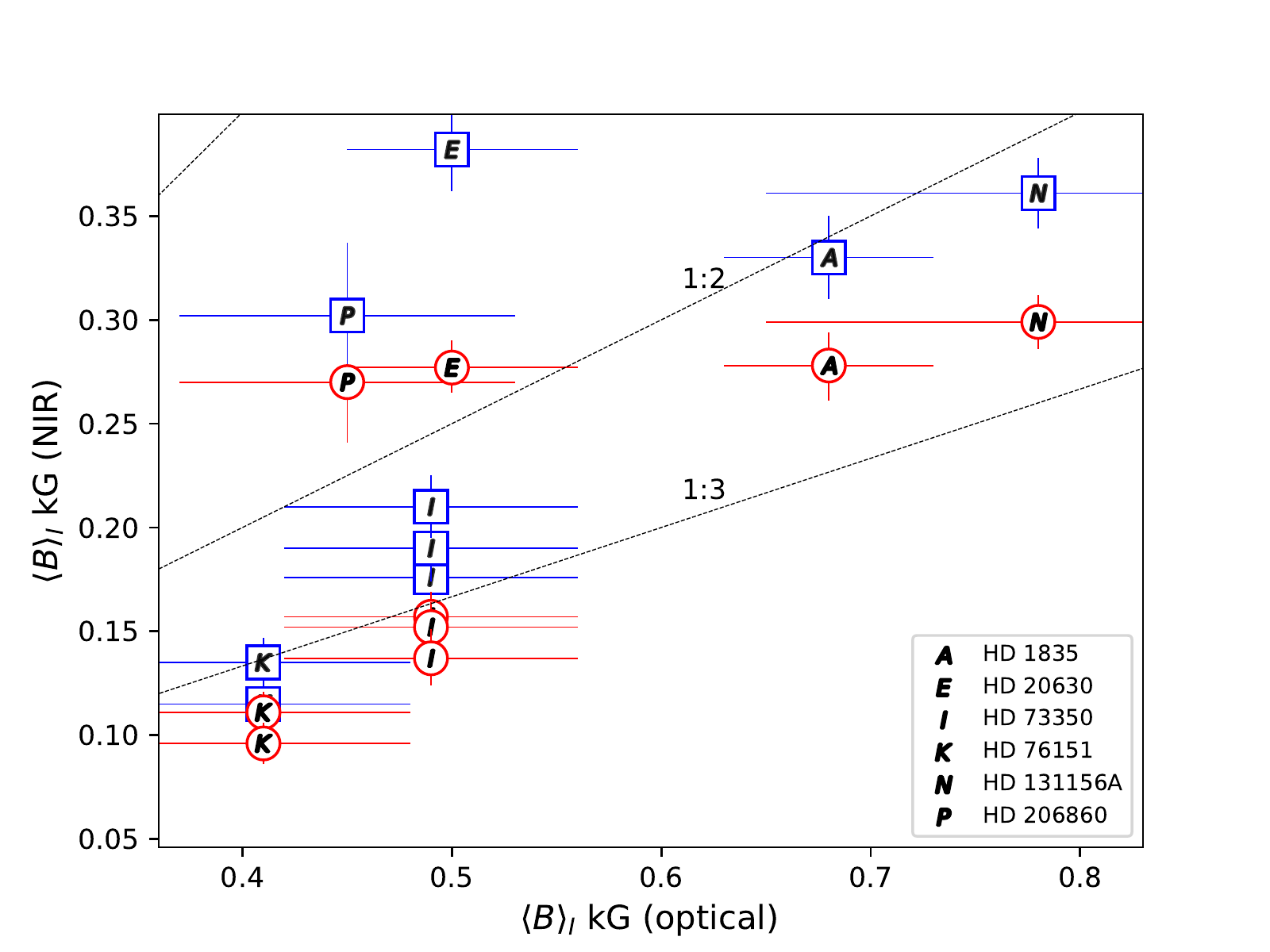}
\caption{Relationship between the small-scale magnetic field strength measured at different wavelengths. The y-axis corresponds to the average magnetic field strength obtained in this study while the x-axis shows the optical measurements taken from \cite{kochukhov:2020a}. Black dotted lines represent fractions between the two results. Stellar labels are the same as in Fig. \ref{fig:model_comp} and \ref{fig:small-vs-large}.}
\label{fig:optical-vs-NIR}
\end{figure}

With the Zeeman broadening becoming a less readily observable signature to detect the presence of magnetic fields at shorter wavelengths, some studies in the optical relied on the Zeeman intensification of spectral lines. While this provides certain advantages, such as a reduced sensitivity to rotation and the possibility to observe small-scale magnetic fields at shorter wavelengths, it also causes the magnetic signal to be more degenerate with other stellar parameters, such as chemical abundance. It is therefore worthwhile to compare results obtained in the two different wavelength ranges to find any systematic differences. Figure \ref{fig:optical-vs-NIR} shows the overlapping sample between this study and the optical study by \cite{kochukhov:2020a}. What can be seen from these results is that the observations at NIR wavelengths produce systematically lower magnetic field strengths compared to the optical, typically by a factor of $\sim2$--$3$. 

Our investigation into the possible impact of effective temperature errors and temperature inhomogeneities described in Sect. \ref{subsec:Teff} showed that these effects do not produce a sufficient variation in the average magnetic field strength to account for the discrepancy. The change in continuum flux in the optical due to a temperature variation of $\pm50$~K is, however, larger, compared to the $<$\,5\,\% in the H-band. Specifically, around 5500\,\AA\, the change in the
continuum flux is about 12\,\%. While this indicates that the magnetic field measurements are more sensitive to temperature at shorter wavelengths, this effect is unlikely to be sufficient to explain the factor of 3 difference in the magnetic field strength inferred for some of the stars. Other aspects, such as variation in microturbulence can also influence the results, but do not seem to change the magnetic field strength by much more than $\sim$\,10\,\% according to previous investigations \citep[e.g.][]{kochukhov:2020a, hahlin:2022}. What can be seen from \cite{kochukhov:2020a} is that the magnetic field strength $B$ is significantly higher than in the present study while the filling factors $f_B$ appear to be more similar. 
For stars present in both samples, the optical filling factors are larger by about 1.7 compared to the NIR measurements.
At the same time, the magnetic field strength measured in the optical reaches to over 3\,kG for most stars, twice the strength compared to the $\sim1.5$\,kG typical of our NIR results.

Lacking additional information on the line profile shape distortion caused by the Zeeman effect, it is possible that the optical results based on Zeeman intensification systematically overestimate magnetic field strength. This issue might be particularly significant for stars with a lower activity as the effect of magnetic field on stellar spectra is quite small even for the very Zeeman-sensitive lines. None of the more active stars from the sample studied by \cite{kochukhov:2020a} was included in the NIR sample observed with CRIRES$^+$. Observations of these more active stars in the NIR would be useful to investigate if this systematic difference persists towards stronger magnetic fields or is only present for weakly active stars. However, more active stars tend to be fast rotators and the technique used here will be more difficult to apply to stars with a $v\sin i$ significantly exceeding $\approx$\,10~km\,s$^{-1}$ as the Zeeman splitting is typically of the order of a few km\,s$^{-1}$ in the H band for moderately active stars.

A possible temporal evolution of the magnetic field in the course of stellar activity cycles might give rise to the observed discrepancies. We do see some variation in our sample and \cite{kochukhov:2020a} also reported trends over time for some of the stars. It is, however, highly unlikely that this would cause a systematic reduction in the measured magnetic field strength for all stars in our sample. If time evolution played a major role, we would expect at least a few stars in our sample showing magnetic fields on a similar level or above the measured optical magnetic field strengths.

\begin{figure}
\includegraphics[width=\linewidth]{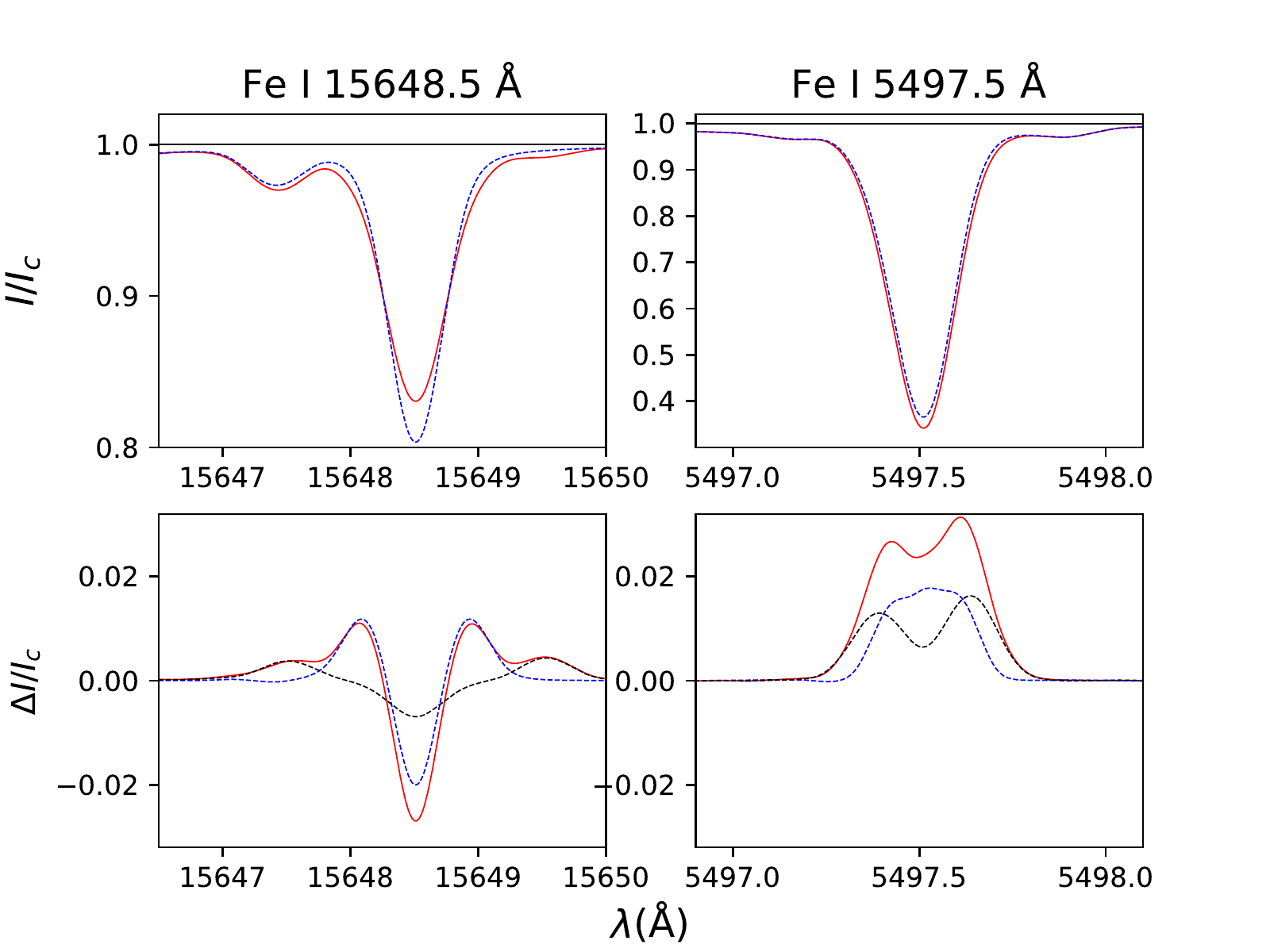}
\caption{Impact of a magnetic field distribution with multiple components on the theoretical profiles of the magnetically sensitive lines \ion{Fe}{I} 15648.5 \AA\ (this study) and \ion{Fe}{I} 5497.5 \AA\ \citep{kochukhov:2020a} for a star with solar parameters rotating with a $v\sin i$ of 5\,km\,s$^{-1}$. In this case, the magnetic field distribution is composed of 0, 1, and 3\,kG fields covering 75, 20, and 5\,\% of the stellar surface respectively. \textit{Top:} non-magnetic spectra in dashed blue and the combined magnetic spectra in red. \textit{Bottom:} difference between the magnetic and non-magnetic spectra in red, contribution to this difference by the 1 and 3\,kG field is shown in dashed blue and dashed black respectively.}
\label{fig:high_sens}
\end{figure}

Interestingly, observations of the Sun aiming to characterise magnetism of the quiet photosphere have encountered a similar issue with discrepant fields found at different wavelengths \citep[see review by][]{bellotrubio:2019}. It appears that when observing the quiet Sun in the NIR, the fields are weaker, $\sim$\,100~G fields are observed compared to kG fields when observing in the optical. One idea to interpret this discrepancy is that the quadratic wavelength dependence of the Zeeman splitting makes the NIR lines much more sensitive to weak fields compared to the optical lines, which are primarily sensitive to strong fields. 
While the technique used in these solar studies relies on circular polarisation, it is possible that similar effects are causing the discrepancy in the analyses of unpolarised spectra. 

We investigated the sensitivity of the optical and NIR spectra to a multi-component magnetic field strength distribution by calculating synthetic spectra and examining the resulting line profile change relative to the non-magnetic model. This was done for the most magnetically sensitive line \ion{Fe}{i} 5497.5~\AA\ employed by \cite{kochukhov:2020a} and for the \ion{Fe}{i} 15648.5~\AA\ line used in this work. An example of our calculations can be seen in Fig. \ref{fig:high_sens}, where spectral contributions of different field strength components are shown. What can be seen from this plot is that the stronger field components are well separated from the line centre for the NIR line, which is not the case for the optical line. 
This complicates analysis of NIR lines as the magnetic signal becomes more spread out in the line profile compared to their shorter wavelength counterparts.
Besides highlighting the importance of including lines with different sensitivities, as less sensitive NIR lines would be less affected by this issue, it also demonstrates a limitation of the two-component model approach if magnetic field strength on the stellar surface varies significantly. The consequence of this is that, when applied to NIR spectra, the two-component approach might suffer from some underestimation of the magnetic signal. This is seen from Fig.~\ref{fig:model_comp}, where the multi-component model produces systematically stronger average fields. While the multi-component approach yields results closer to the optical field strength measurements for all stars in the overlapping sample, it is not sufficient to reach an agreement. Therefore, the magnetic model prescription is unlikely to be solely responsible for the optical vs. NIR field strength discrepancy.

Finally, it cannot be excluded that vastly different formation heights of the optical and NIR diagnostic lines are, at least partially, responsible for the observed field strength discrepancy. Calculating the depth of formation according to \citet{Achmad:1991}, we found that the core of the \ion{Fe}{i} 15648.5~\AA\ line is formed at $\log \tau_{5000}\approx-1$ whereas the core of the \ion{Fe}{i} 5497.5~\AA\ line reaches to over $\log \tau_{5000}\approx-4$, primarily due to its much larger equivalent width. The lines will therefore sample a different magnetic field structure due to this formation height difference. Solar studies \citep[e.g.][]{Morosin:2020} suggest that magnetic fields outside major active regions occupy a smaller area at lower heights. Higher up, the field lines expand, occupying a larger area and forming a canopy. It is possible that optical lines reach into this canopy, revealing the full range of fields present at the stellar surface. In contrast, the stronger low filling factor fields in deeper layers of the atmosphere might partially escape detection in the NIR lines due to the difficulties mentioned above and because these fields might be associated with much cooler regions. 

\section{Conclusions}
We have investigated the small-scale magnetic field of 16 Sun-like stars by measuring Zeeman broadening within six \ion{Fe}{I} lines in the H-band of CRIRES$^+$.
CRIRES$^+$ has great potential for the investigation of magnetic properties of cool stars. With this instrument it becomes possible to measure magnetic fields on weakly active stars. Thanks to the very high spectral resolution reached by CRIRES$^+$ and a high sensitivity to Zeeman broadening in the NIR, even small changes of the line shapes can be reliably detected. When using single observations with S/N in the range of 100--300 we are able to detect average magnetic field strengths on the order of $\sim$\,100~G without a significant sensitivity to uncertainties in stellar temperature. This limit could likely be improved by using higher S/N spectra, either with longer observations or by taking advantage of the intensity spectra associated with high-quality spectropolarimetric observations employed to study large-scale stellar magnetic fields. 

Comparing with other recent NIR magnetic field measurements, our findings are compatible with the results of the analysis of HD 22049 \cite{petit:2021} using SPIRou 
observations. This shows that the magnetic broadening signal extracted from NIR spectra is not particularly sensitive to assumptions in the spectroscopic modelling, as we used both NLTE and different van der Waals broadening parameters but still obtained a similar average magnetic field strength. 
The use of NLTE is also uncommon in studies of stellar magnetic fields and given that NLTE can have significant impact on \ion{Fe}{I} lines, this omission could become problematic as NLTE becomes more common in the analysis of stars. 

We did, however, find a significant dependence of the results on the type of model prescription used to describe the magnetic field strength distribution. For some stars, a multi-component model with magnetic field strengths of 0, 1, 2, and 3\,kG produced significantly stronger mean field strengths compared to the two-component model. This highlights a shortcoming of the two-component model's ability to describe magnetic field distributions with a large range of field strengths. \cla{In a realistic environment it is unlikely that the field strength on a stellar surface will have a single homogeneous value. Particularly when magnetic broadening is more significant, such as in the NIR, this assumption is likely to break down.} In addition, \cla{when using the multi-component approach,} it is still important to be careful when increasing the number of filling factors as small contributions from strong fields could significantly change the average magnetic field strength without having a large impact on the overall fit quality. 
As the difference between the two- and multi-component models are significant, comparison using results obtained with the different models 
should be made with caution if it has not been shown that the methods produce similar results for the stars in question. \cla{In general it does seem that the discrepancy of the two methods is reduced for more rapid rotation, due to the line shape information being smeared out by the rotational broadening. This indicates that the two-component model is indistinguishable from a multi-component model 
as long as the magnetic broadening is significantly smaller than other sources of line broadening.}

When comparing with the small-scale field measurement results obtained in the optical for the same stars, a systematic difference is evident. The optical measurements obtained by \cite{kochukhov:2020a} using Zeeman intensification inferred stronger magnetic fields by a factor of $\sim2$--$3$ when comparing with our NIR results. This difference can not be explained by systematics related to stellar parameters as this typically affects the results by $\lesssim10$\,\%. Other possible sources of this discrepancy were discussed in Sect. \ref{sec:wl-comp}. The optical vs. NIR field strength discrepancy,  previously discussed in the context of the Sun  \citep{bellotrubio:2019}, highlights potential problems when trying to compare magnetic field measurements obtained at different wavelengths. Similar comparisons should also be carried out for more active stars in order to investigate if this discrepancy extends to higher field strengths or is only present for weakly active stars such as those studied here. This is important to investigate as the persistent discrepancy between Zeeman intensification and broadening could complicate magnetic diagnostic of rapidly rotating stars for which only intensification measurements can be done.

\begin{acknowledgements}
A.H. and O.K. acknowledge support by the Swedish Research Council (project 2019-03548).

D.S. acknowledges funding from project PID2021-126365NB-C21(MCI/AEI/FEDER, UE) and financial support from the grant CEX2021-001131-S funded by MCIN/AEI/ 10.13039/501100011033 

Based on observations collected at the European Organisation for Astronomical Research in the Southern Hemisphere under ESO programme 0108.D-0659(A).

CRIRES+ is an ESO upgrade project carried out by Th\"uringer Landessternwarte Tautenburg, Georg-August Universit\"at G\"ottingen, and Uppsala University. The project is funded by the Federal Ministry of Education and Research (Germany) through Grants 05A11MG3, 05A14MG4, 05A17MG2 and the Knut and Alice Wallenberg Foundation.

Analysis of this work has also used {\tt matplotlib} \citep{hunter:2007}, {\tt numpy} \citep{harris:2020}, and {\tt astropy} \citep{astropycollaboration:2013}

\end{acknowledgements}

%\bibliographystyle{aa}
%\bibliography{references}

%APPENDIX
\onecolumn
\begin{appendix}
\section{Observation log}
\begin{table}[h!]
\centering
\caption{Observations used in this work.}
\label{tab:obslog}
\begin{tabular}{lcccc}
\hline\hline
Object &  Night &  MJD &  Exposure time (s) &  SNR \\
\hline
HD 1835 & 2021-10-30 & 59518.1183 & 40.0 & 270\\
HD 3651 & 2021-10-30 & 59518.1295 & 20.0 & 312\\
HD 9986 & 2021-10-30 & 59518.1562 & 60.0 & 274\\
HD 10476 & 2021-10-30 & 59518.1414 & 10.0 & 296\\
HD 20630 & 2022-02-07 & 59618.0550 & 20.0 & 298\\
HD 22049 & 2022-02-09 & 59620.0208 & 4.0 & 336\\
HD 59967 & 2021-10-30 & 59518.2675 & 60.0 & 293\\
HD 59967 & 2022-02-09 & 59620.0412 & 60.0 & 278\\
HD 59967 & 2022-03-23 & 59662.0063 & 60.0 & 278\\
HD 59967 & 2022-03-24 & 59662.9939 & 60.0 & 272\\
HD 59967 & 2022-03-30 & 59668.9914 & 60.0 & 287\\
HD 73256 & 2021-10-08 & 59496.3658 & 90.0 & 155\\
HD 73256 & 2021-10-30 & 59518.3290 & 90.0 & 149\\
HD 73256 & 2022-02-07 & 59618.0678 & 90.0 & 250\\
HD 73256 & 2022-03-30 & 59669.0666 & 90.0 & 243\\
HD 73350 & 2022-03-31 & 59669.0860 & 60.0 & 264\\
HD 73350 & 2022-02-09 & 59620.0543 & 60.0 & 274\\
HD 73350 & 2022-03-10 & 59648.9979 & 60.0 & 264\\
HD 75732 & 2022-03-23 & 59662.0751 & 30.0 & 285\\
HD 76151 & 2022-03-30 & 59669.1026 & 40.0 & 260\\
HD 76151 & 2022-02-07 & 59618.2143 & 40.0 & 258\\
HD 102195 & 2022-03-11 & 59650.2736 & 90.0 & 232\\
HD 130322 & 2022-03-23 & 59662.3306 & 90.0 & 223\\
HD 131156A & 2022-03-23 & 59662.3453 & 4.0 & 262\\
HD 179949 & 2021-10-08 & 59495.9816 & 120.0 & 284\\
HD 206860 & 2021-10-28 & 59516.0424 & 30.0 & 271\\

\hline
\end{tabular}
\tablefoot{From left to right: Target name, date at the start of the observation night (relates to magnetic field results in Table \ref{tab:results}), modified Julian date(MJD) in the middle of the observation, exposure times , median SNR per pixel.}
\end{table}
\section{Model spectral fits and posterior distributions}
\label{sec:complete_fits}
The following section includes the best fits and resulting posterior distributions obtained from the MCMC inference described in Sect. \ref{sec:inference} for the remaining stars.

\subsection{HD 1835}
\vspace{-\abovedisplayskip}
\begin{minipage}{1.0\textwidth}
  \centering
  \includegraphics[width=\textwidth]{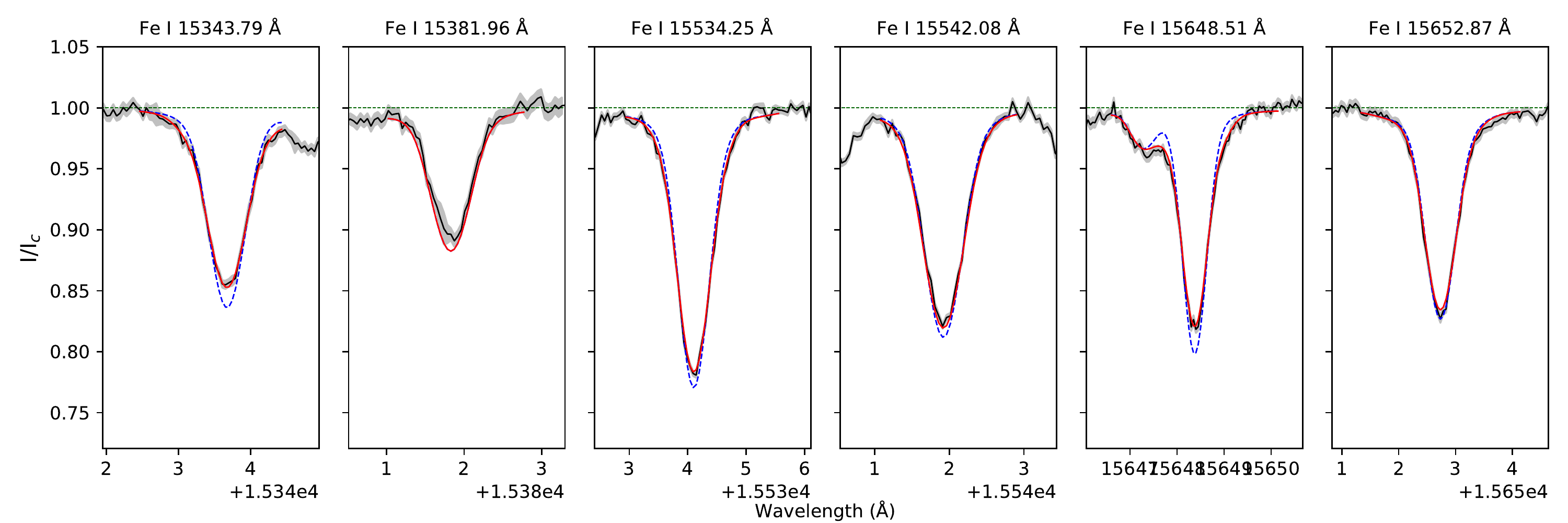}
  
  \includegraphics[width=0.49\textwidth]{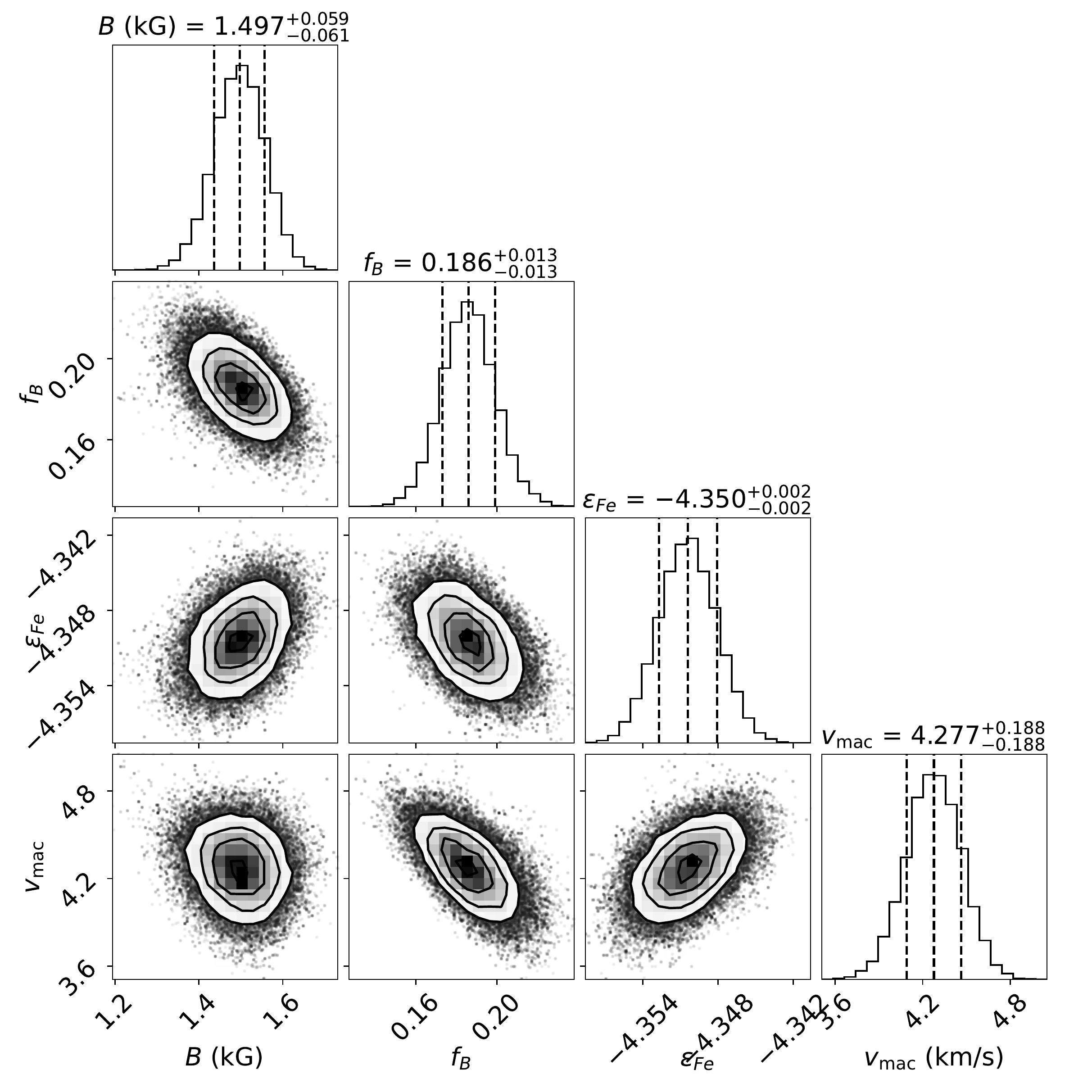}
  \includegraphics[width=0.49\textwidth]{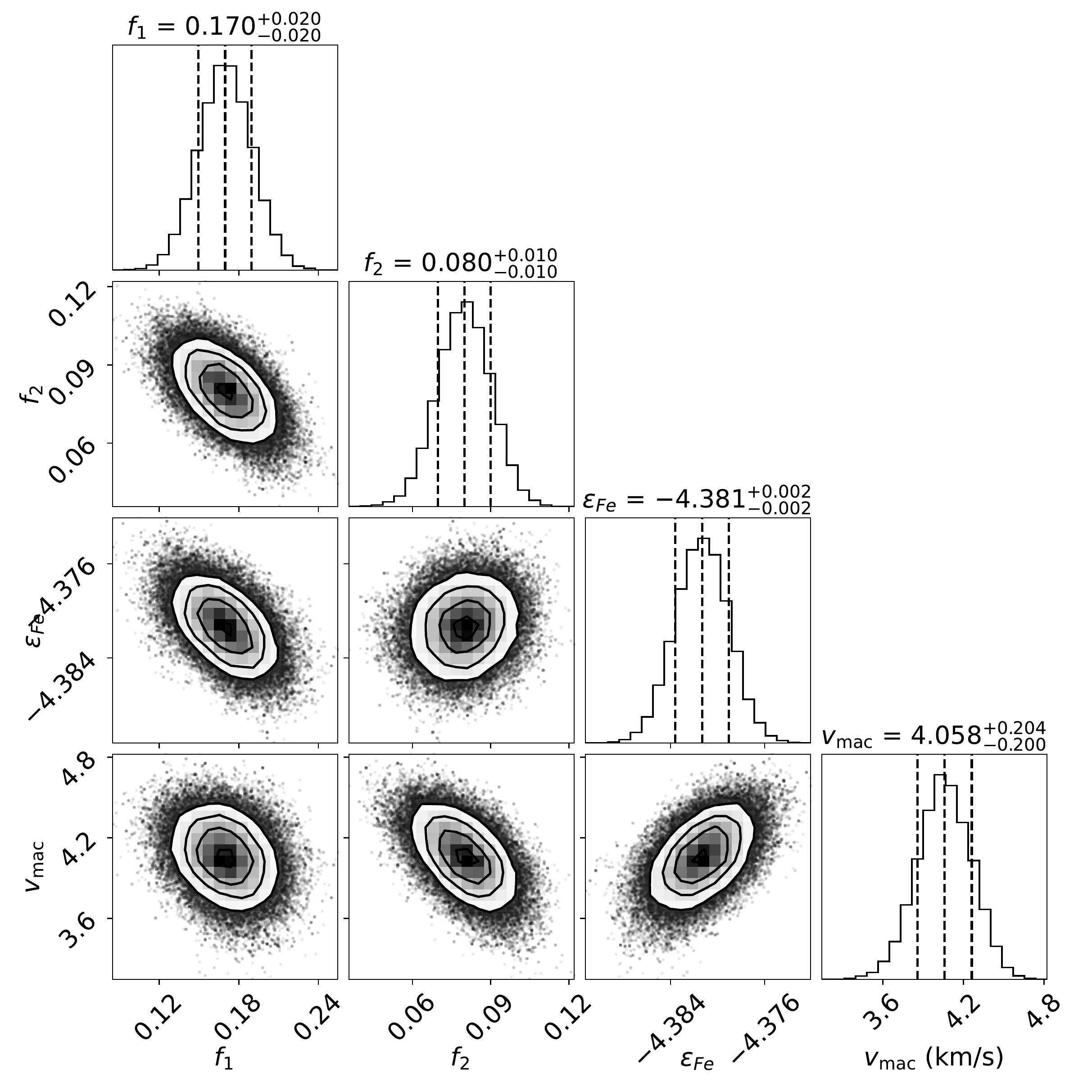}
  \includegraphics[width=0.45\textwidth]{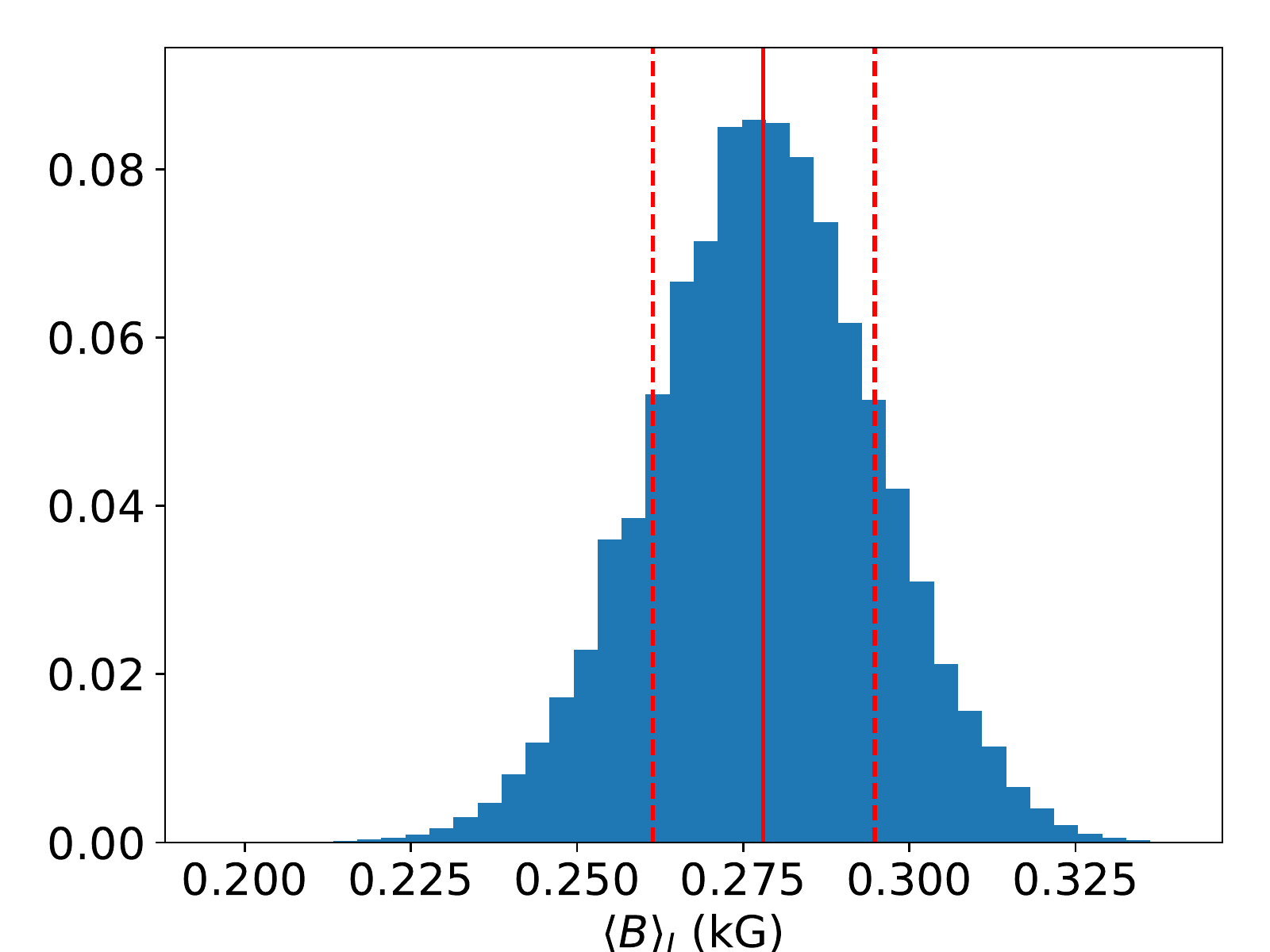}
  \includegraphics[width=0.45\textwidth]{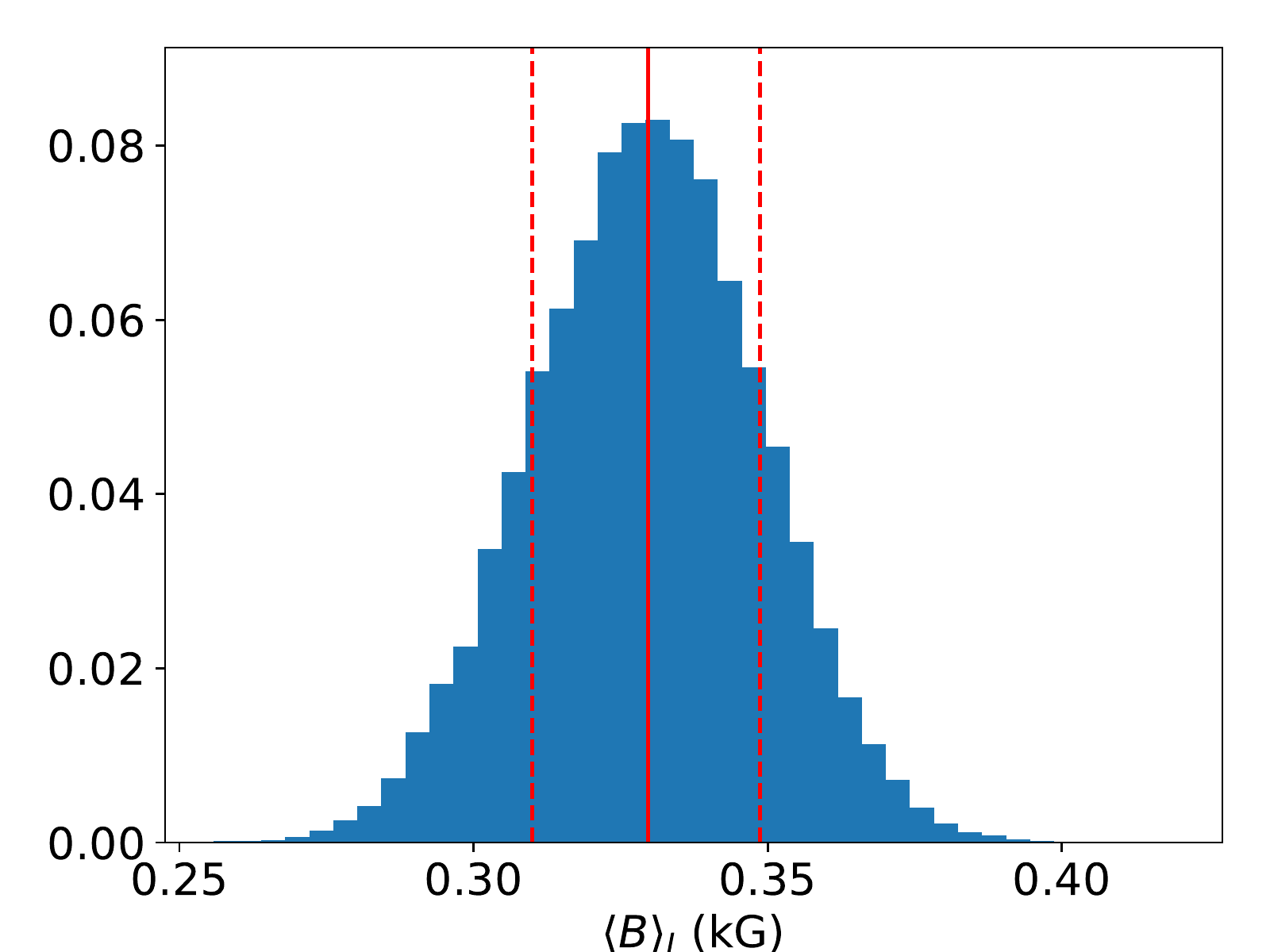}
  \captionof{figure}{Same as Fig. \ref{fig:mcmc_fit} and \ref{fig:mcmc_corner} but for HD 1835. \cla{\textit{Left.} Two-component results. \textit{Right.} Multi-component results.}}
  \label{fig:HD1835_obs}
\end{minipage}
\vspace{0.3\textheight}
\subsection{HD 3651}
\begin{minipage}{1.0\textwidth}
  \centering
  \includegraphics[width=\textwidth]{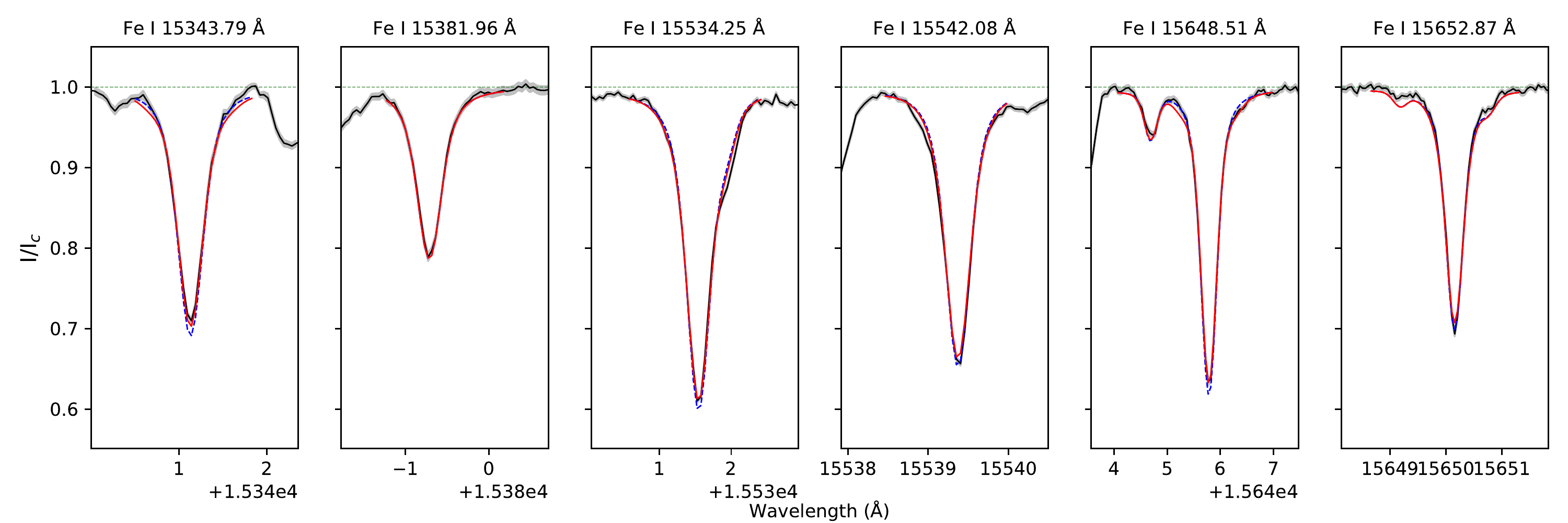}
  \includegraphics[width=0.49\textwidth]{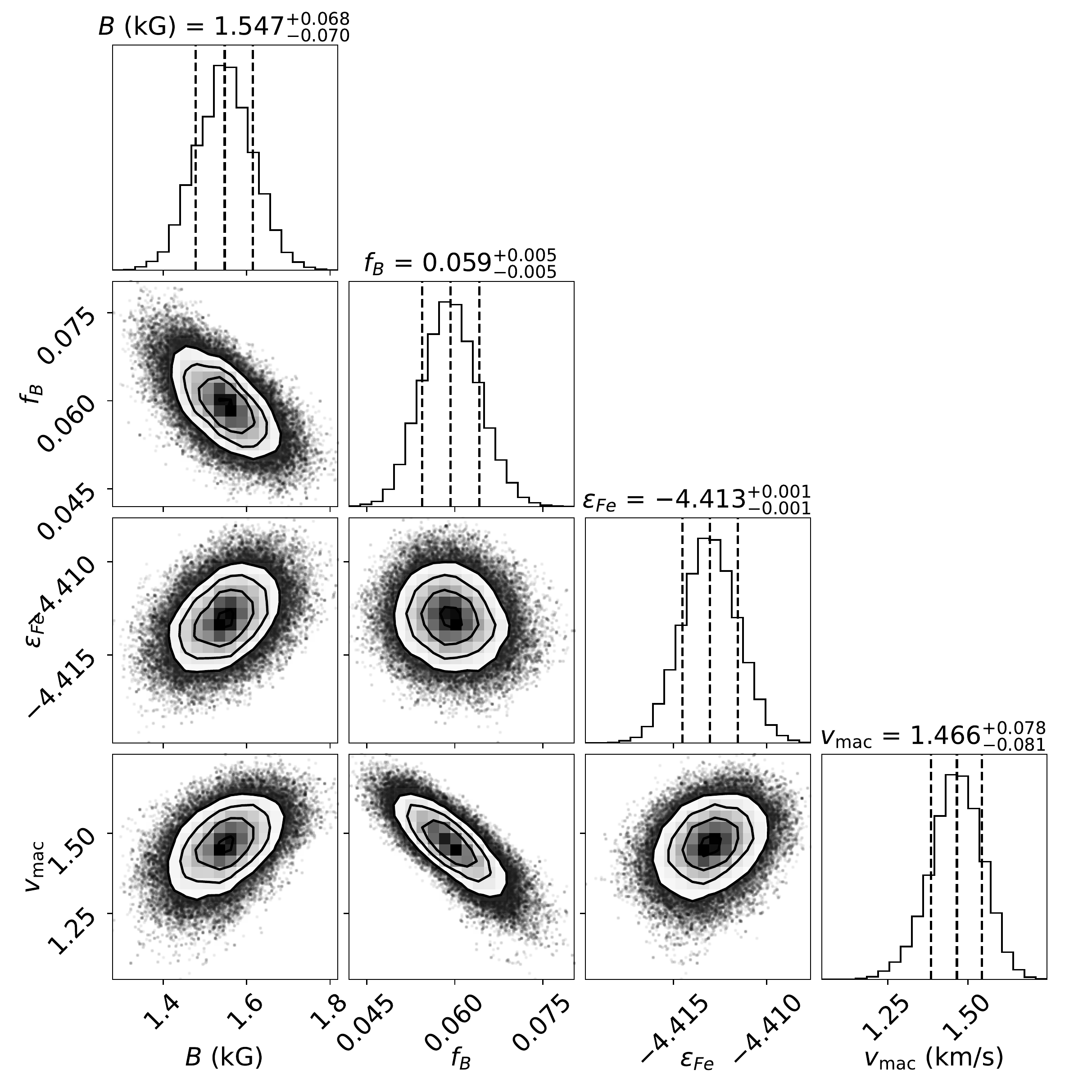}
  \includegraphics[width=0.49\textwidth]{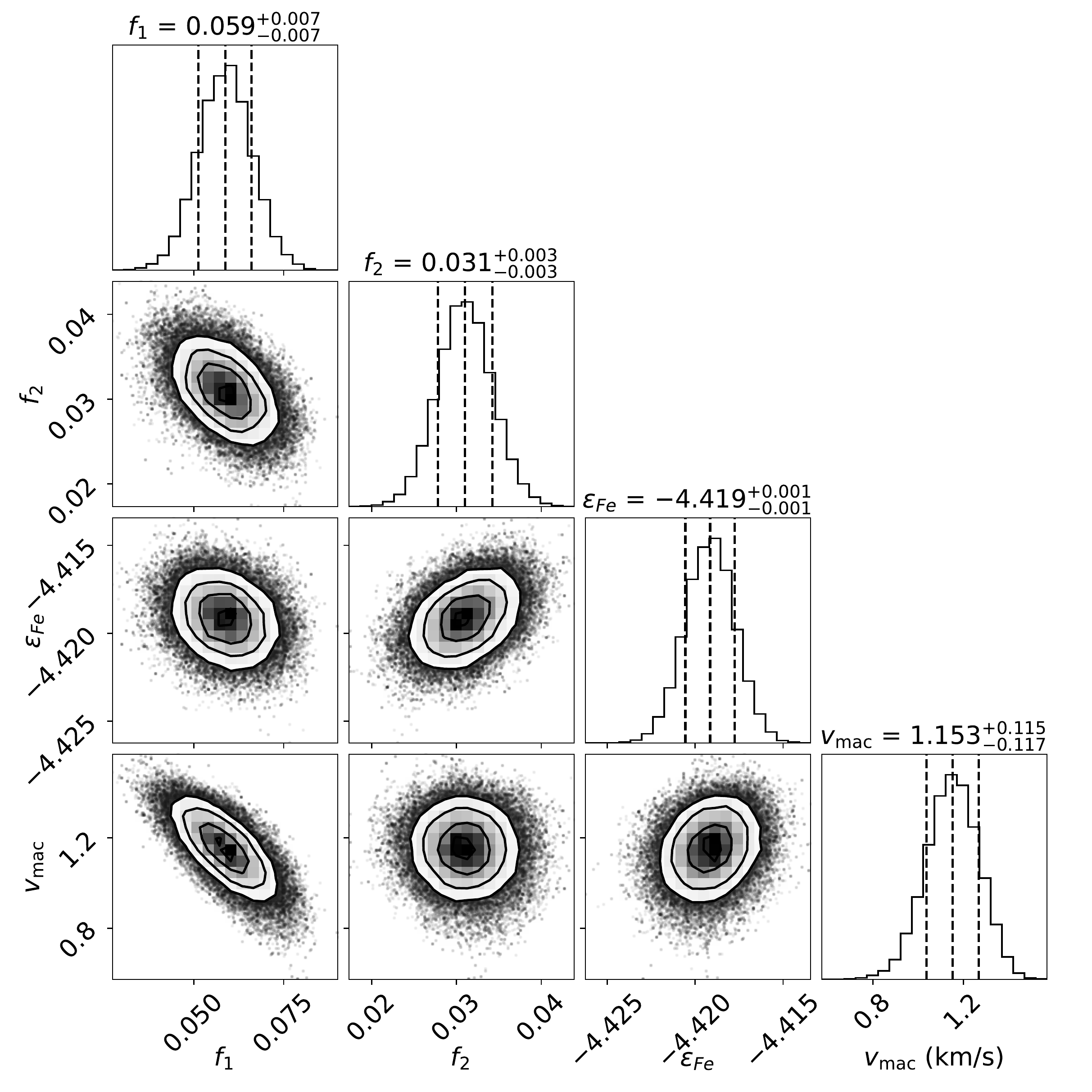}
  \includegraphics[width=0.45\textwidth]{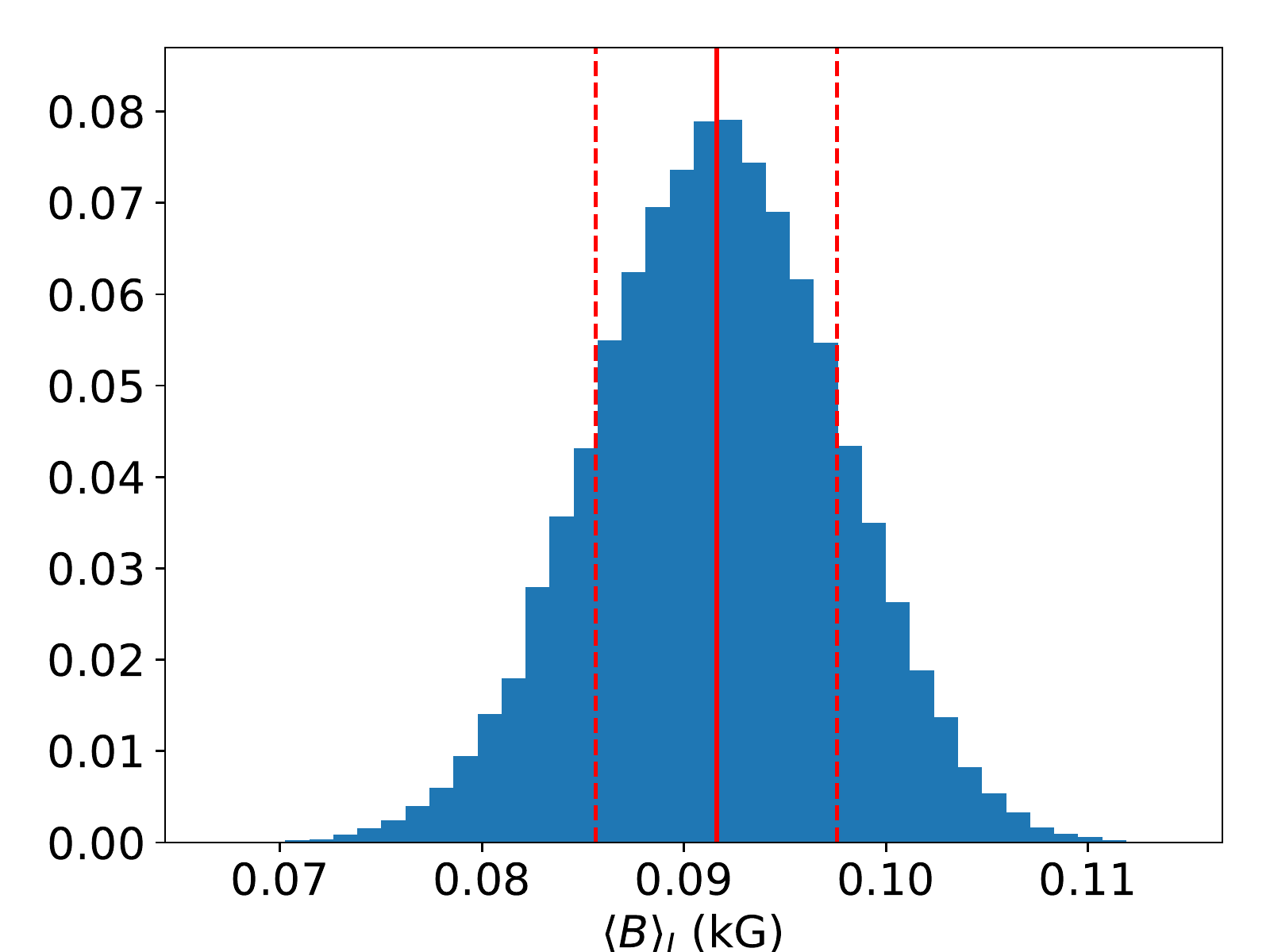}
  \includegraphics[width=0.45\textwidth]{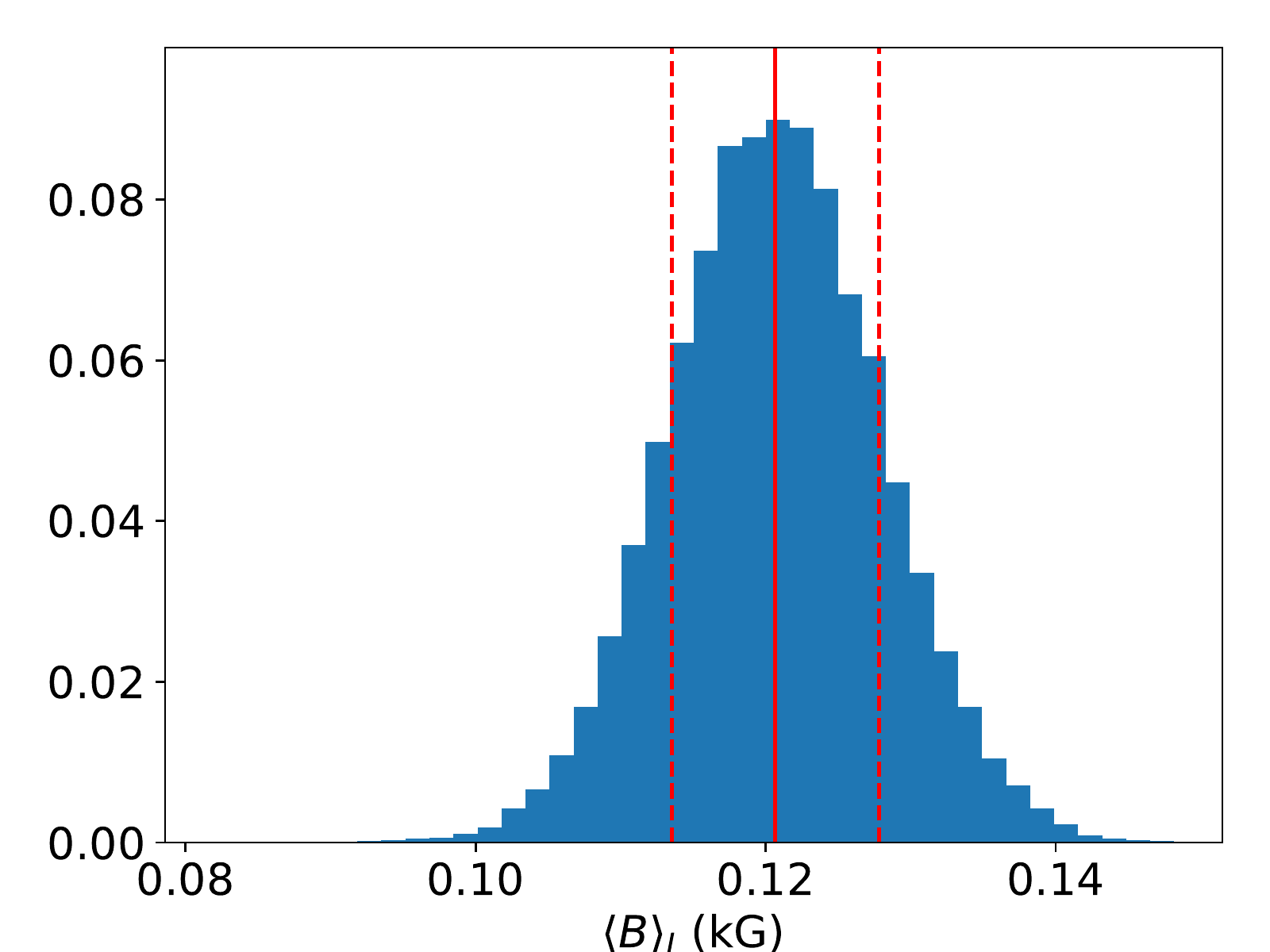}
  \captionof{figure}{Same as Fig. \ref{fig:HD1835_obs} but for HD 3651.}
  \label{fig:HD3651_obs}
\end{minipage}
\vspace{0.3\textheight}
\subsection{HD 9986}
\begin{minipage}{1.0\textwidth}
    \centering
    \includegraphics[width=\textwidth]{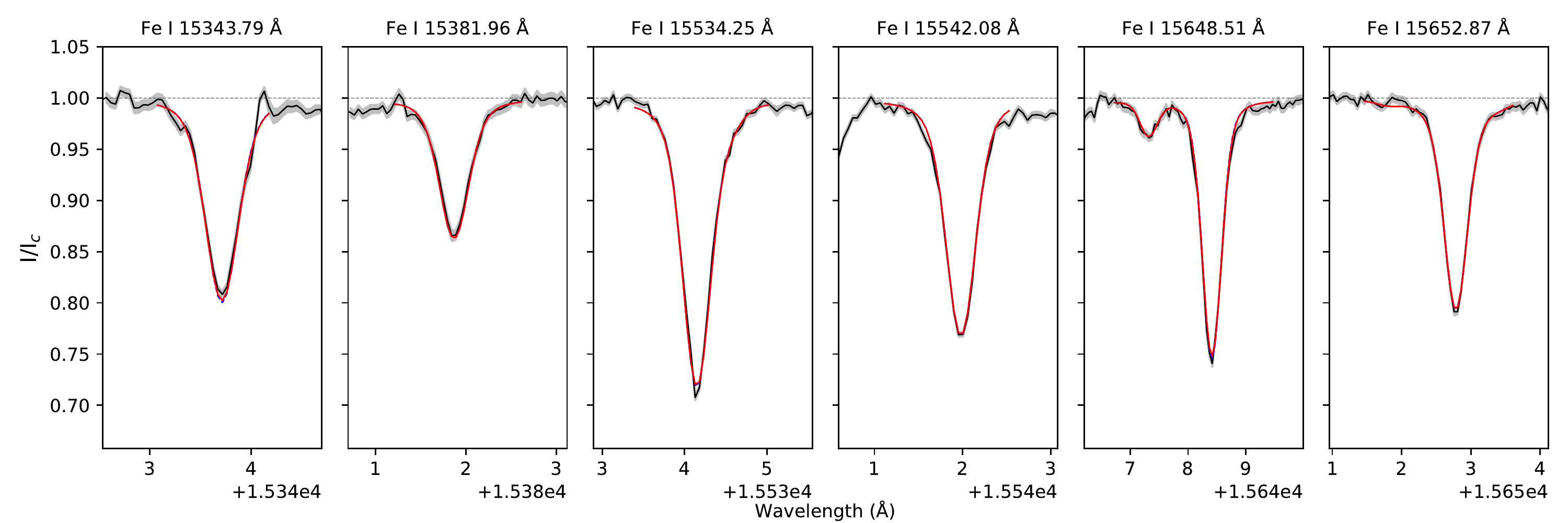}
    \includegraphics[width=0.49\textwidth]{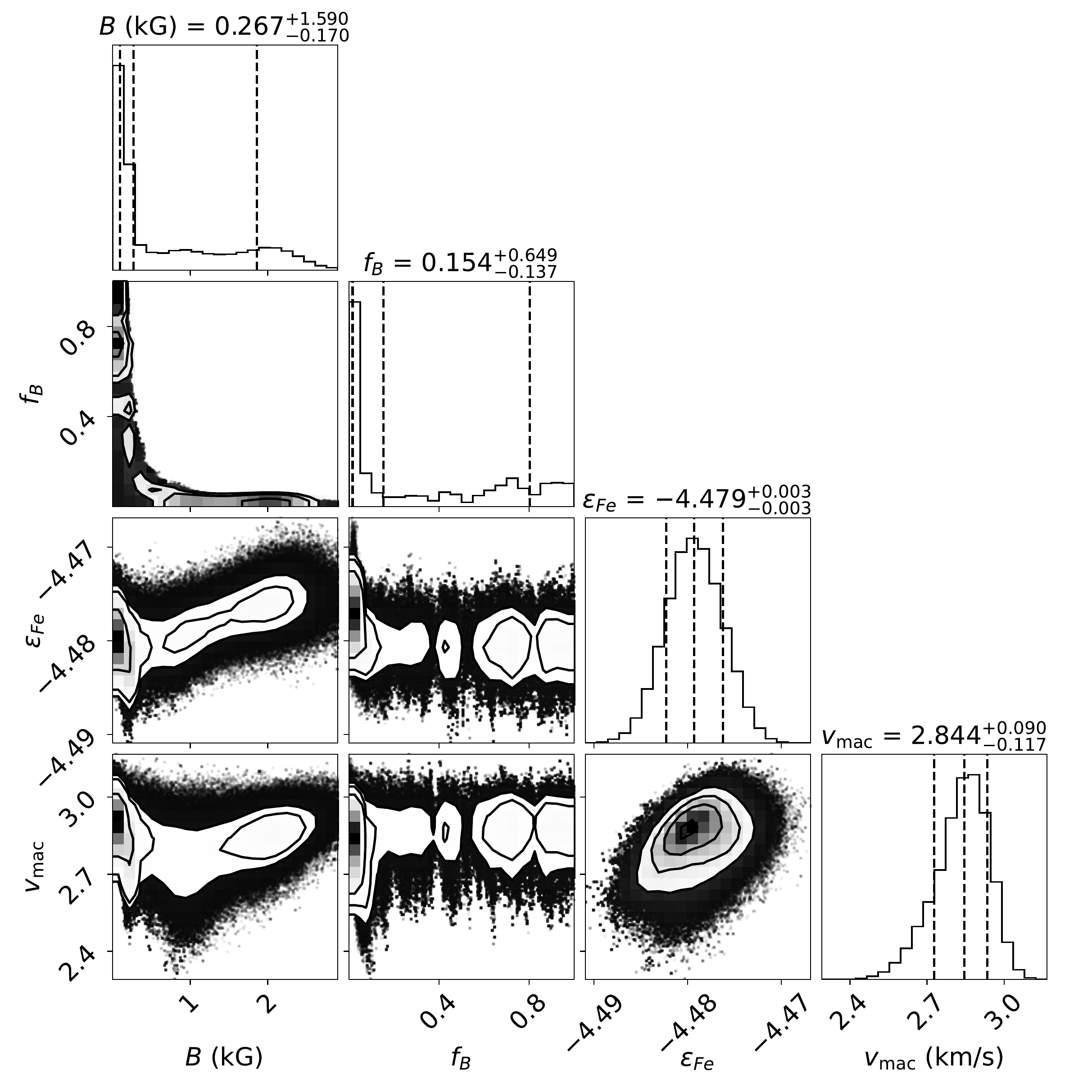}
    \includegraphics[width=0.49\textwidth]{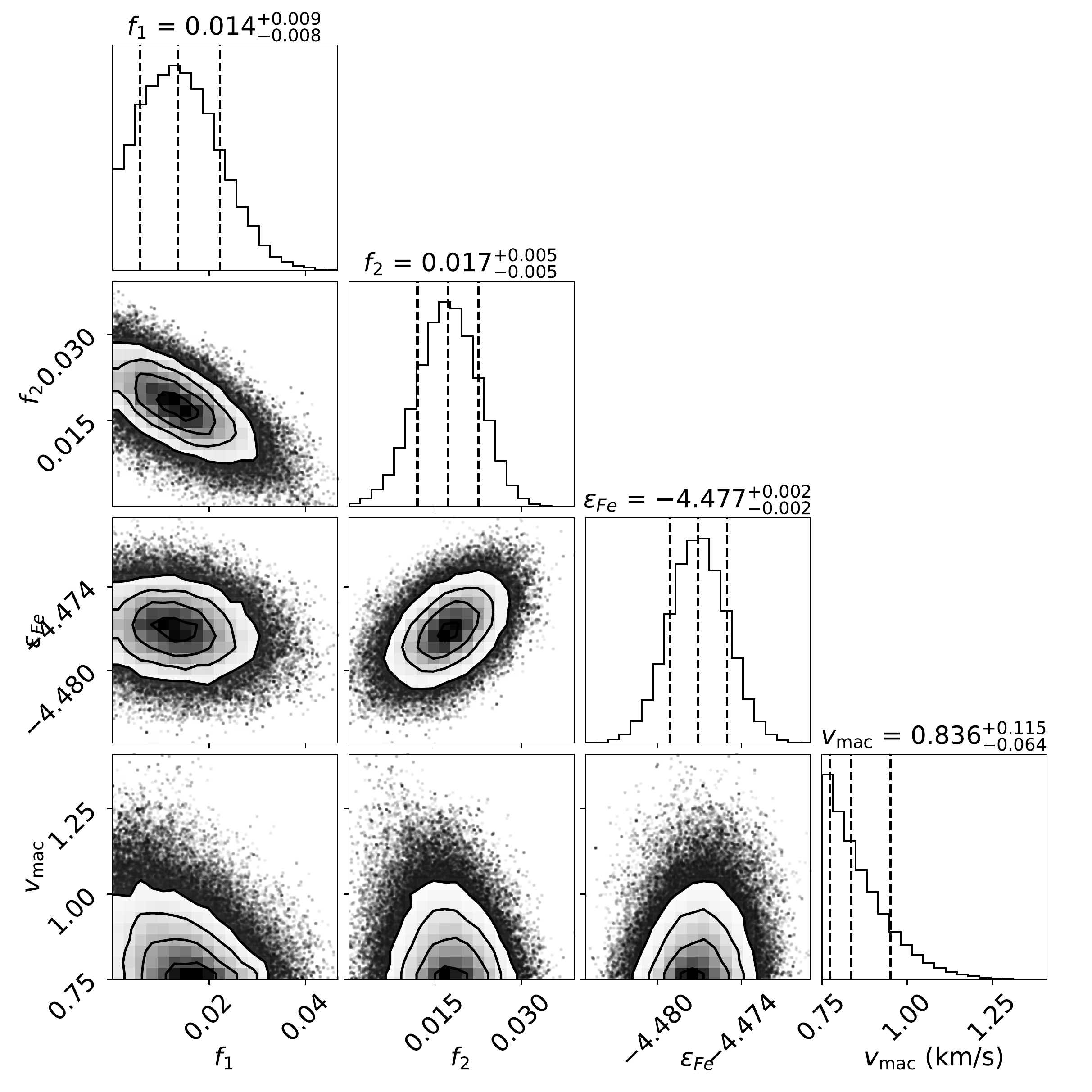}
    \includegraphics[width=0.45\textwidth]{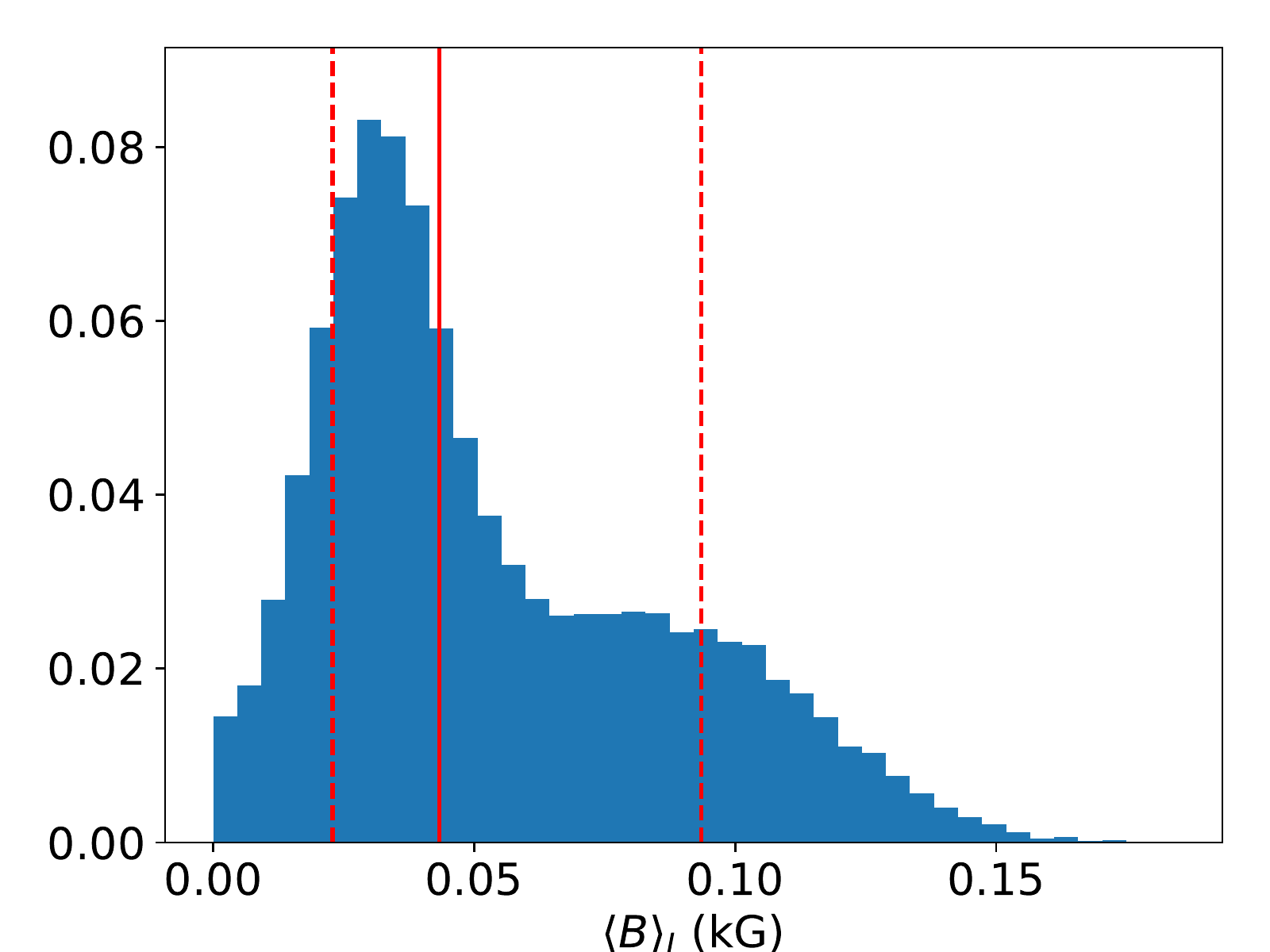}   
    \includegraphics[width=0.45\textwidth]{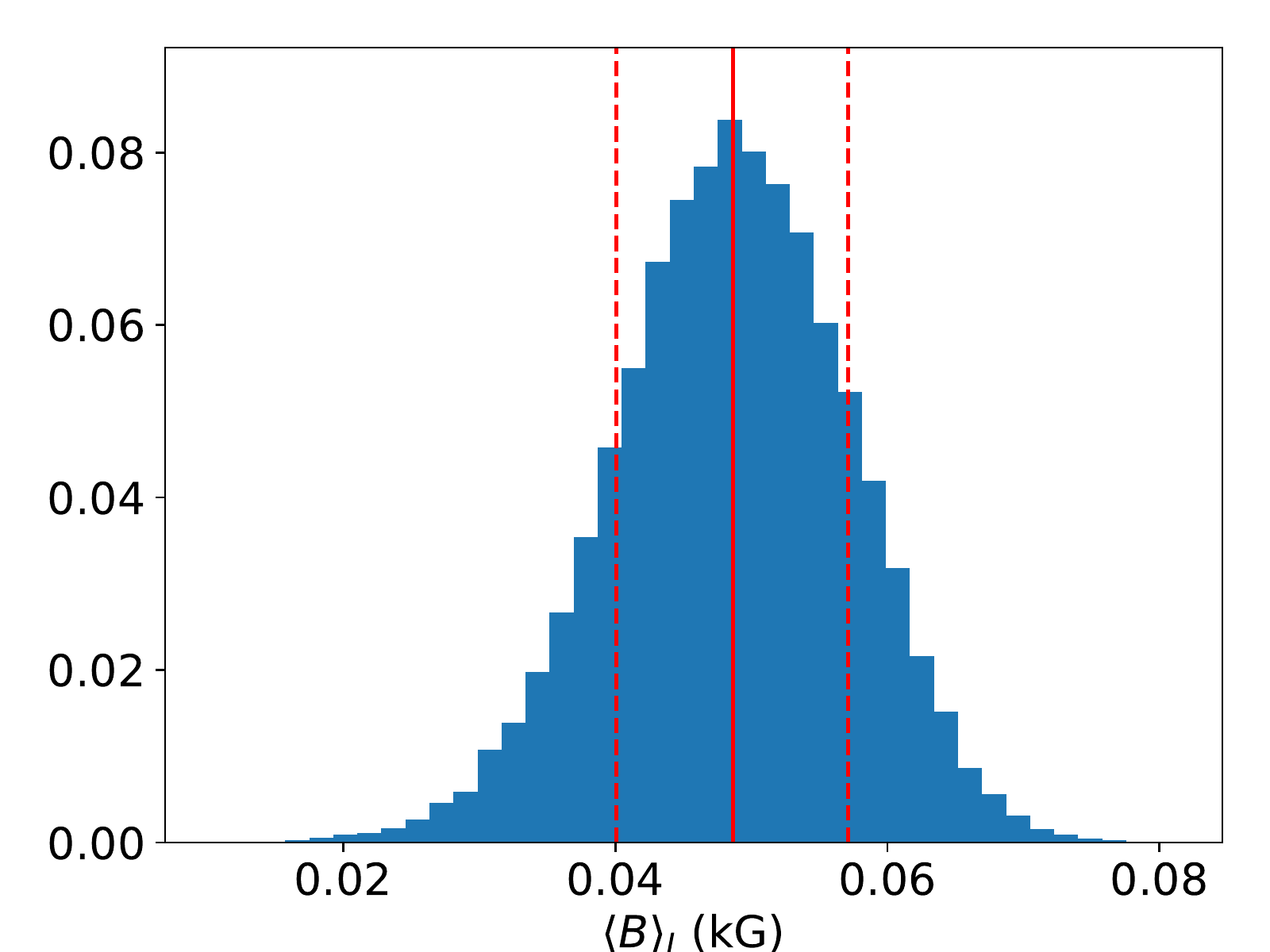}
    \captionof{figure}{Same as Fig. \ref{fig:HD1835_obs} but for HD 9986.}
    \label{fig:HD9986_obs}
\end{minipage}
\vspace{0.3\textheight}
\subsection{HD 10476}
\vspace{-\abovedisplayskip}
\begin{minipage}{1.0\textwidth}
    \centering
    \includegraphics[width=\textwidth]{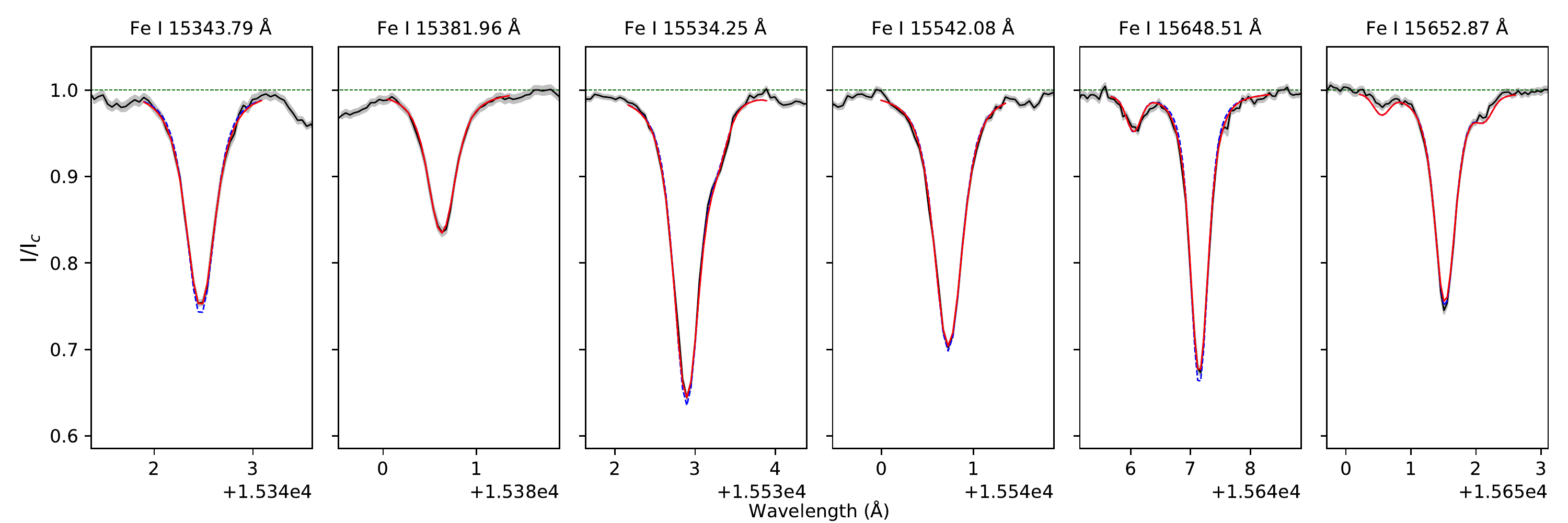}
    \includegraphics[width=0.49\textwidth]{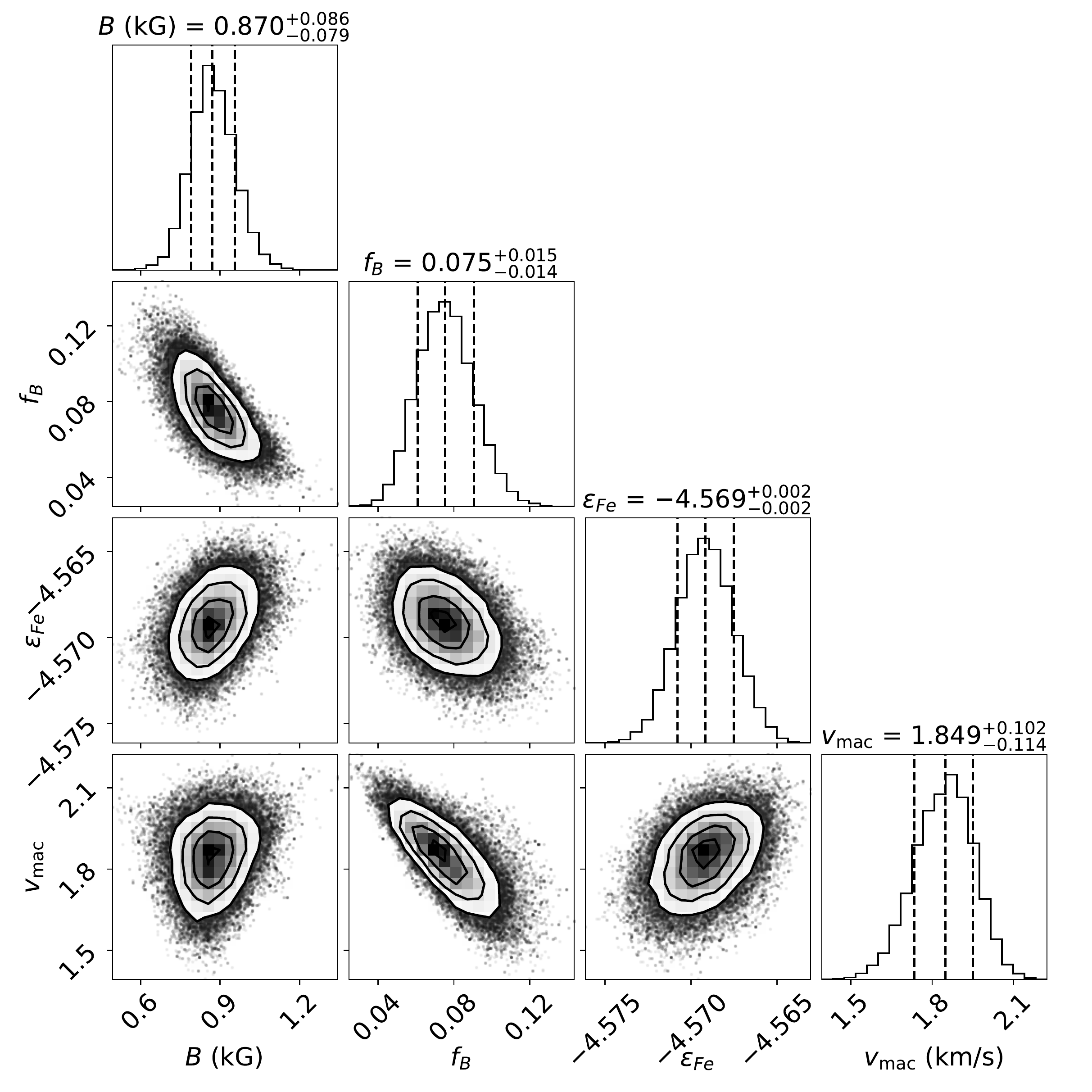}
    \includegraphics[width=0.49\textwidth]{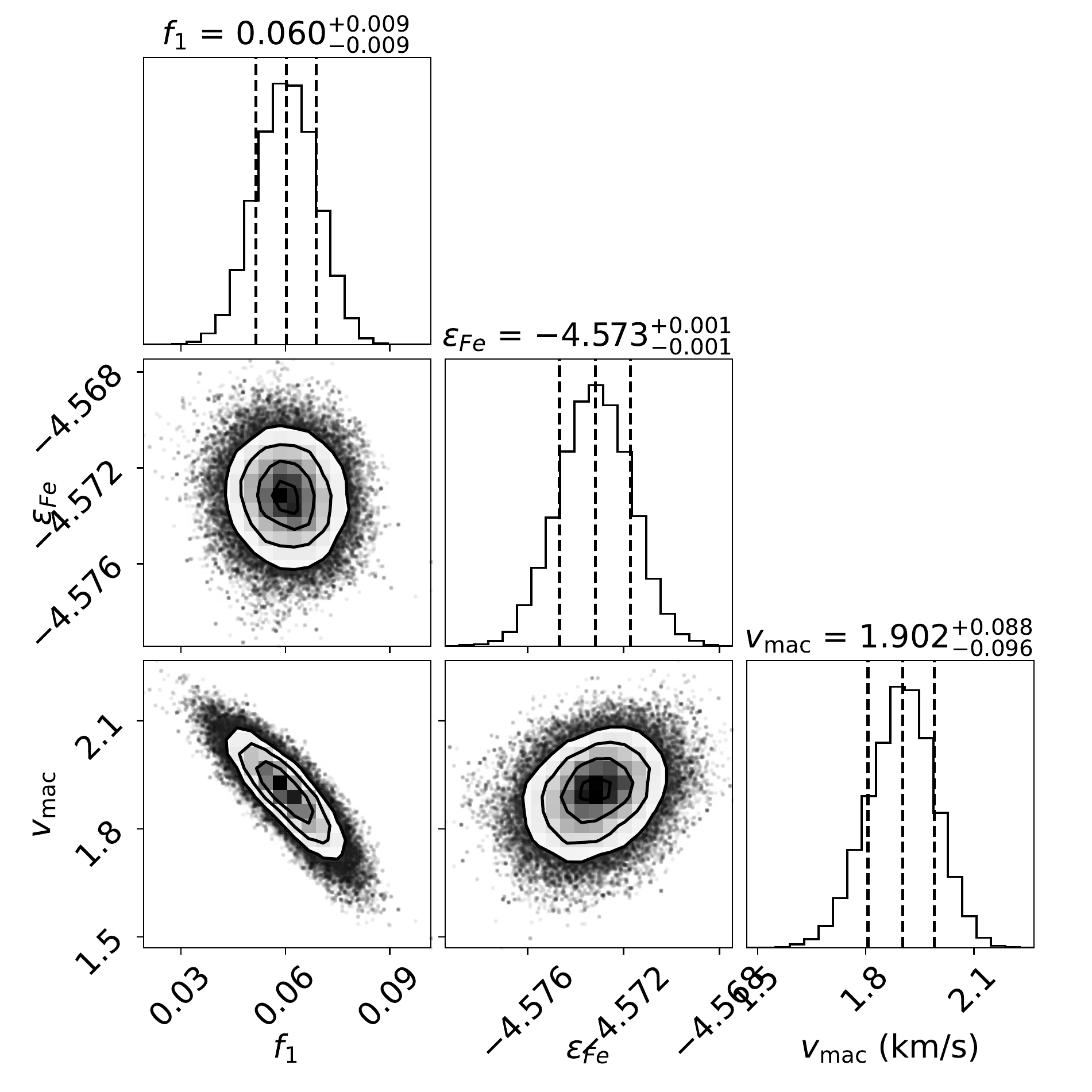}
    \includegraphics[width=0.45\textwidth]{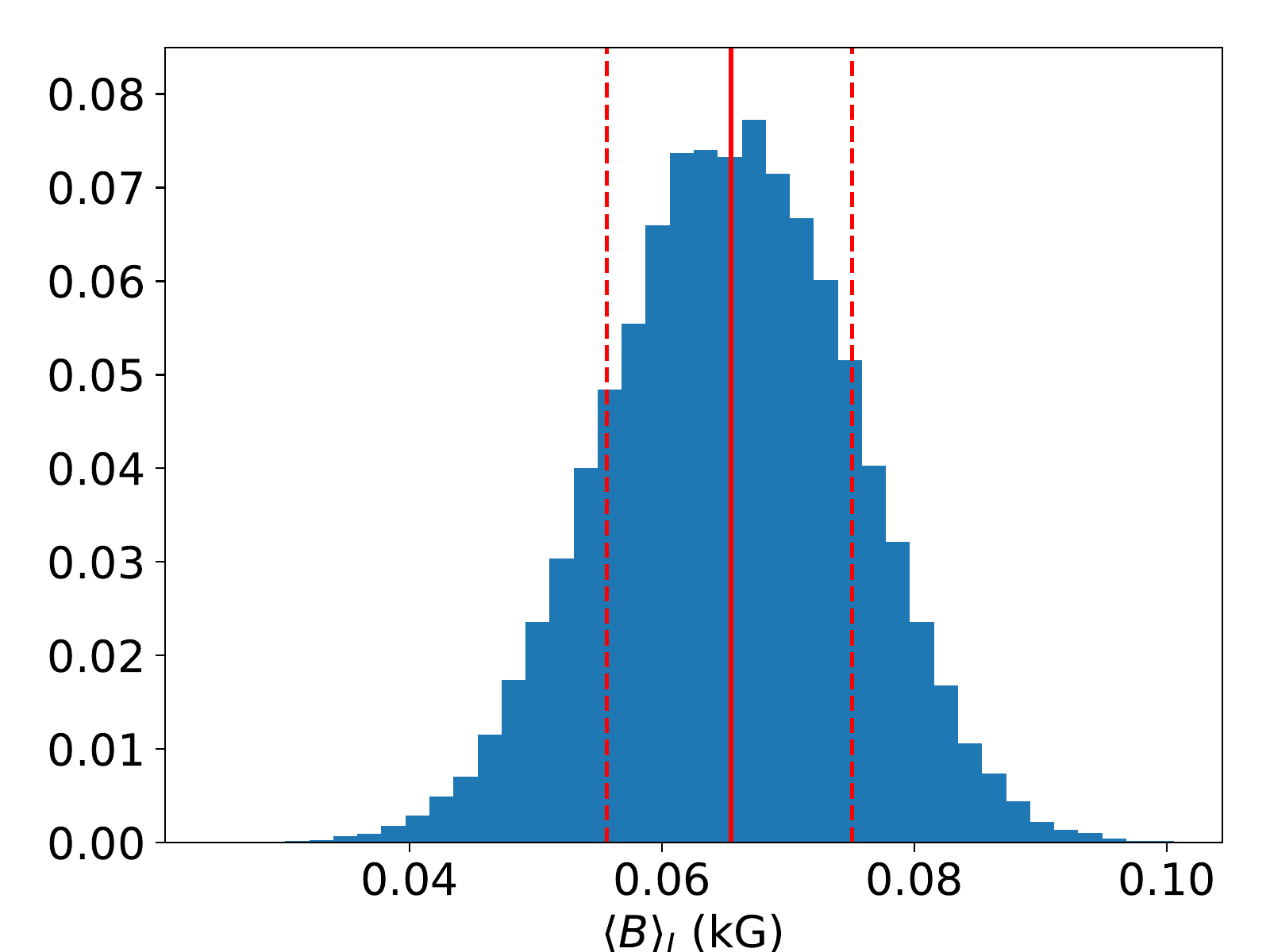}
    \includegraphics[width=0.45\textwidth]{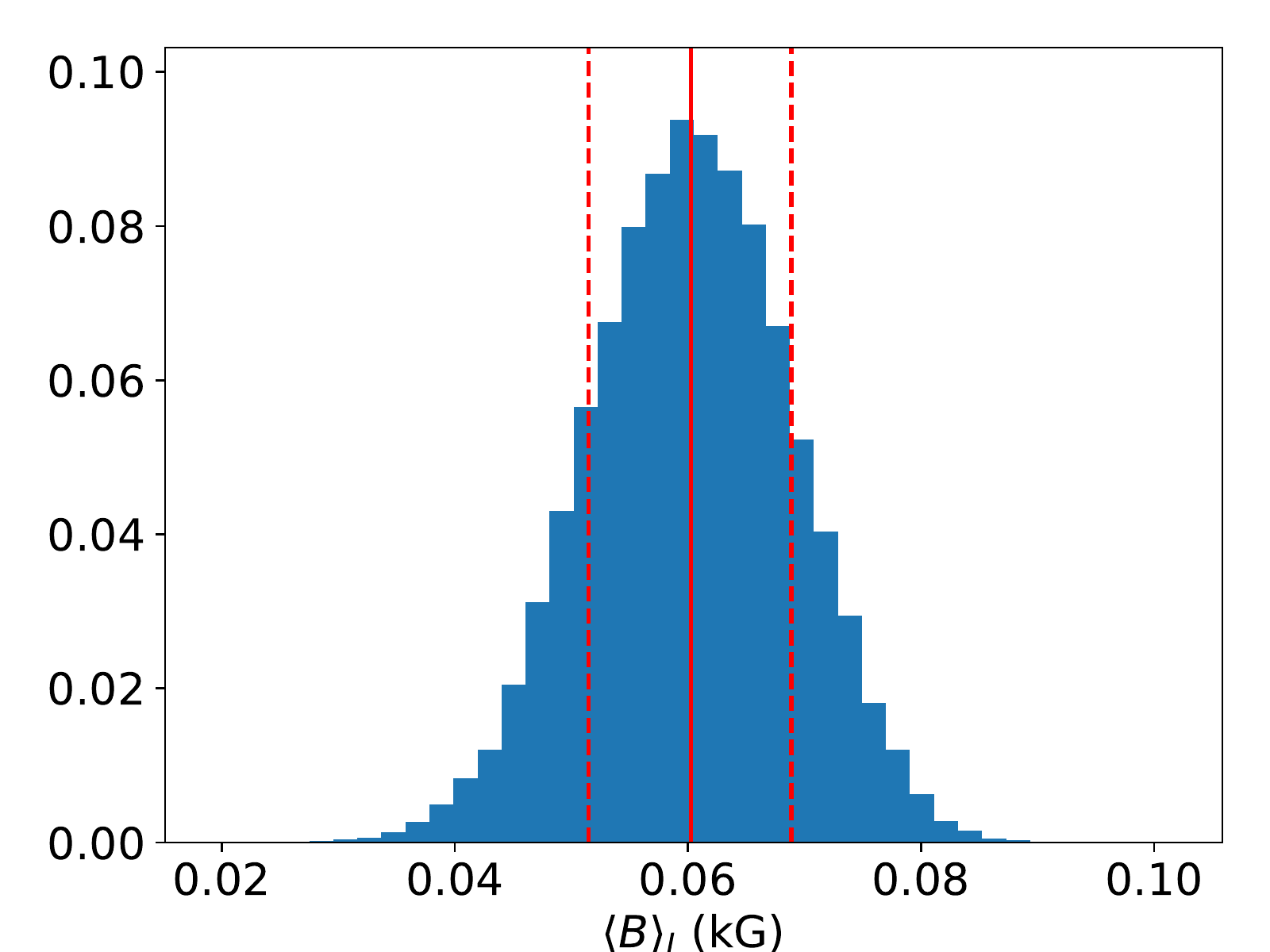}
    \captionof{figure}{Same as Fig. \ref{fig:HD1835_obs} but for HD 10476.}
    \label{fig:HD10476_obs}
\end{minipage}
\vspace{0.3\textheight}
\subsection{HD 22049}
\begin{minipage}{1.0\textwidth}
    \centering
    \includegraphics[width=\textwidth]{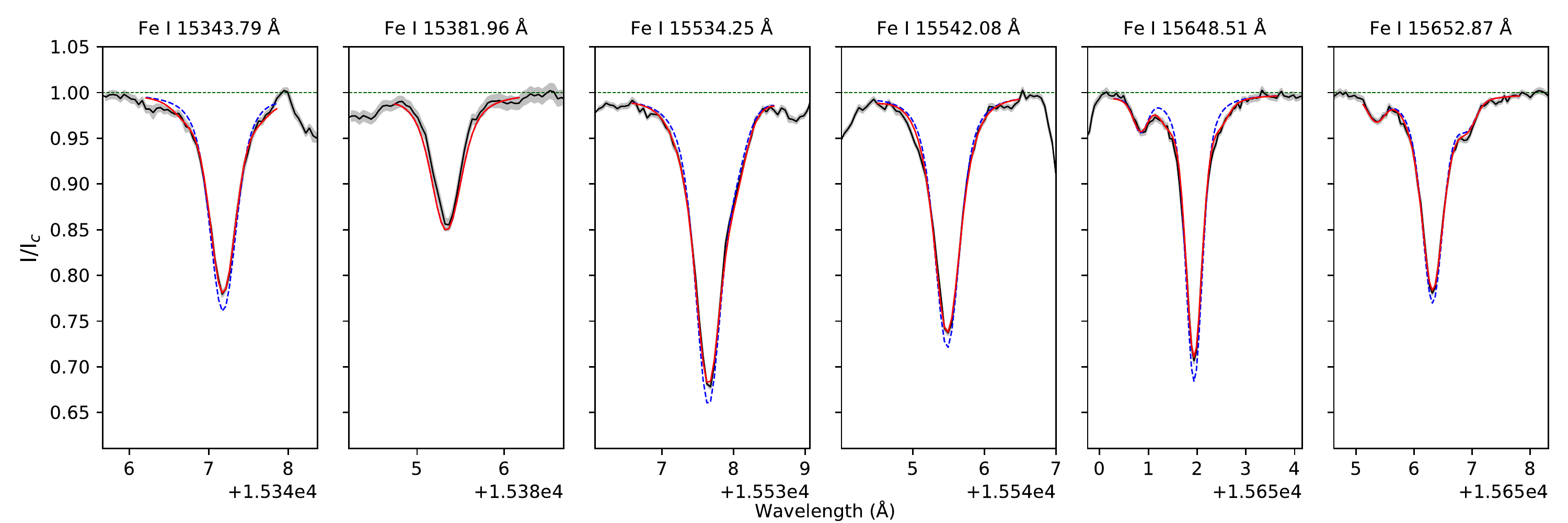}
    \includegraphics[width=0.49\textwidth]{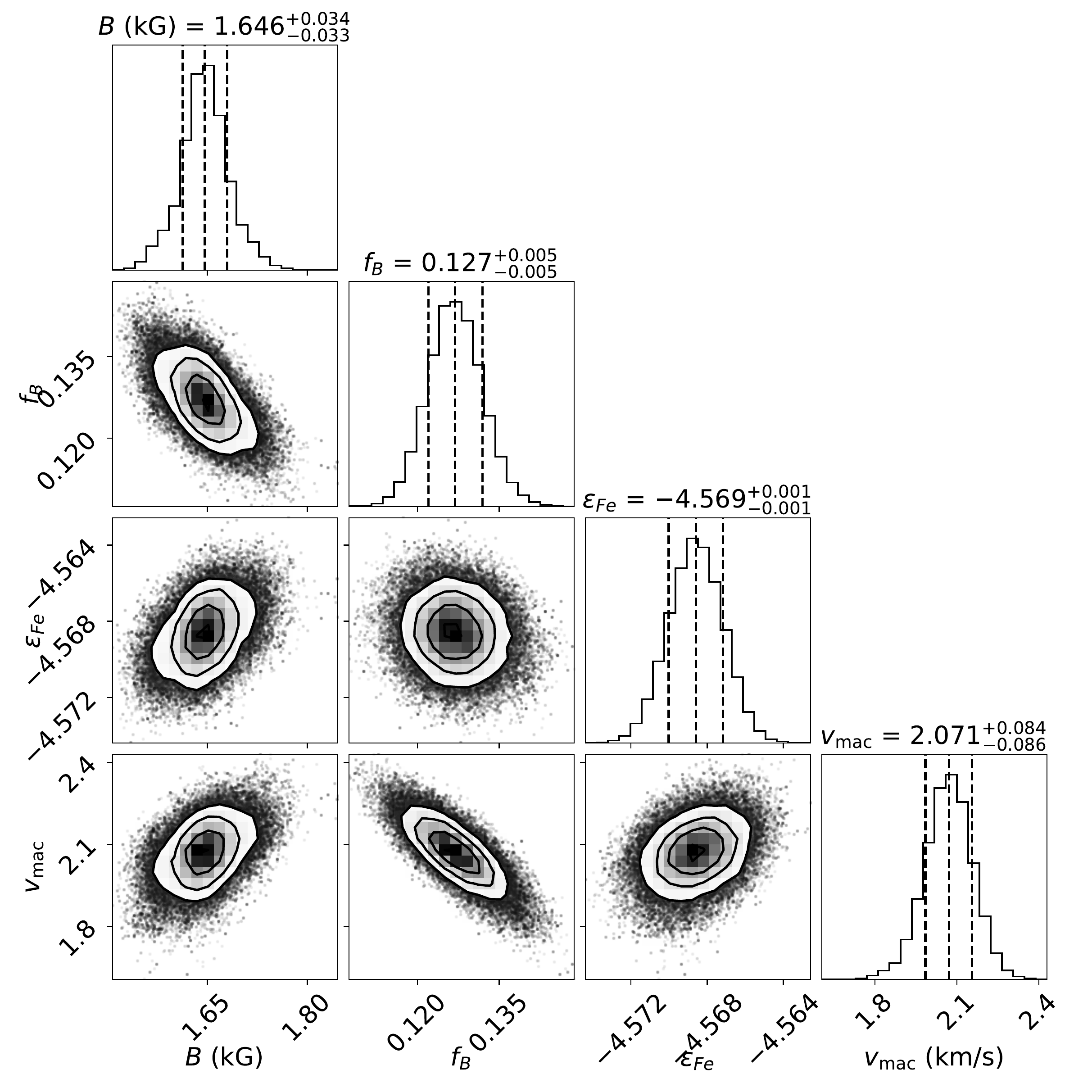}
    \includegraphics[width=0.49\textwidth]{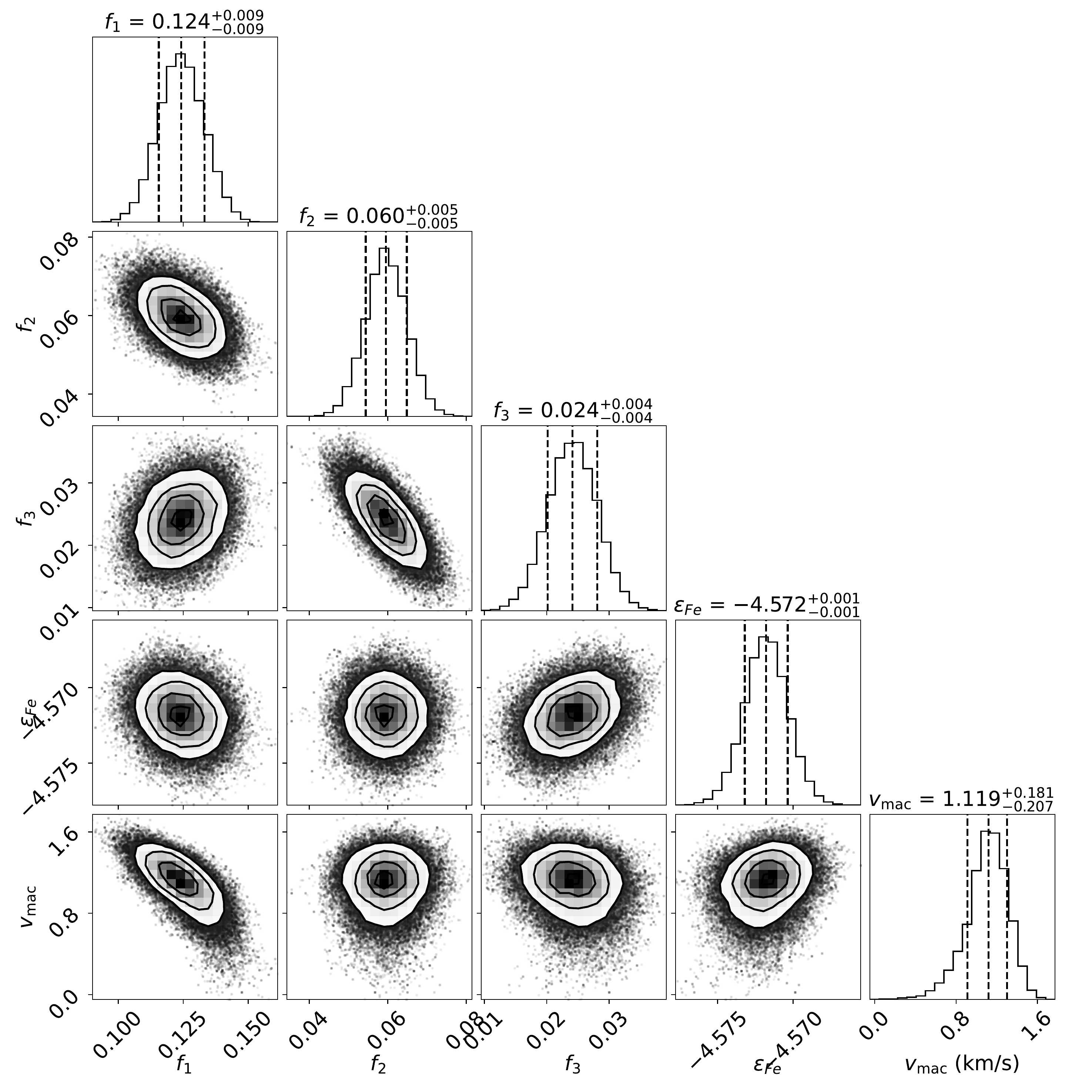}
    \includegraphics[width=0.45\textwidth]{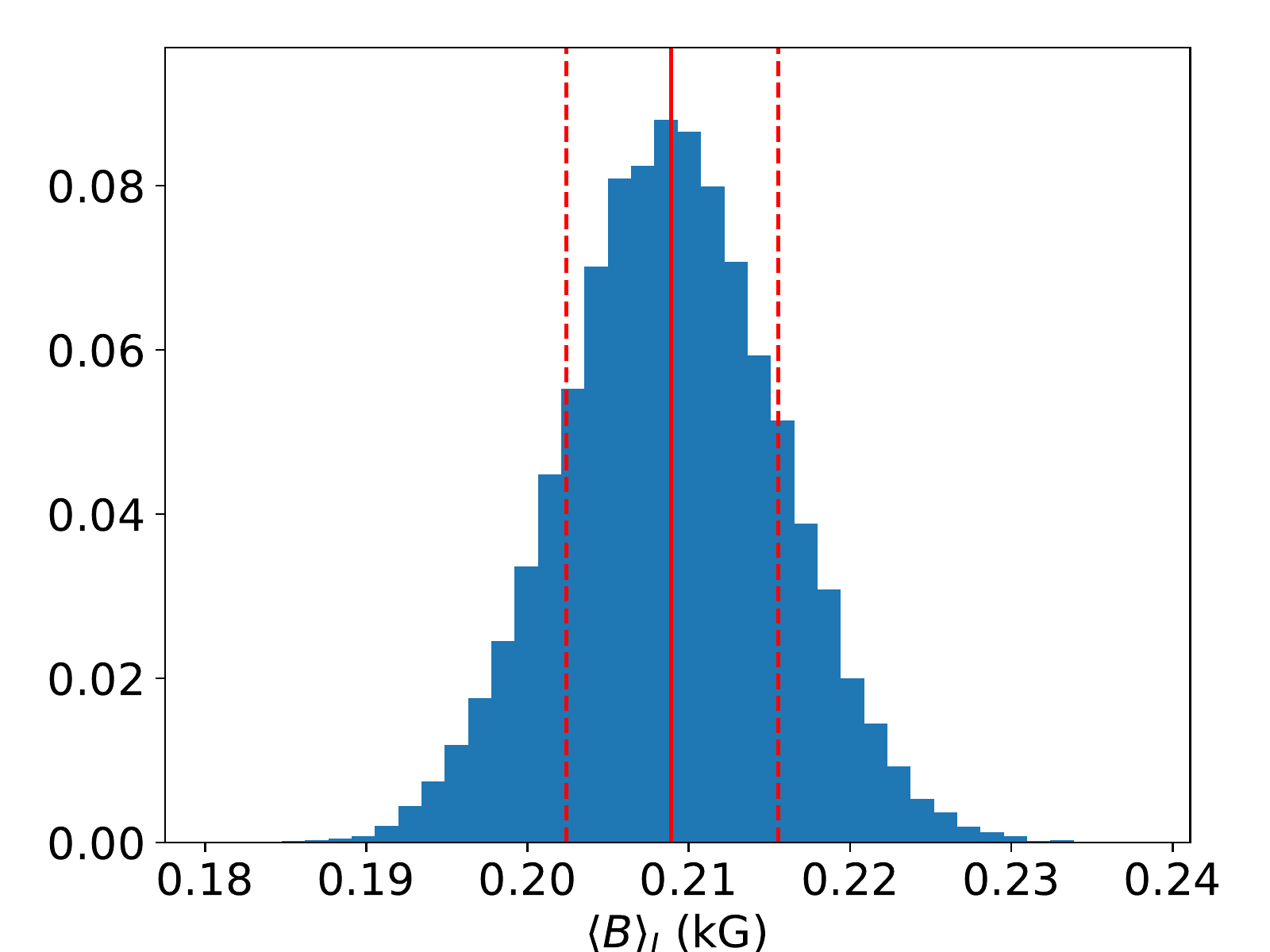}
    \includegraphics[width=0.45\textwidth]{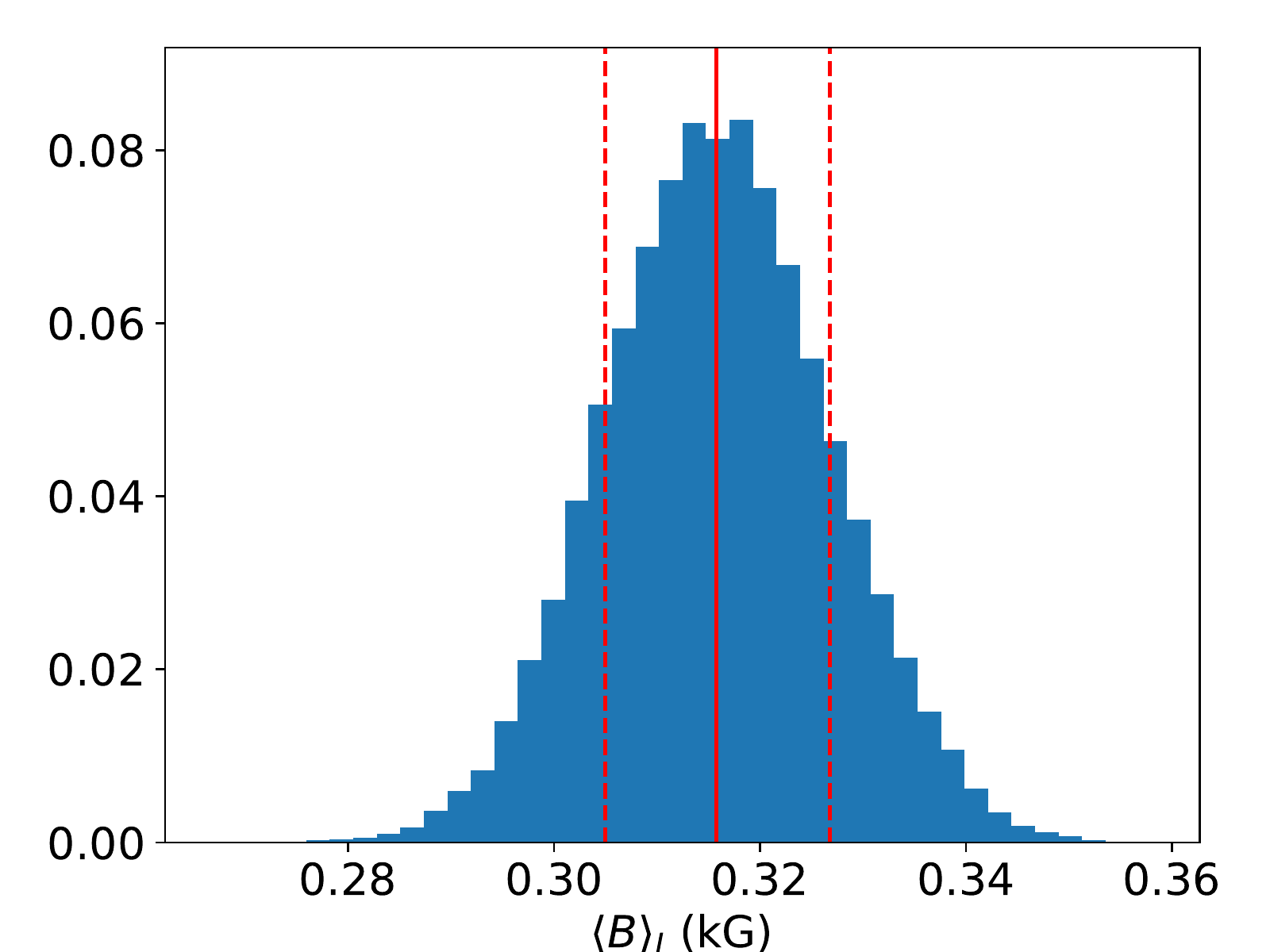}
    \captionof{figure}{Same as Fig. \ref{fig:HD1835_obs} but for HD 22049.}
    \label{fig:HD22049_obs}
\end{minipage}
\vspace{0.3\textheight}
\subsection{HD 59967}
\begin{minipage}{1.0\textwidth}
    \centering
    \includegraphics[width=\textwidth]{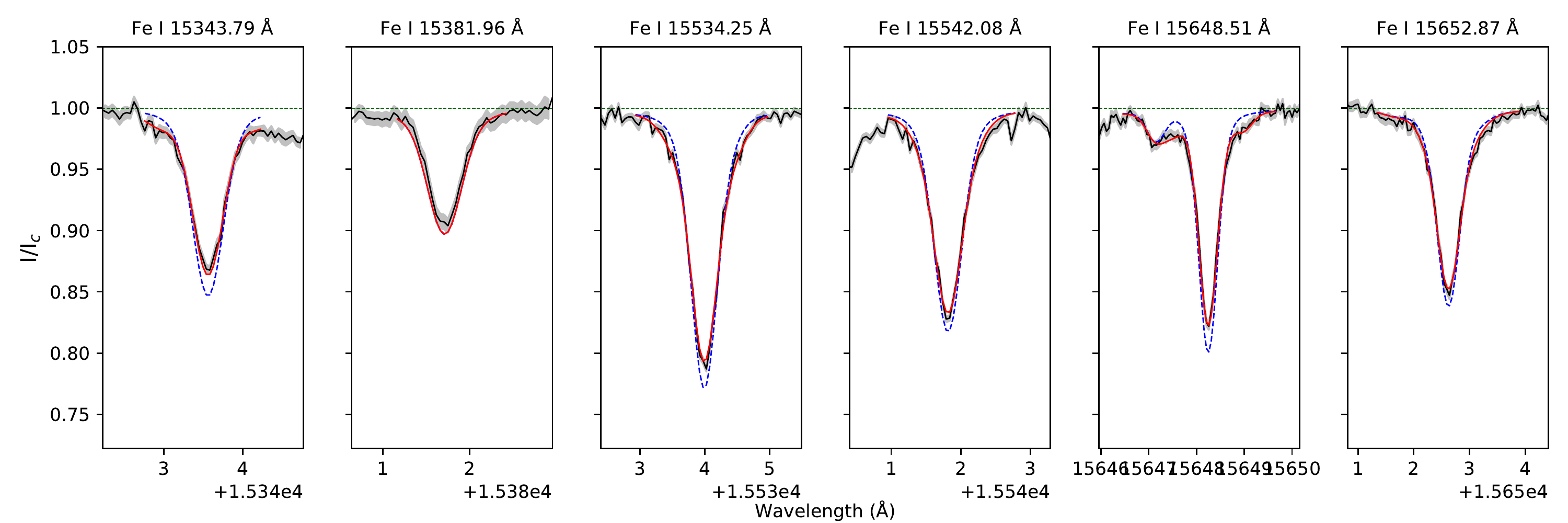}
    \includegraphics[width=0.49\textwidth]{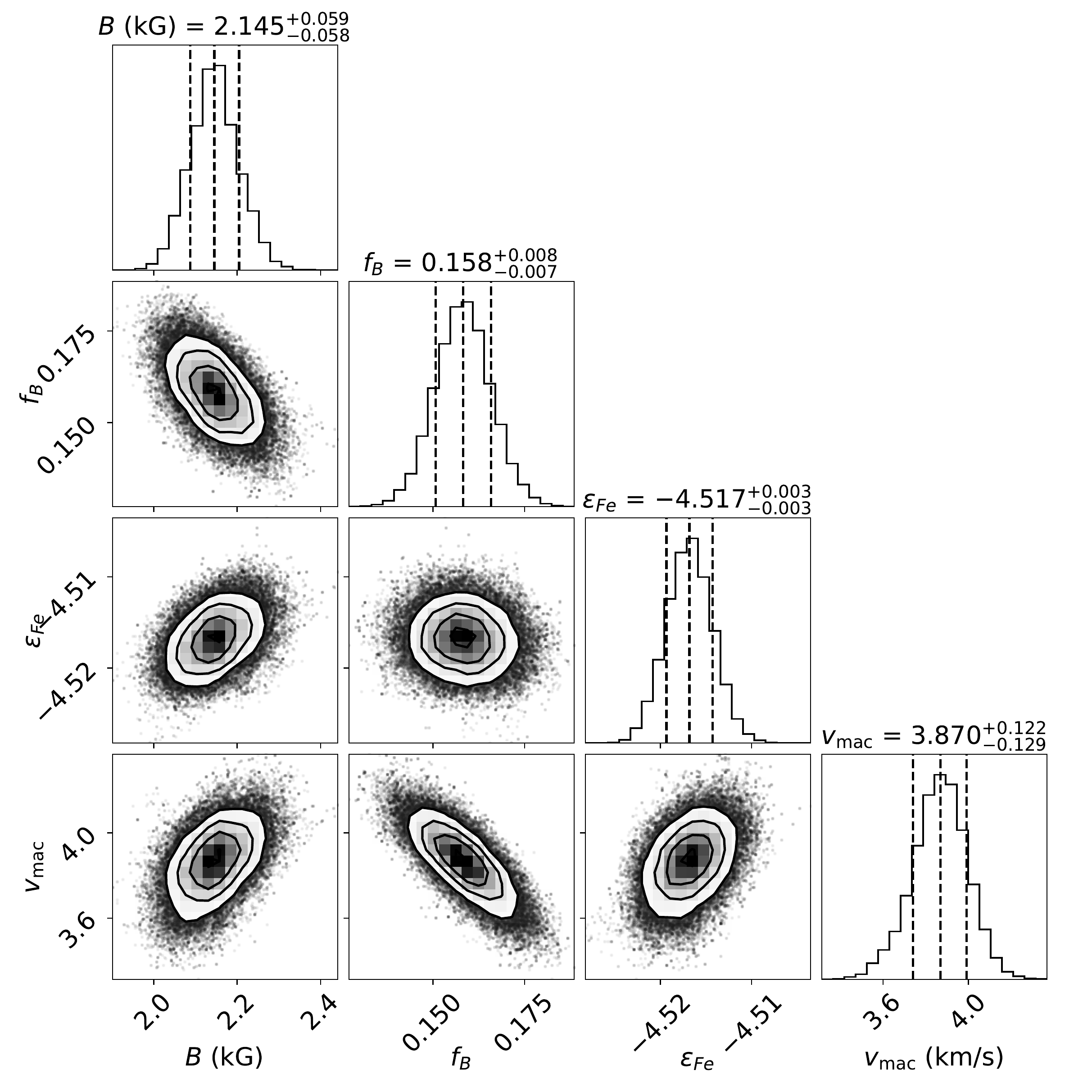}
    \includegraphics[width=0.49\textwidth]{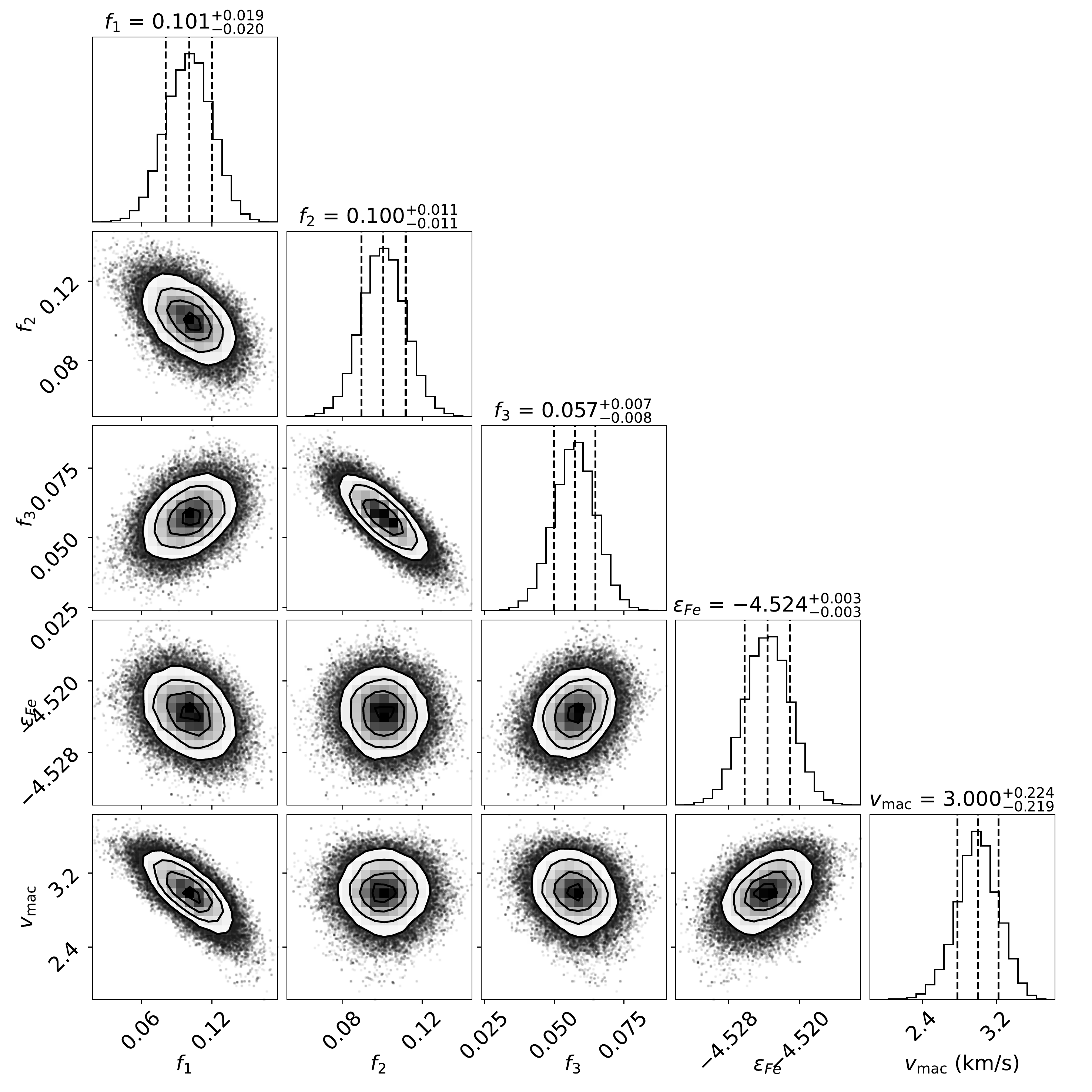}
    \includegraphics[width=0.45\textwidth]{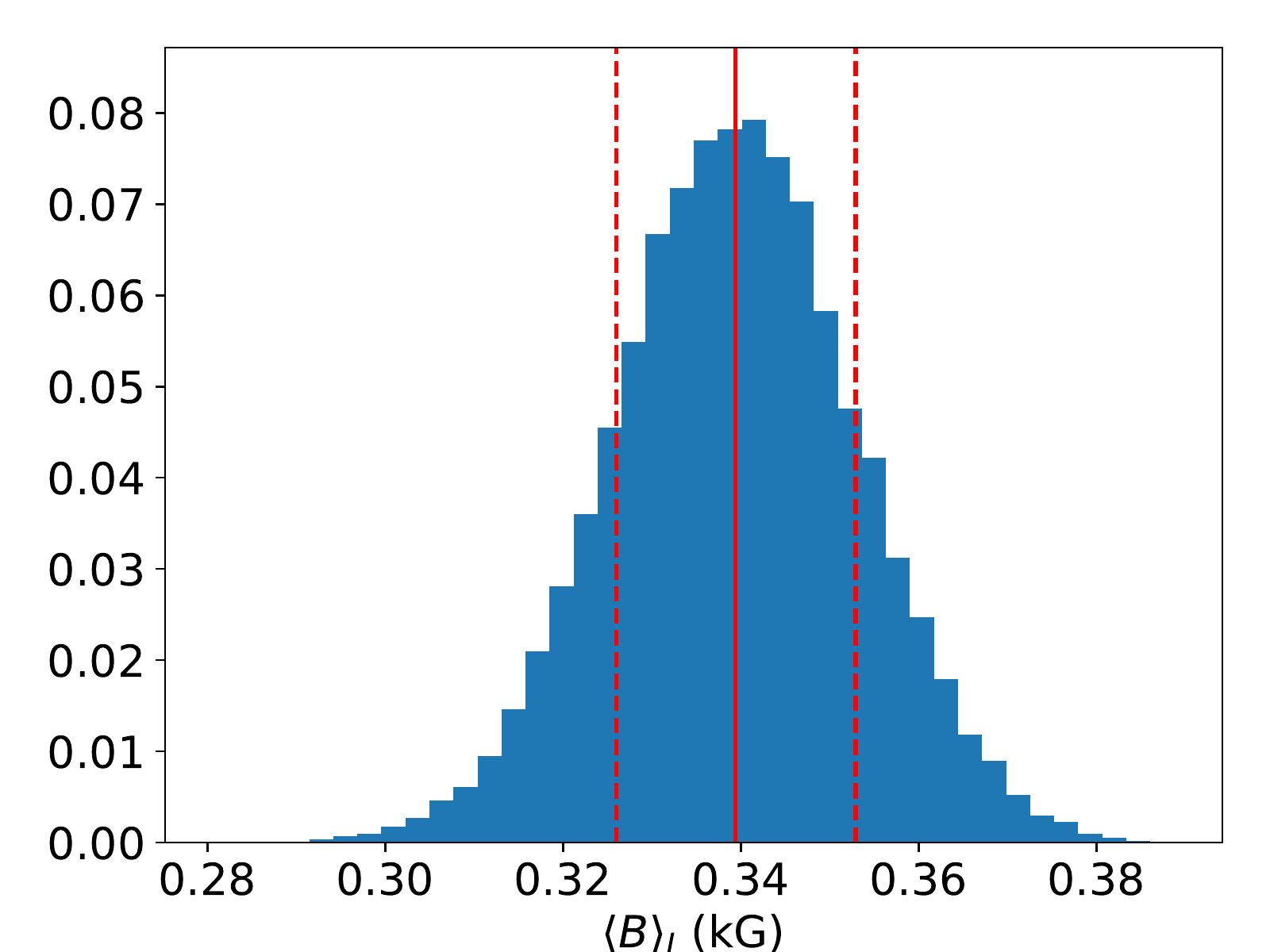}
    \includegraphics[width=0.45\textwidth]{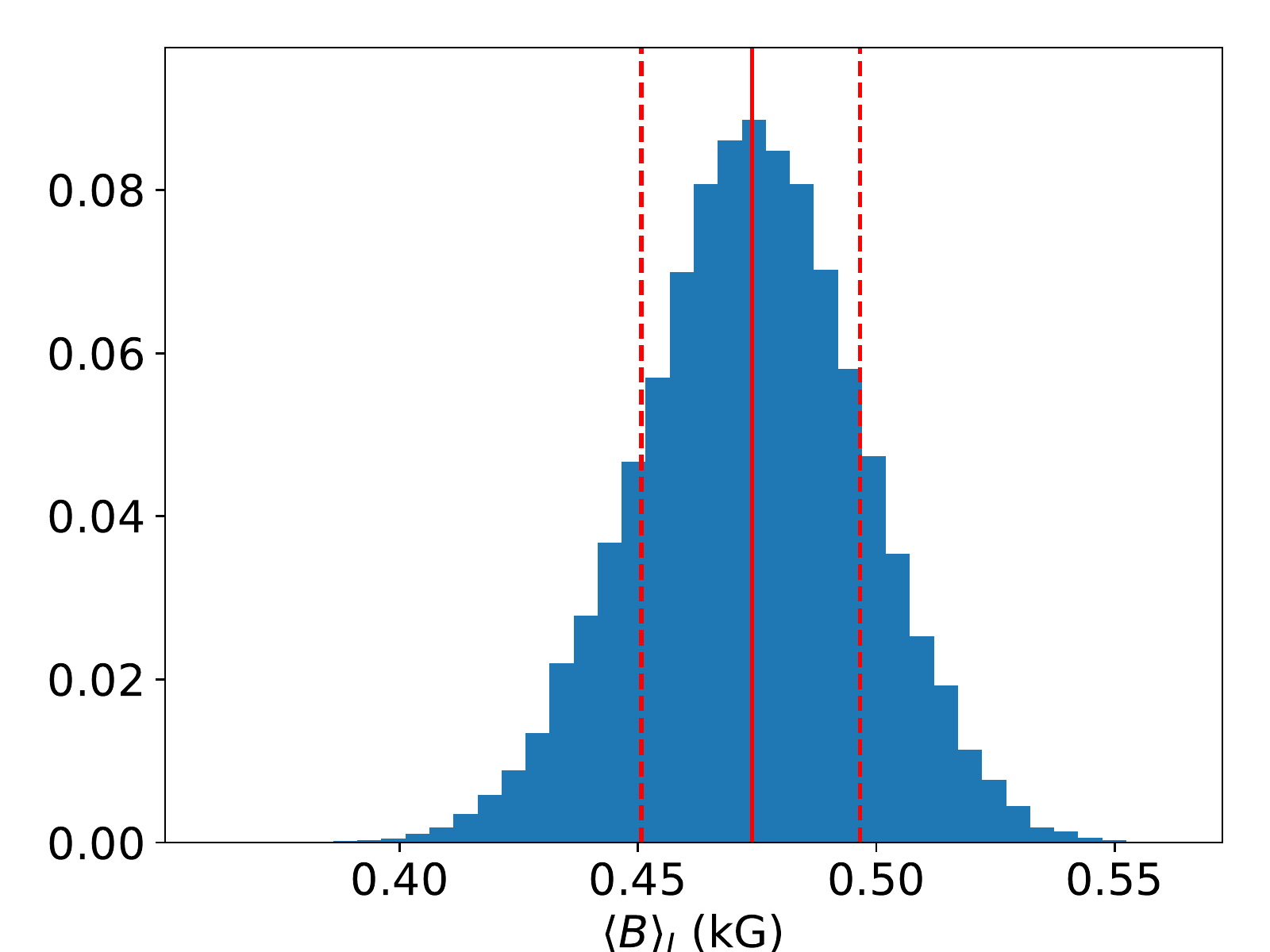}
    \captionof{figure}{Same as Fig. \ref{fig:HD1835_obs} but for HD 59967 observed on 2022 March 24.}
    \label{fig:HD59967_obs}
\end{minipage}
\vspace{0.3\textheight}
\subsection{HD 73256}
\begin{minipage}{1.0\textwidth}
    \centering
    \includegraphics[width=\textwidth]{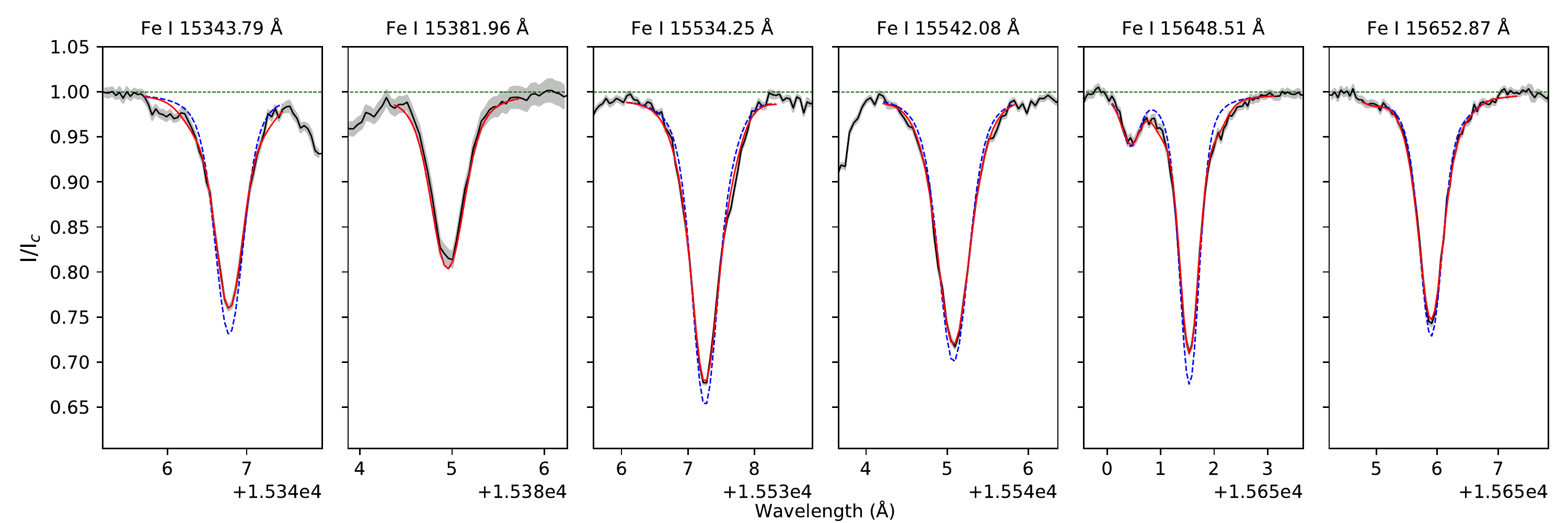}
    \includegraphics[width=0.49\textwidth]{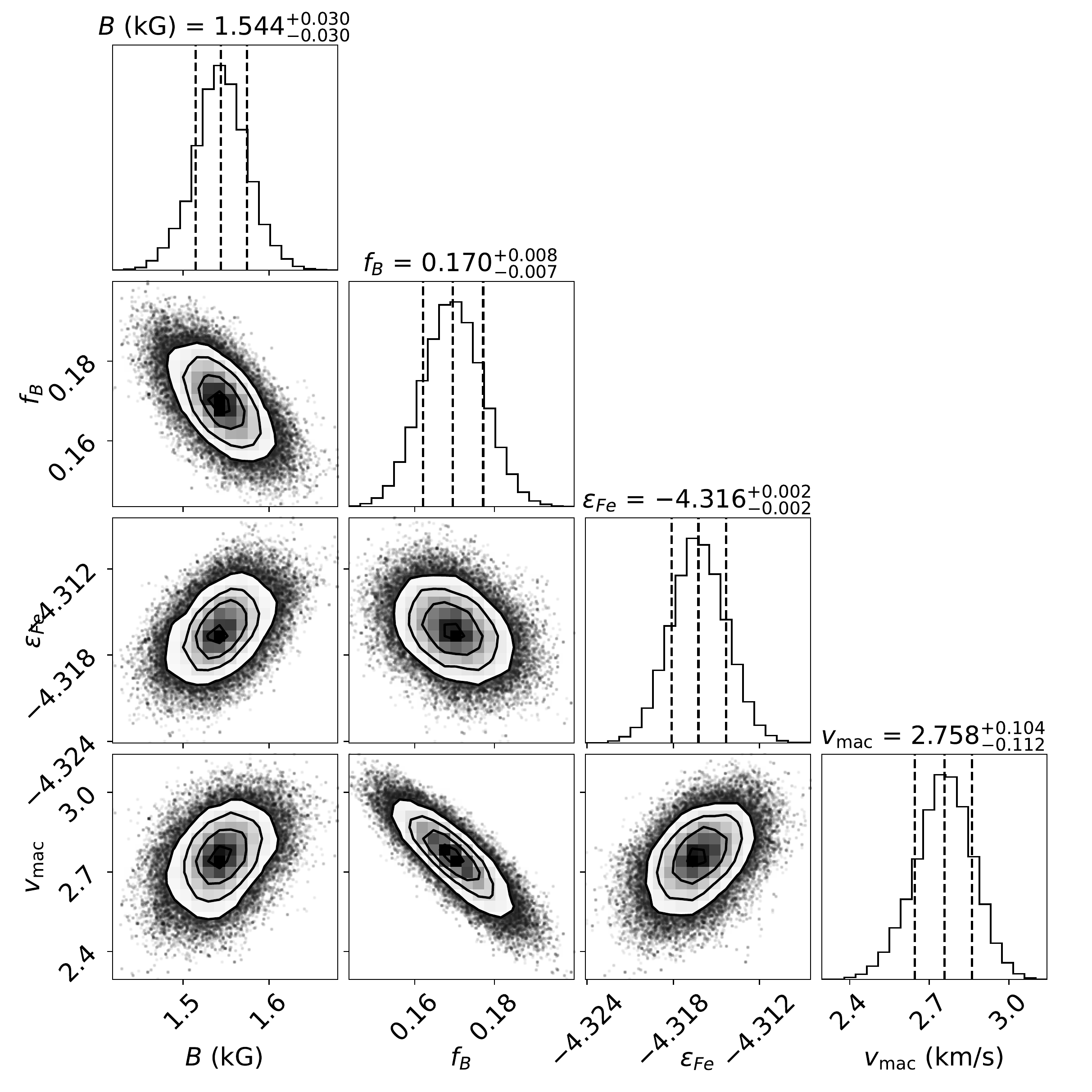}
    \includegraphics[width=0.49\textwidth]{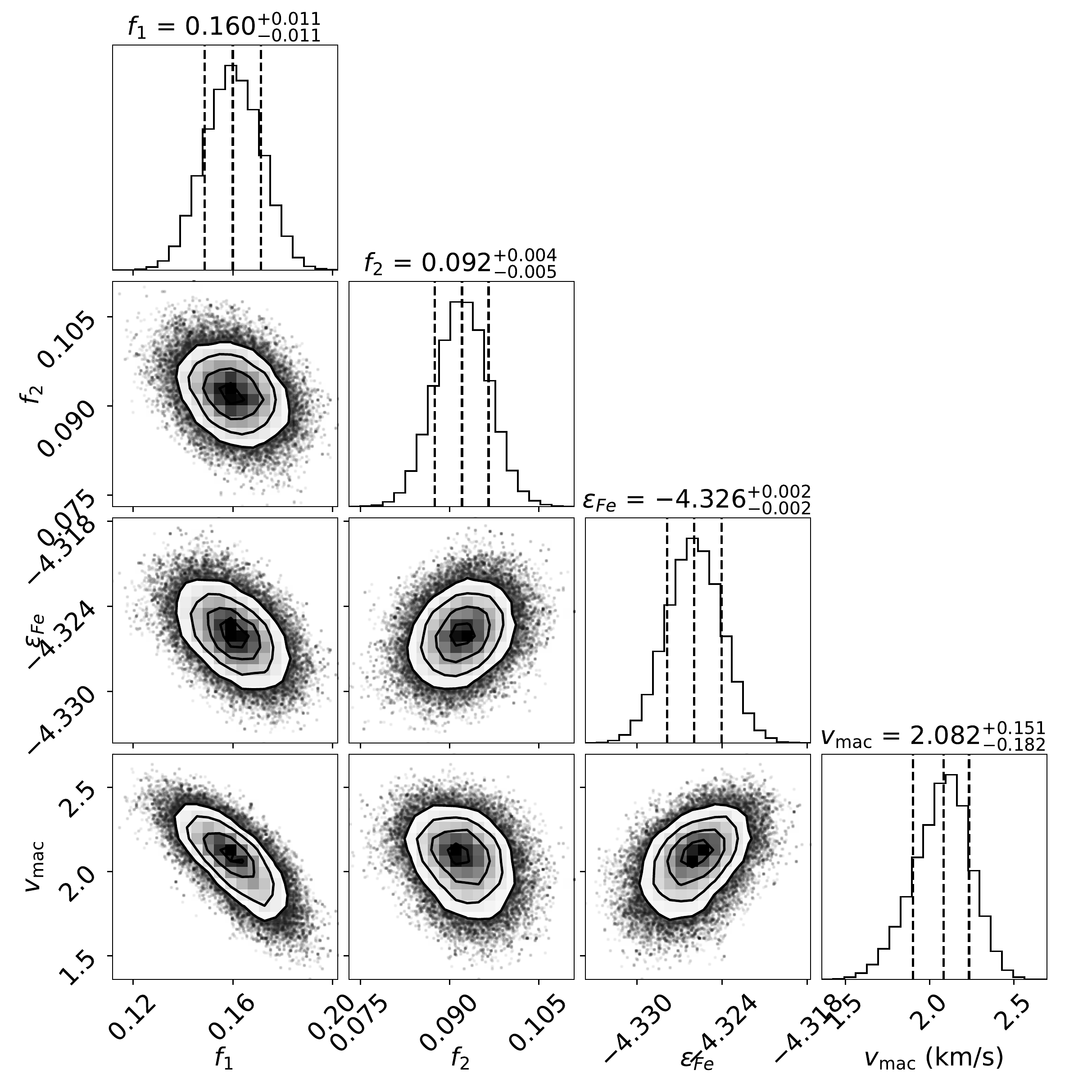}
    \includegraphics[width=0.45\textwidth]{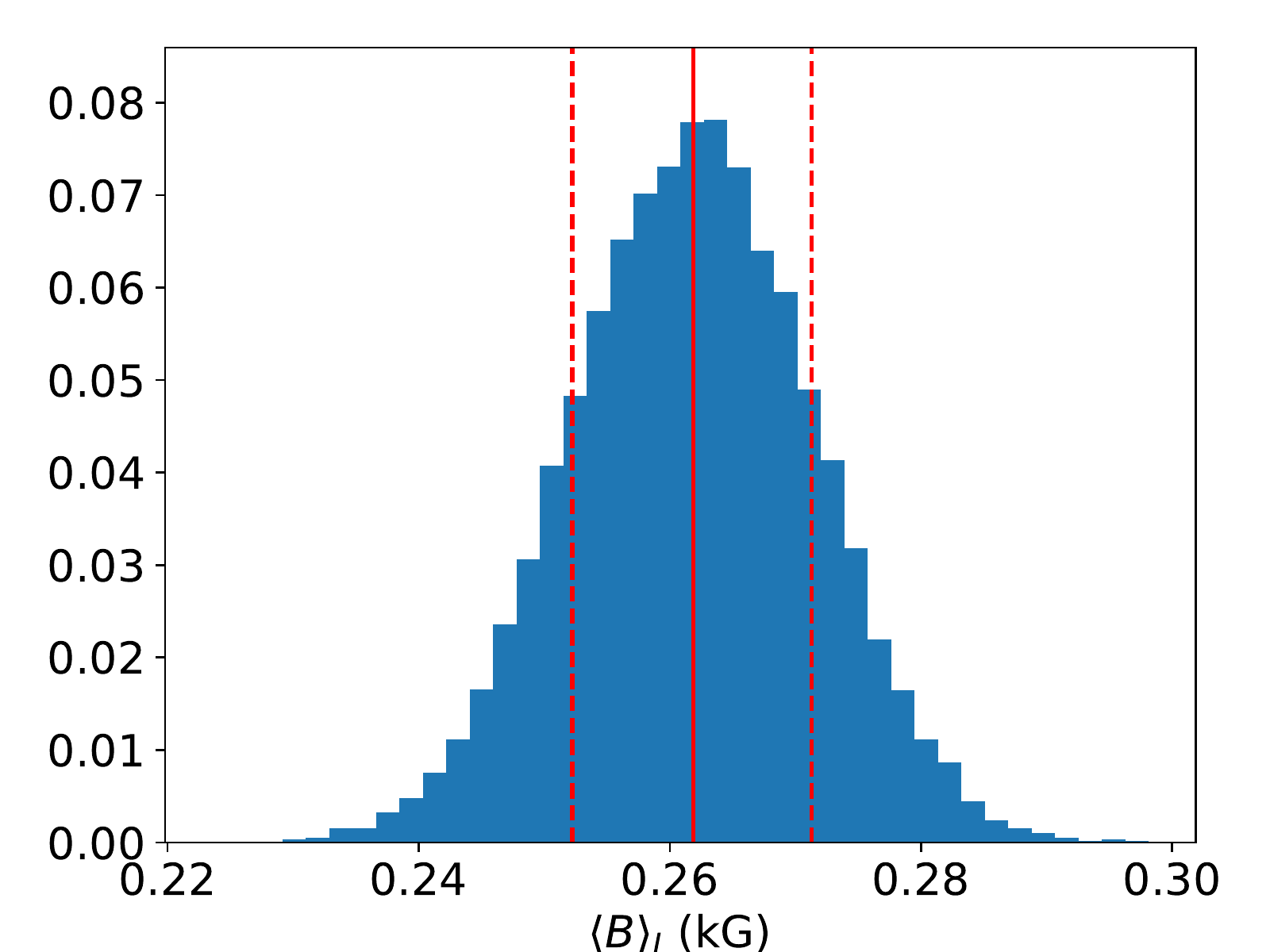}
    \includegraphics[width=0.45\textwidth]{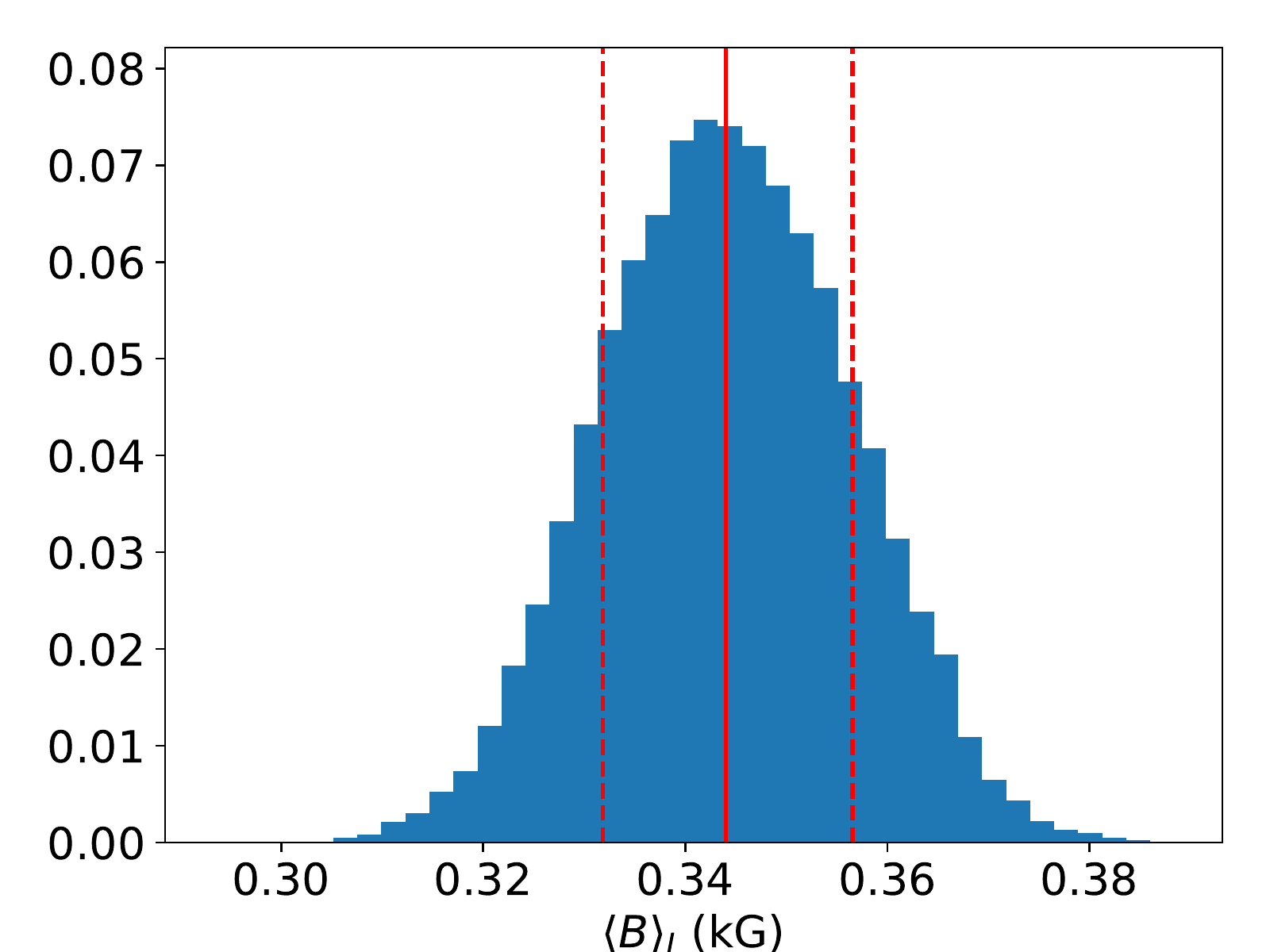}
    \captionof{figure}{Same as Fig. \ref{fig:HD1835_obs} but for HD 73256 observed on 2022 March 30.}
    \label{fig:HD73256_obs}
\end{minipage}
\vspace{0.3\textheight}
\subsection{HD 73350}
\begin{minipage}{1.0\textwidth}
    \centering
    \includegraphics[width=\textwidth]{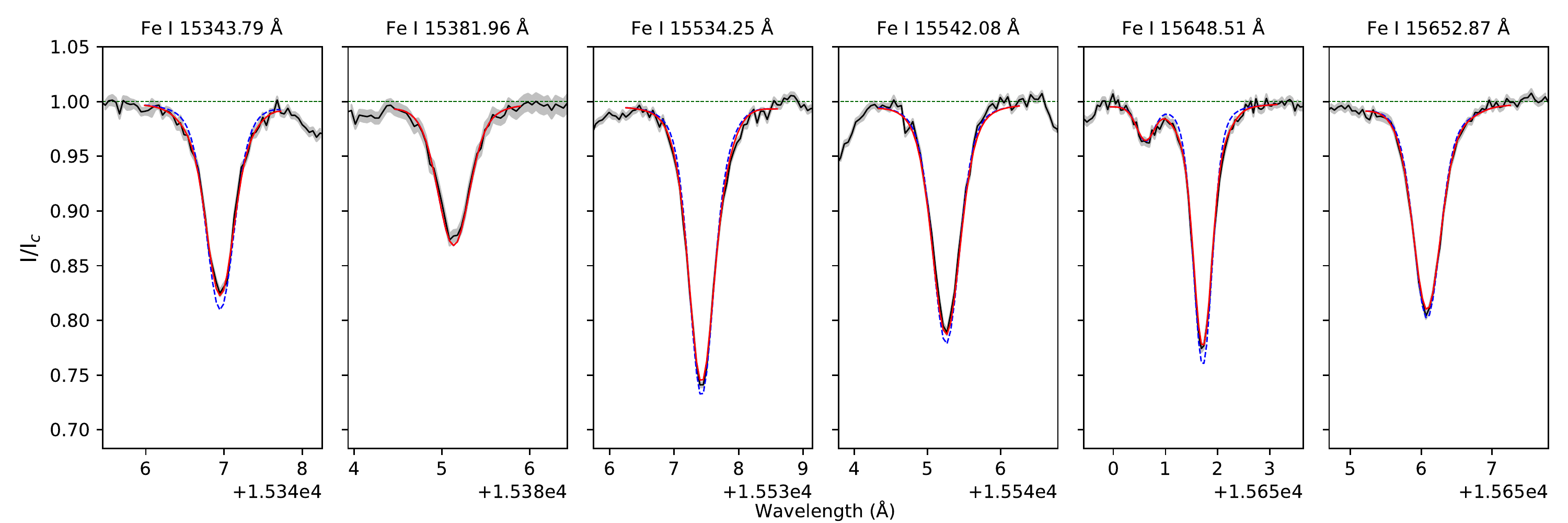}
    \includegraphics[width=0.49\textwidth]{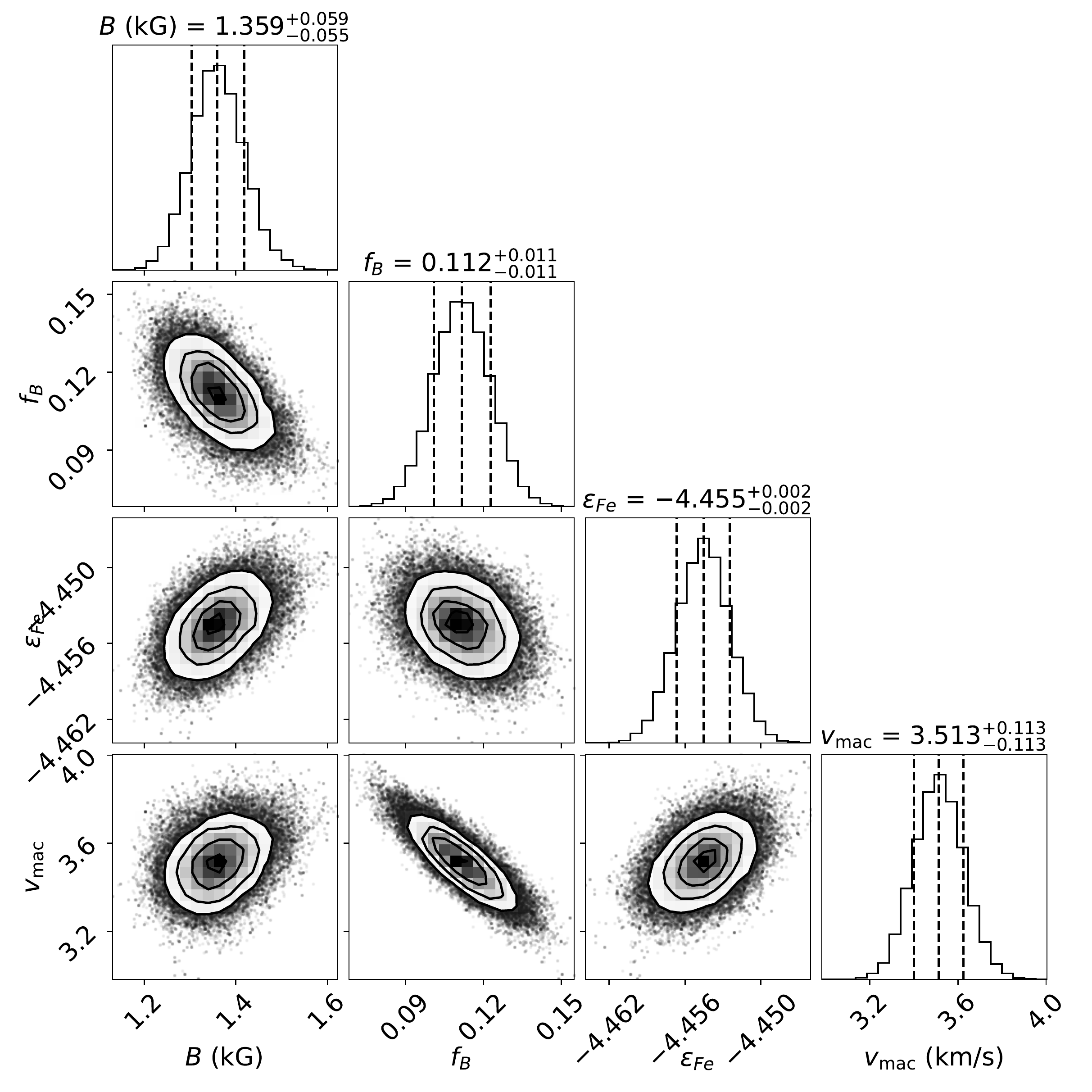}
    \includegraphics[width=0.49\textwidth]{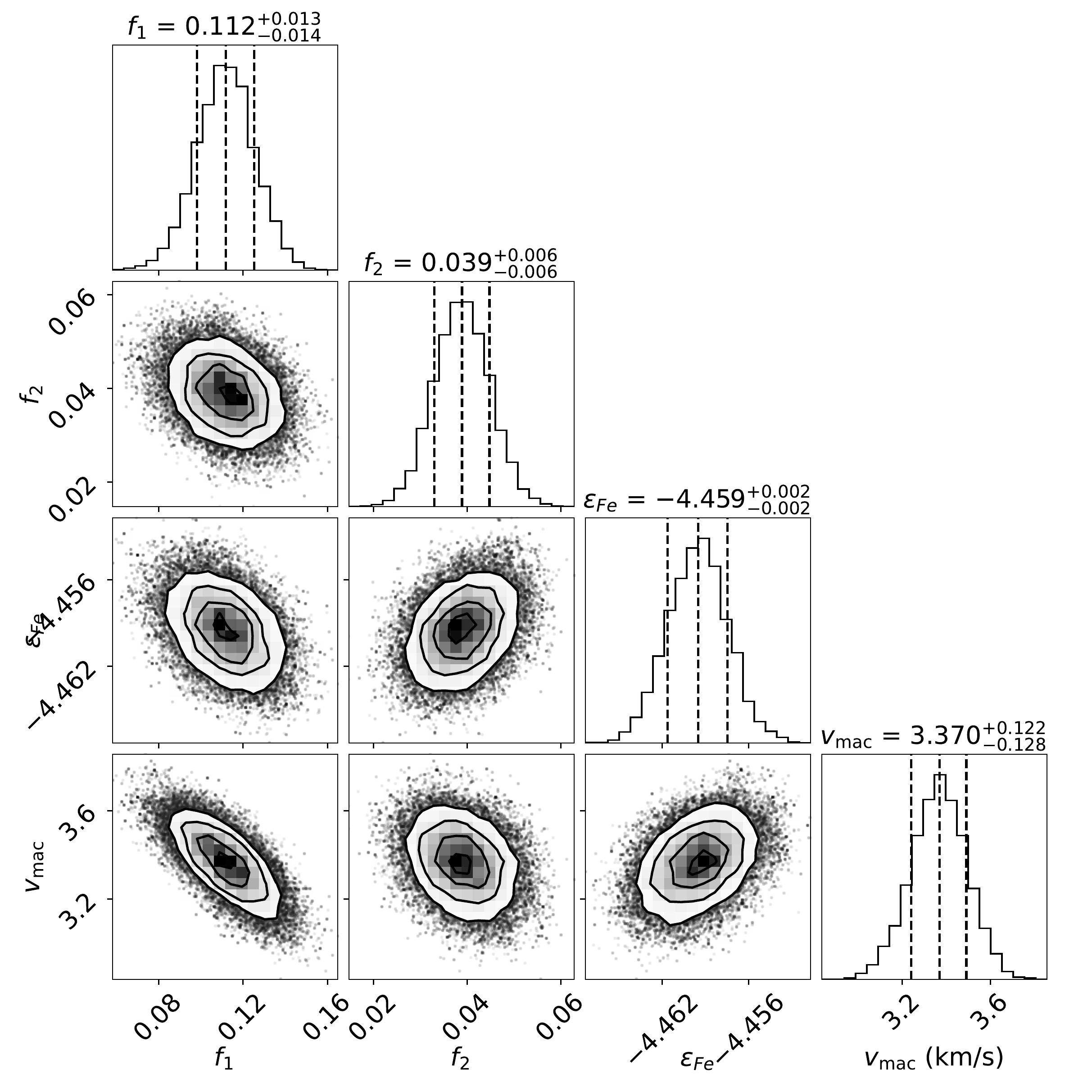}
    \includegraphics[width=0.45\textwidth]{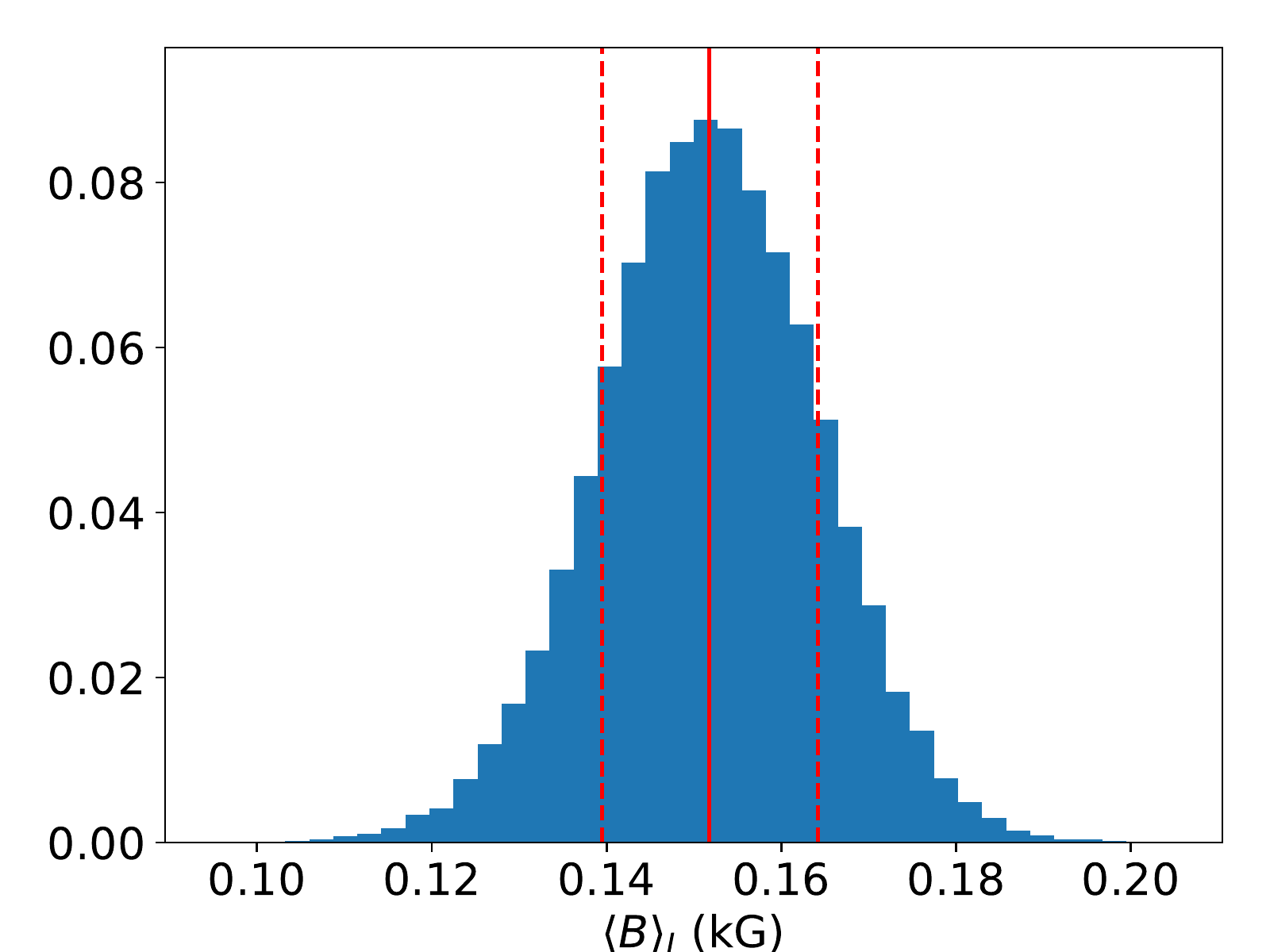}
    \includegraphics[width=0.45\textwidth]{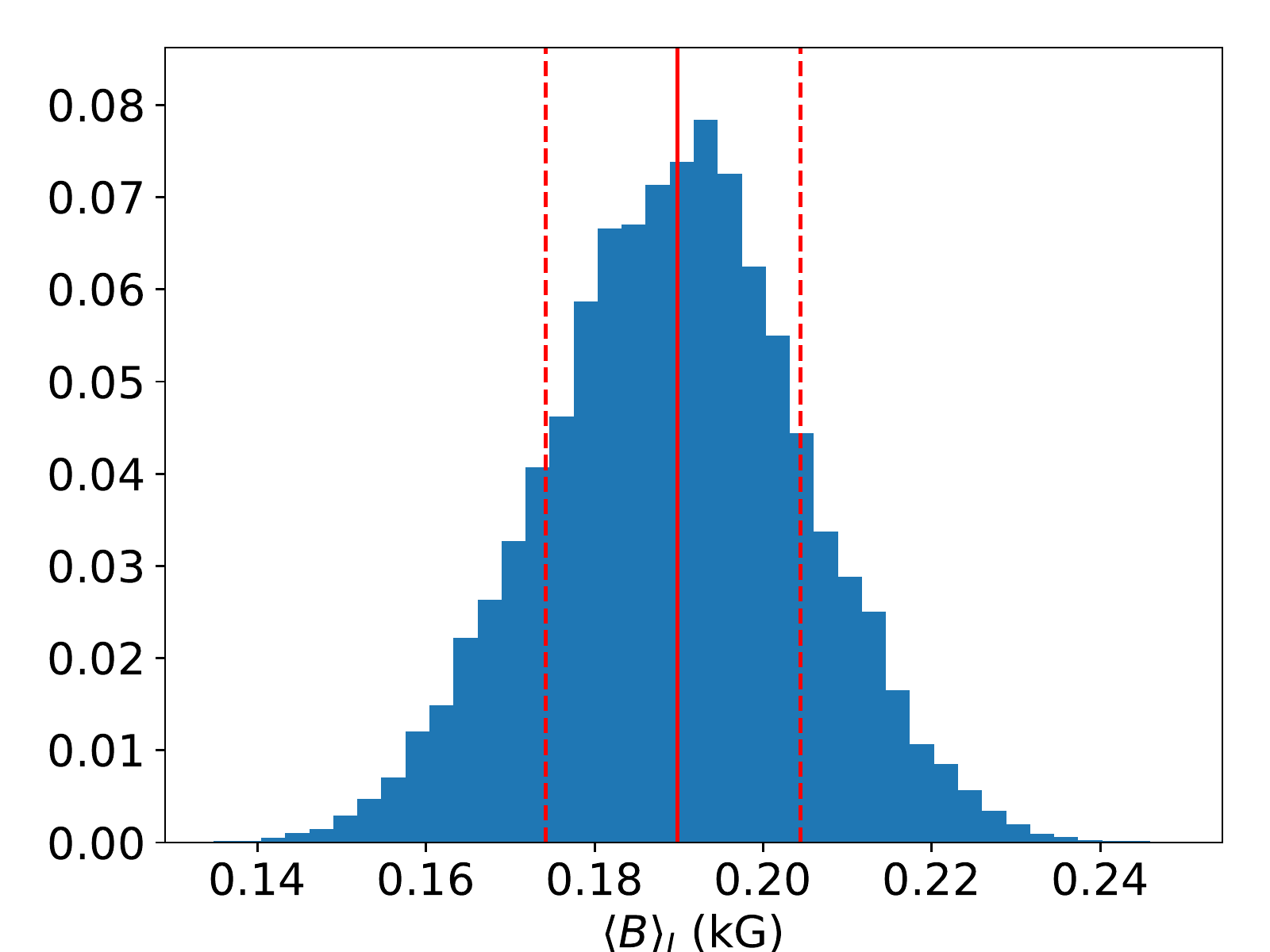}
    \captionof{figure}{Same as Fig. \ref{fig:HD1835_obs} but for HD 73350 observed on 2022 March 10.}
    \label{fig:HD73350_obs}
\end{minipage}
\subsection{HD 75732}
\begin{minipage}{1.0\textwidth}
    \centering
    \includegraphics[width=\textwidth]{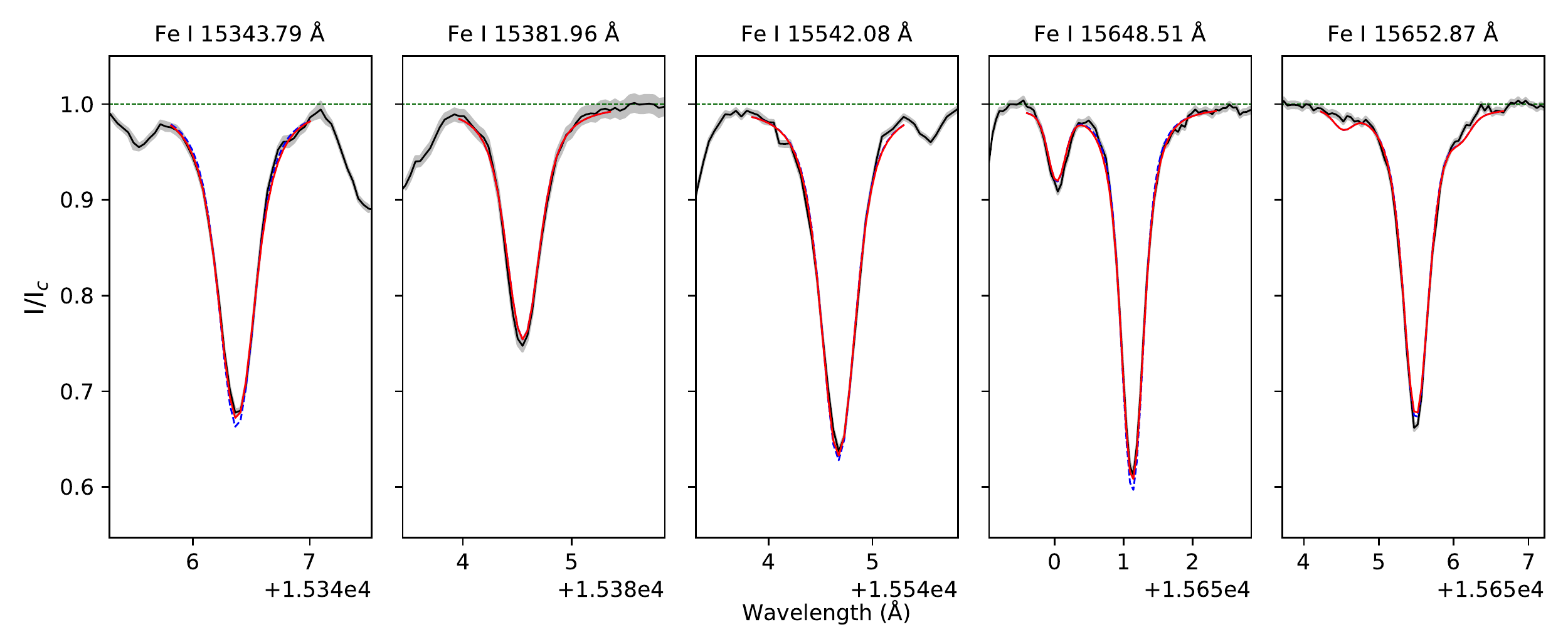}
    \includegraphics[width=0.49\textwidth]{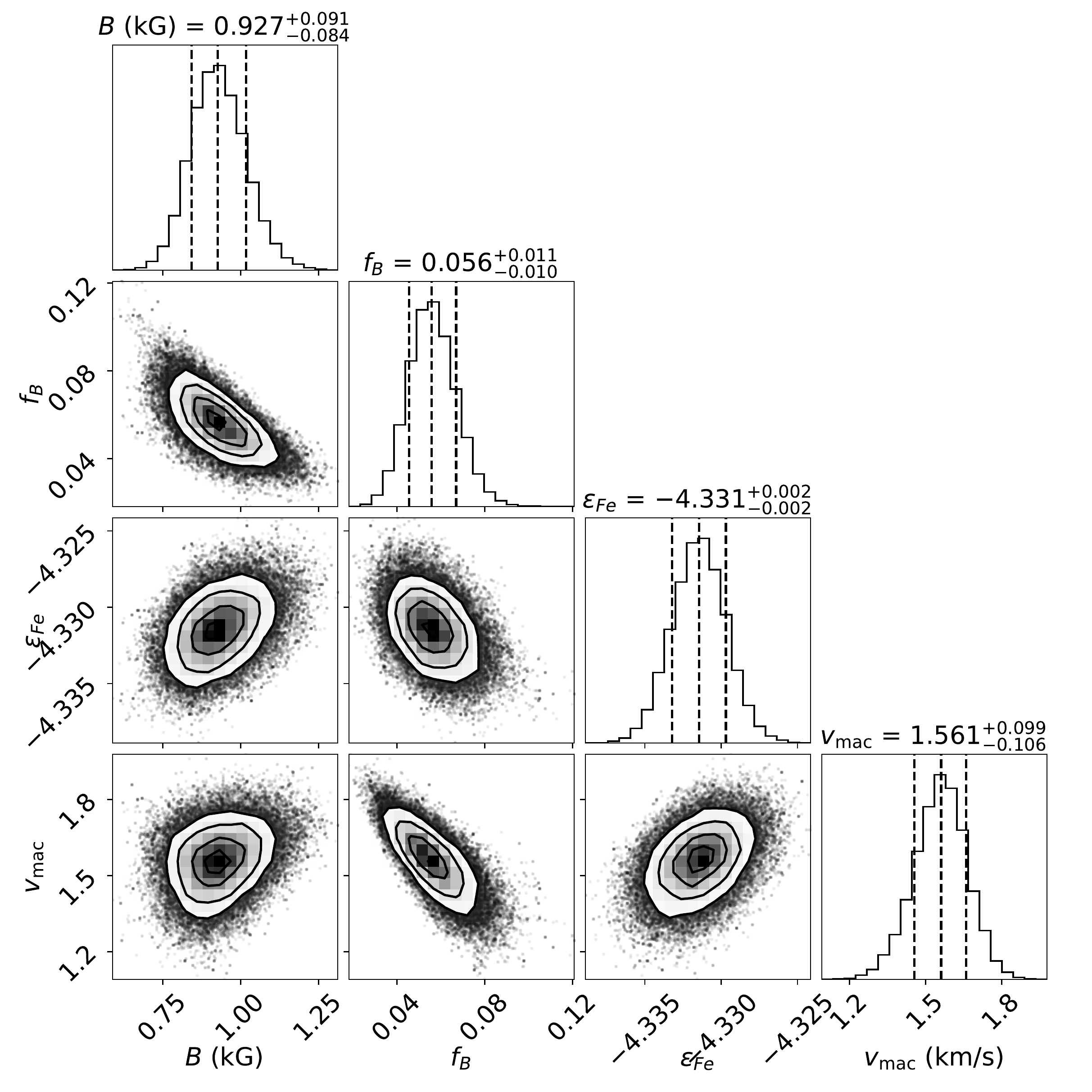}
    \includegraphics[width=0.49\textwidth]{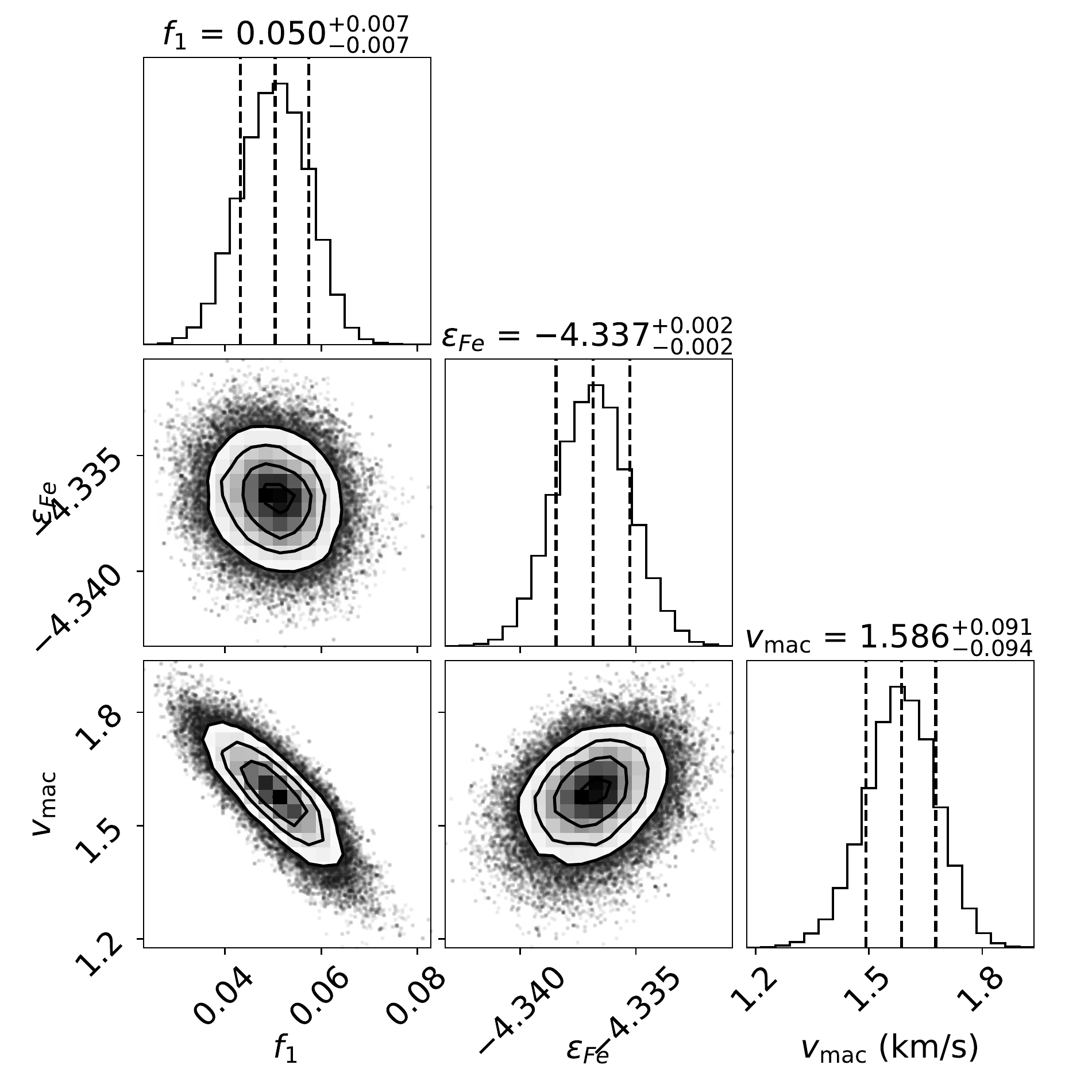}
    \includegraphics[width=0.45\textwidth]{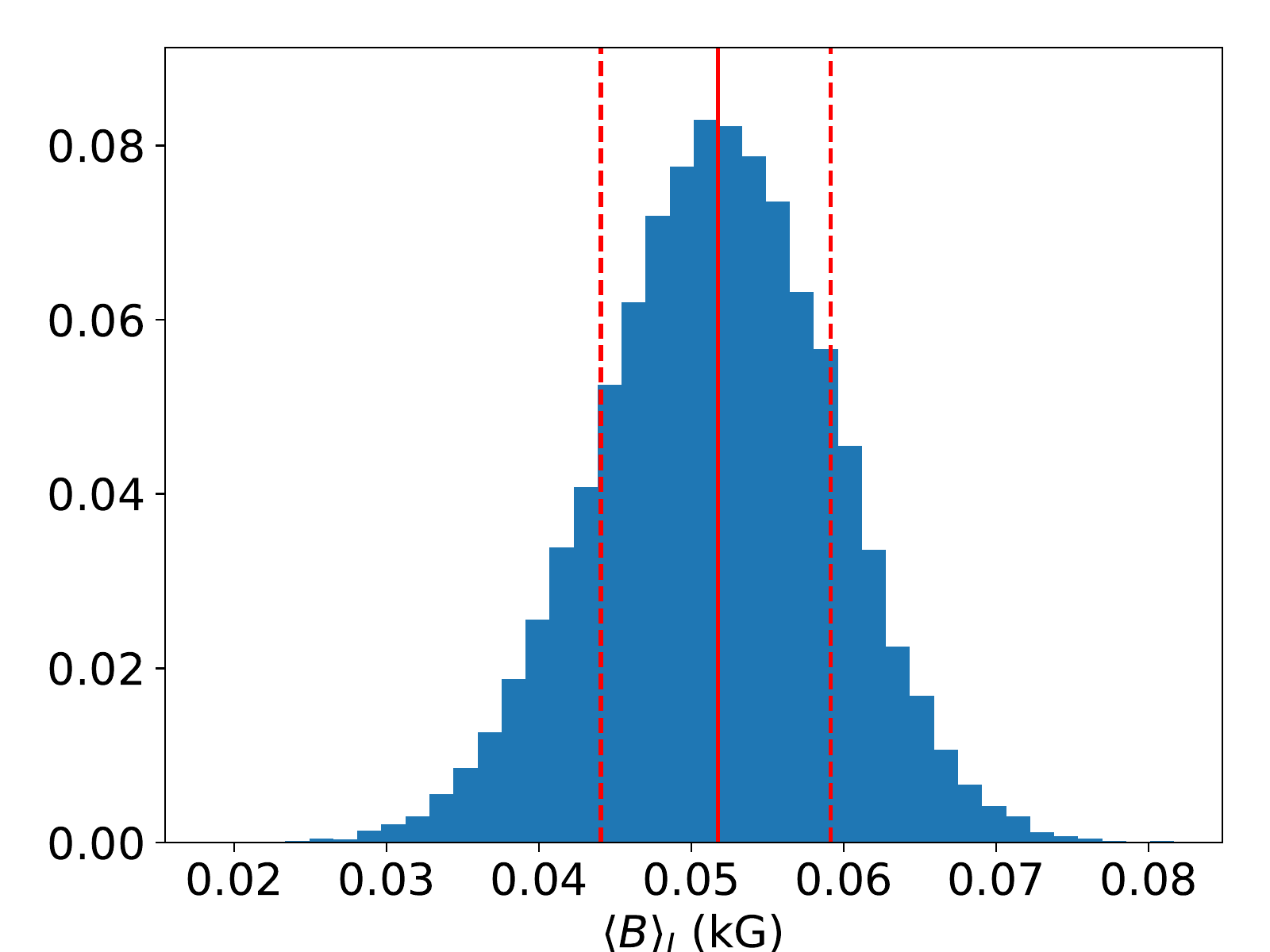}
    \includegraphics[width=0.45\textwidth]{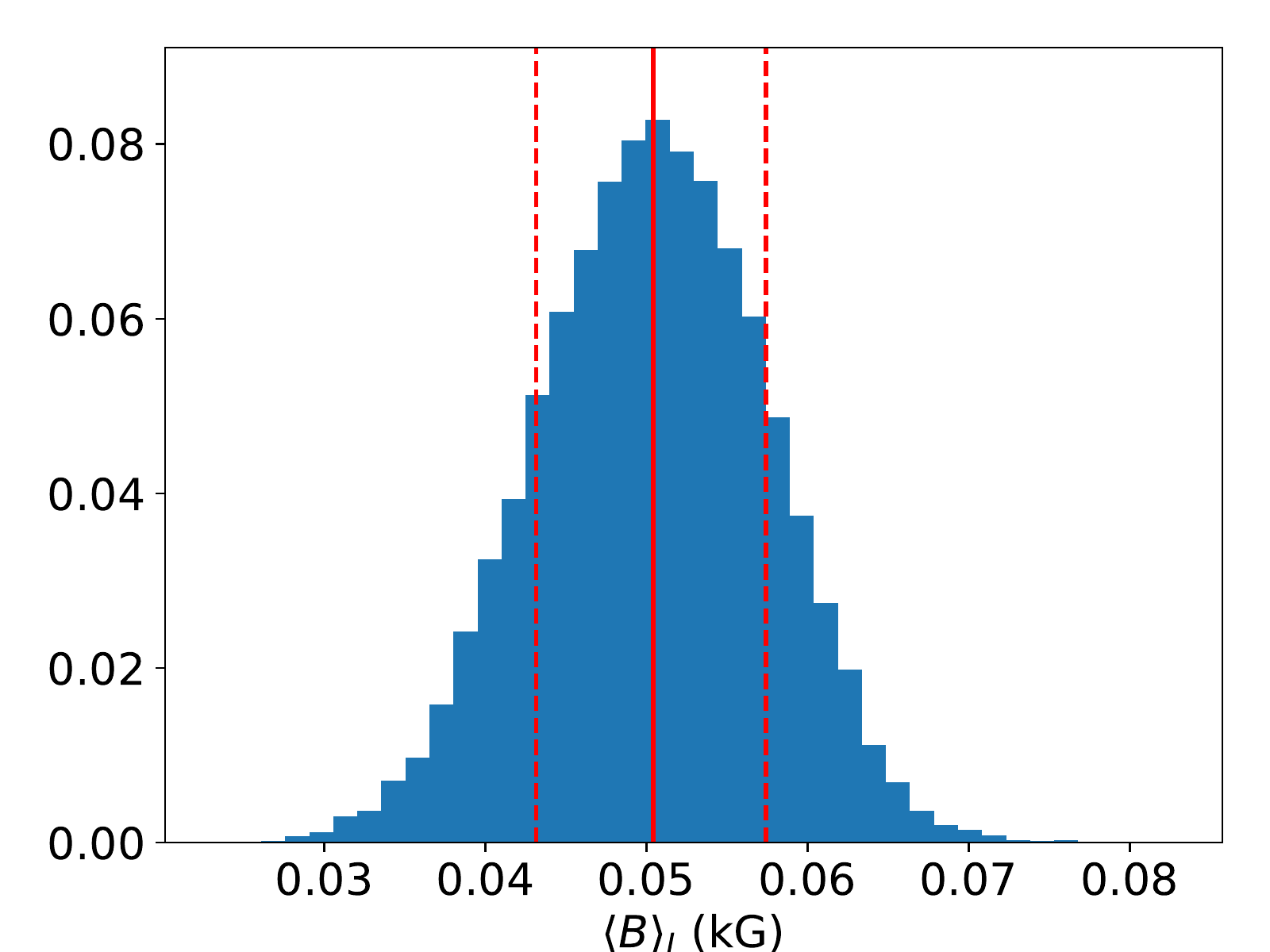}
    \captionof{figure}{Same as Fig. \ref{fig:HD1835_obs} but for HD 75732.}
    \label{fig:HD75732_obs}
\end{minipage}
\vspace{0.3\textheight}
\subsection{HD 76151}
\begin{minipage}{1.0\textwidth}
    \centering
    \includegraphics[width=\textwidth]{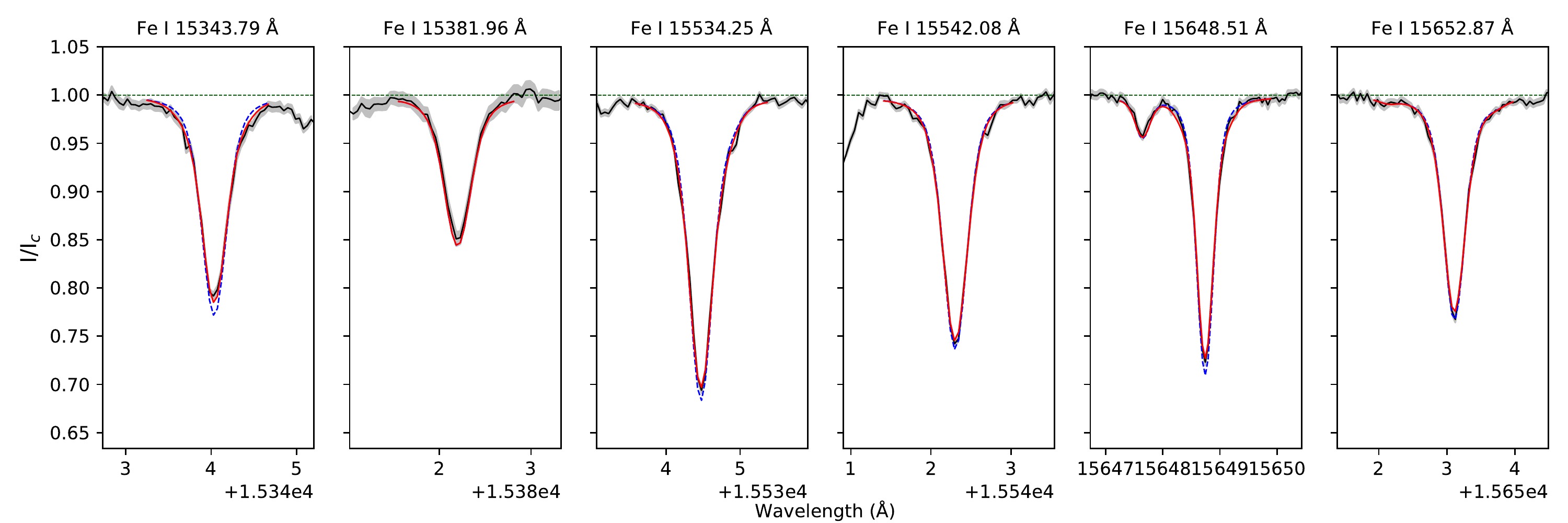}
    \includegraphics[width=0.49\textwidth]{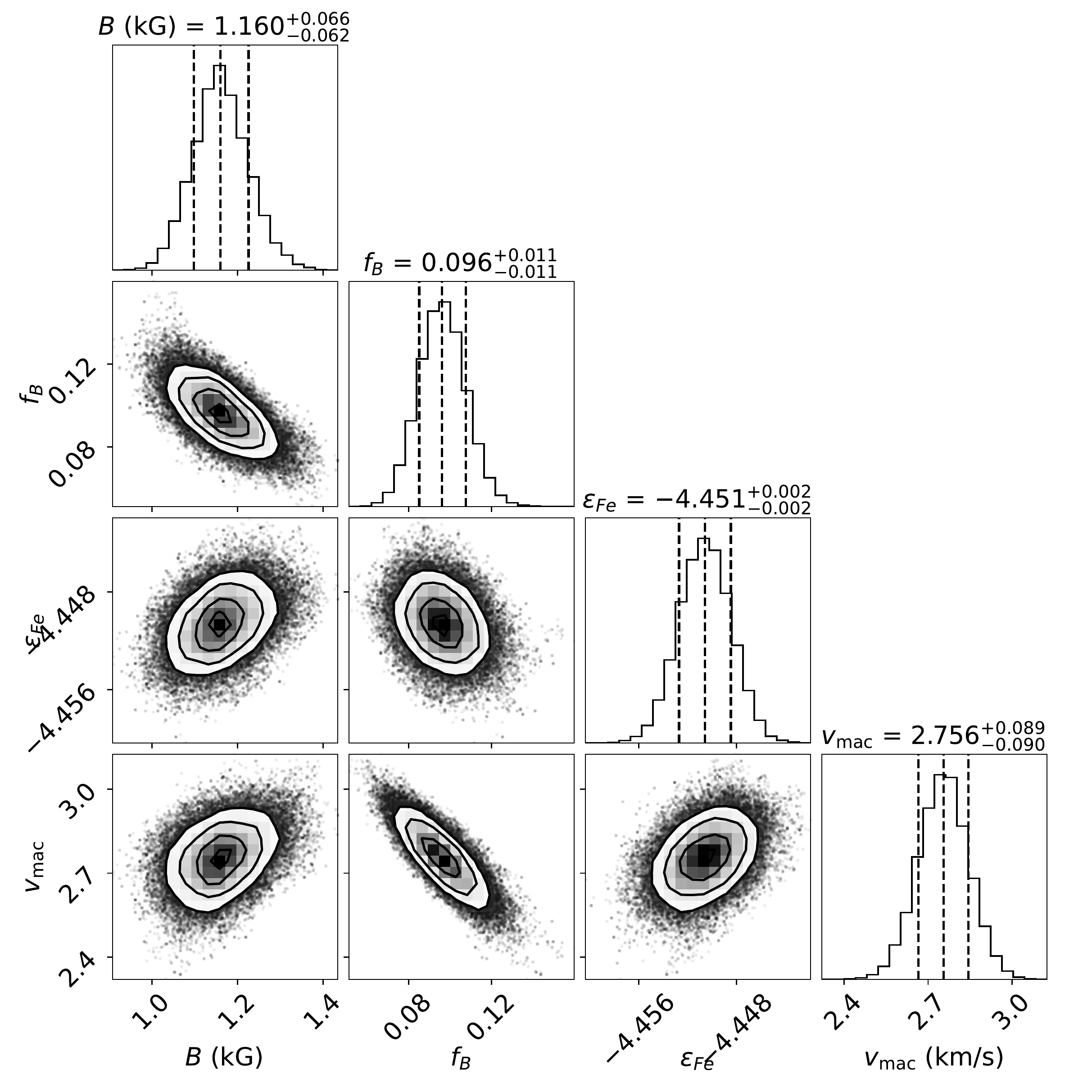}
    \includegraphics[width=0.49\textwidth]{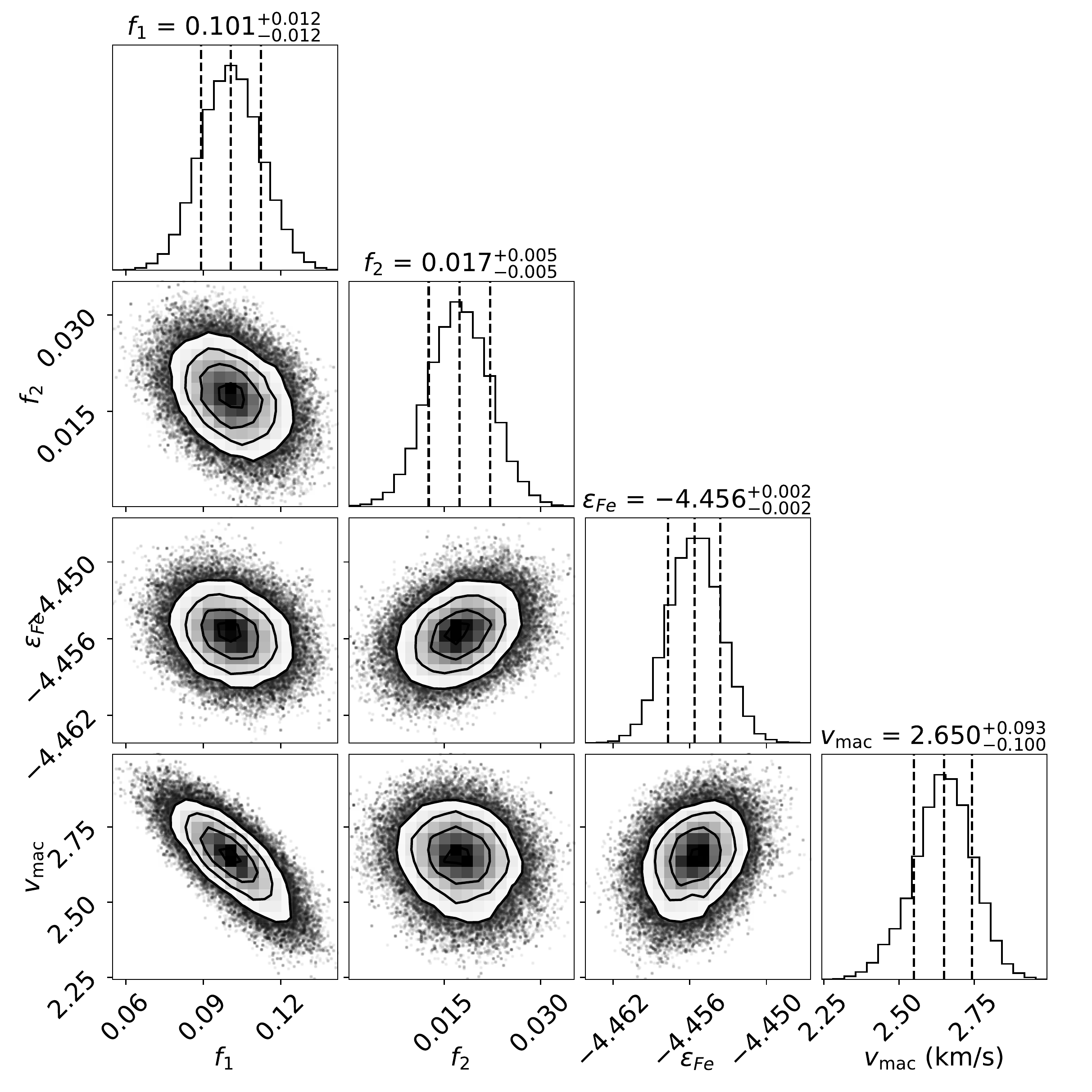}
    \includegraphics[width=0.45\textwidth]{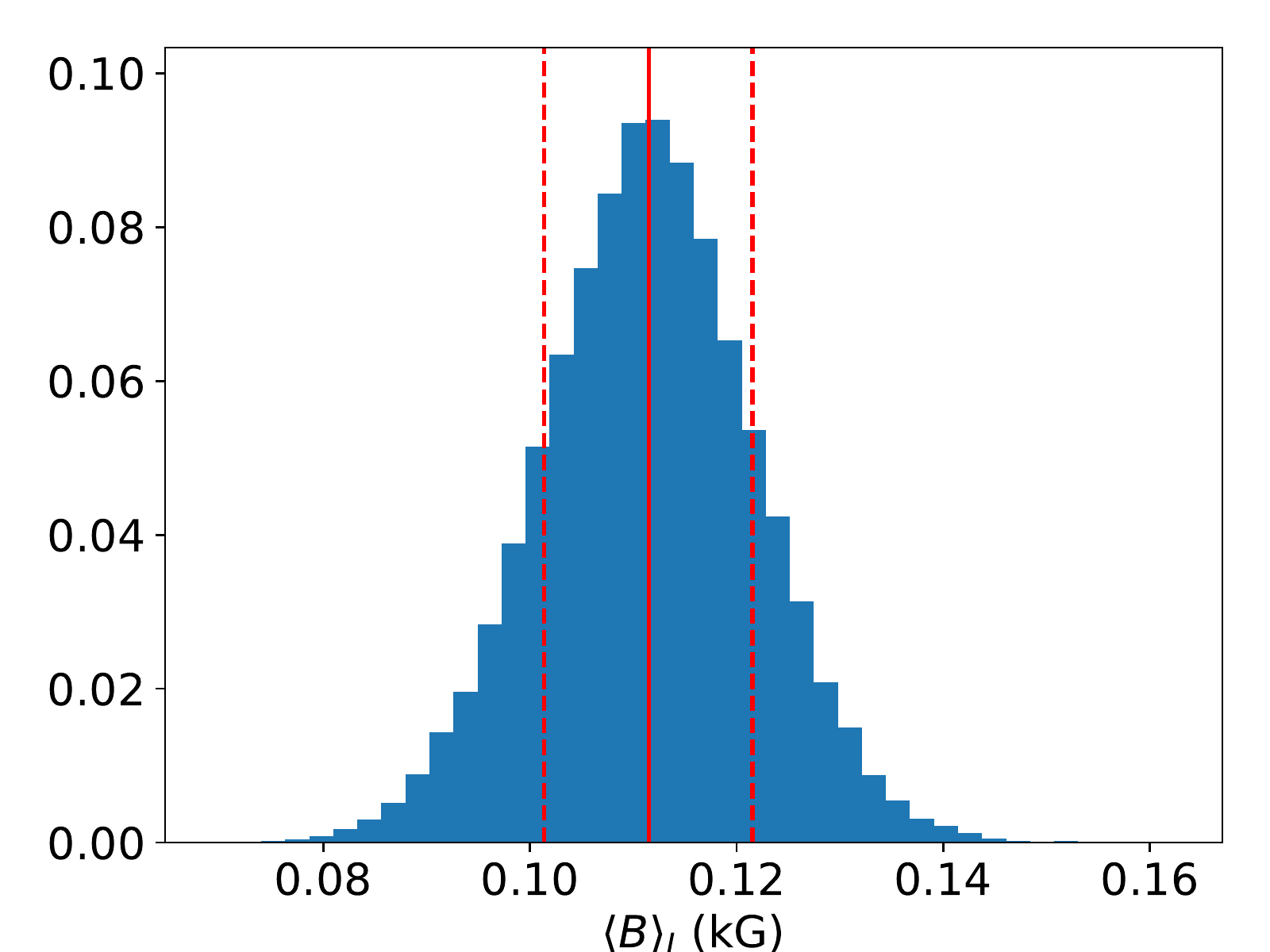}
    \includegraphics[width=0.45\textwidth]{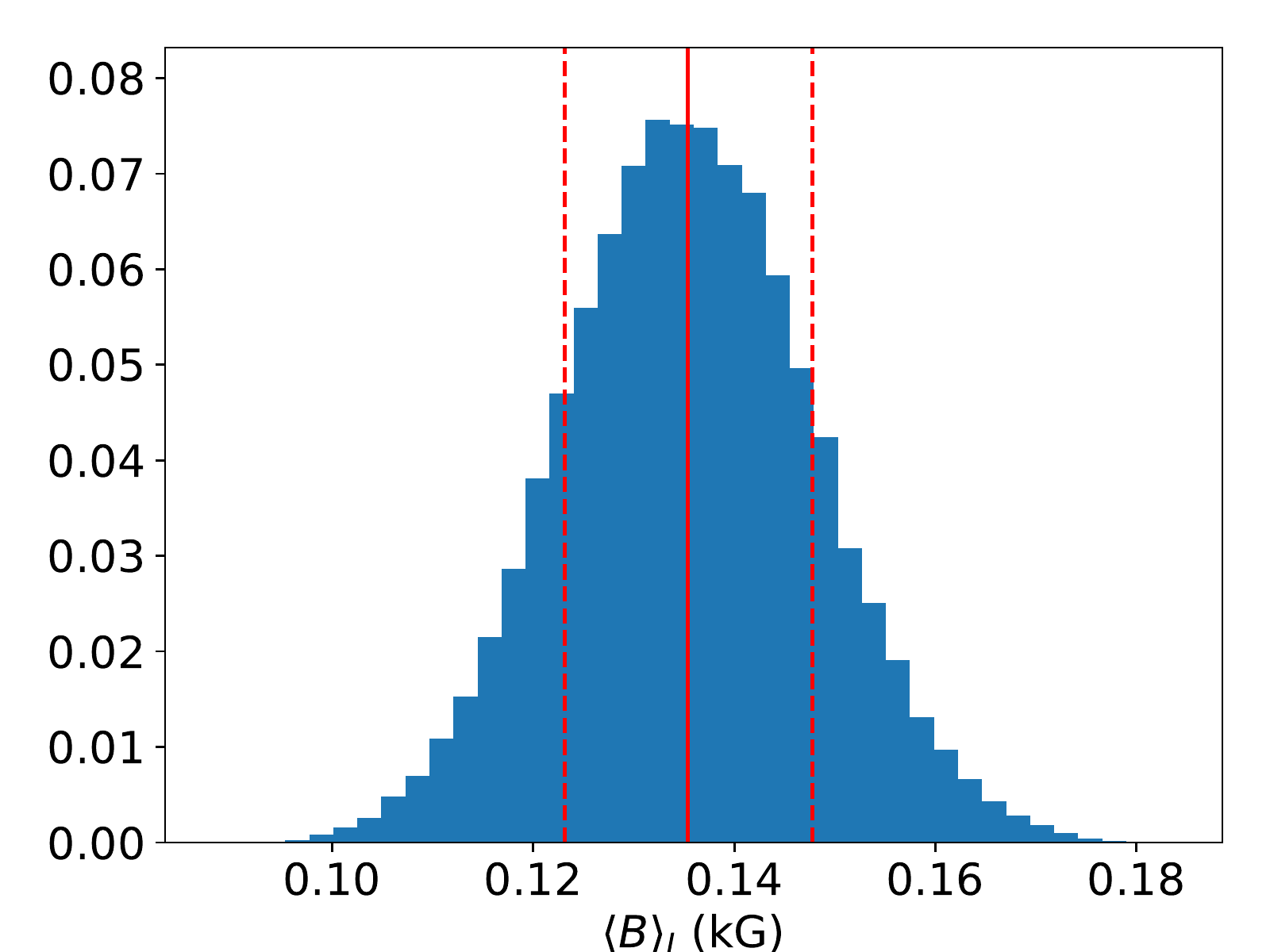}
    \captionof{figure}{Same as Fig. \ref{fig:HD1835_obs} but for HD 76151 observed on 2022 February 7.}
    \label{fig:HD76151_obs}
\end{minipage}
\vspace{0.3\textheight}
\subsection{HD 102195}
\begin{minipage}{1.0\textwidth}
    \centering
    \includegraphics[width=\textwidth]{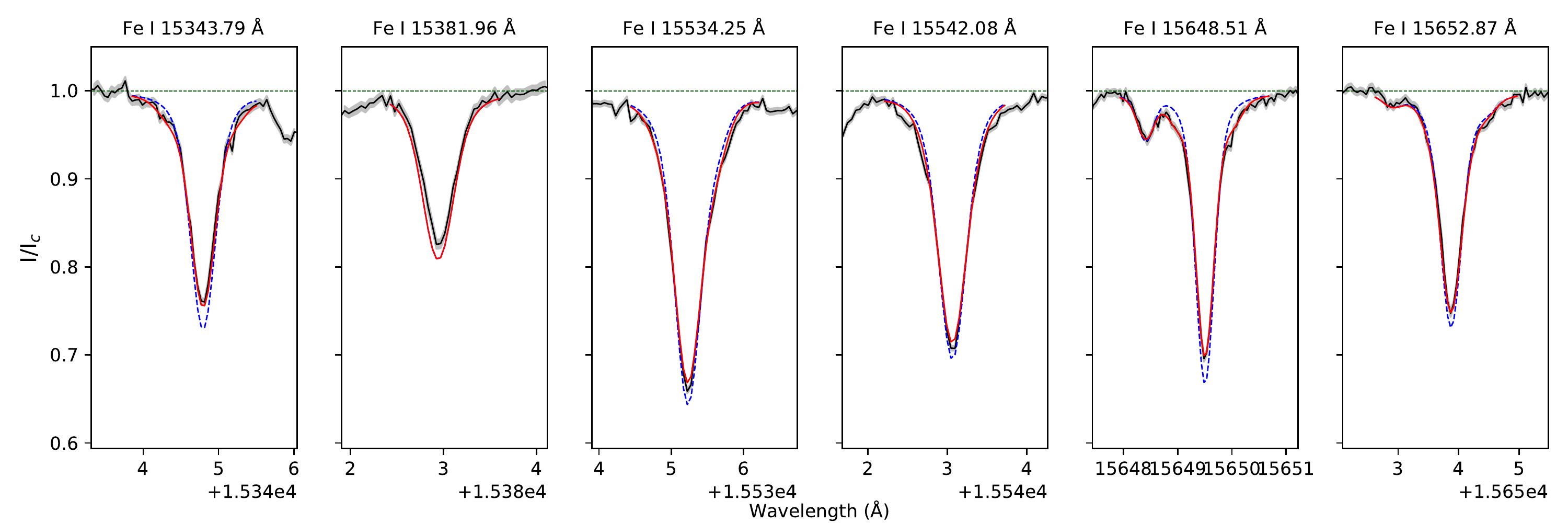}
    \includegraphics[width=0.49\textwidth]{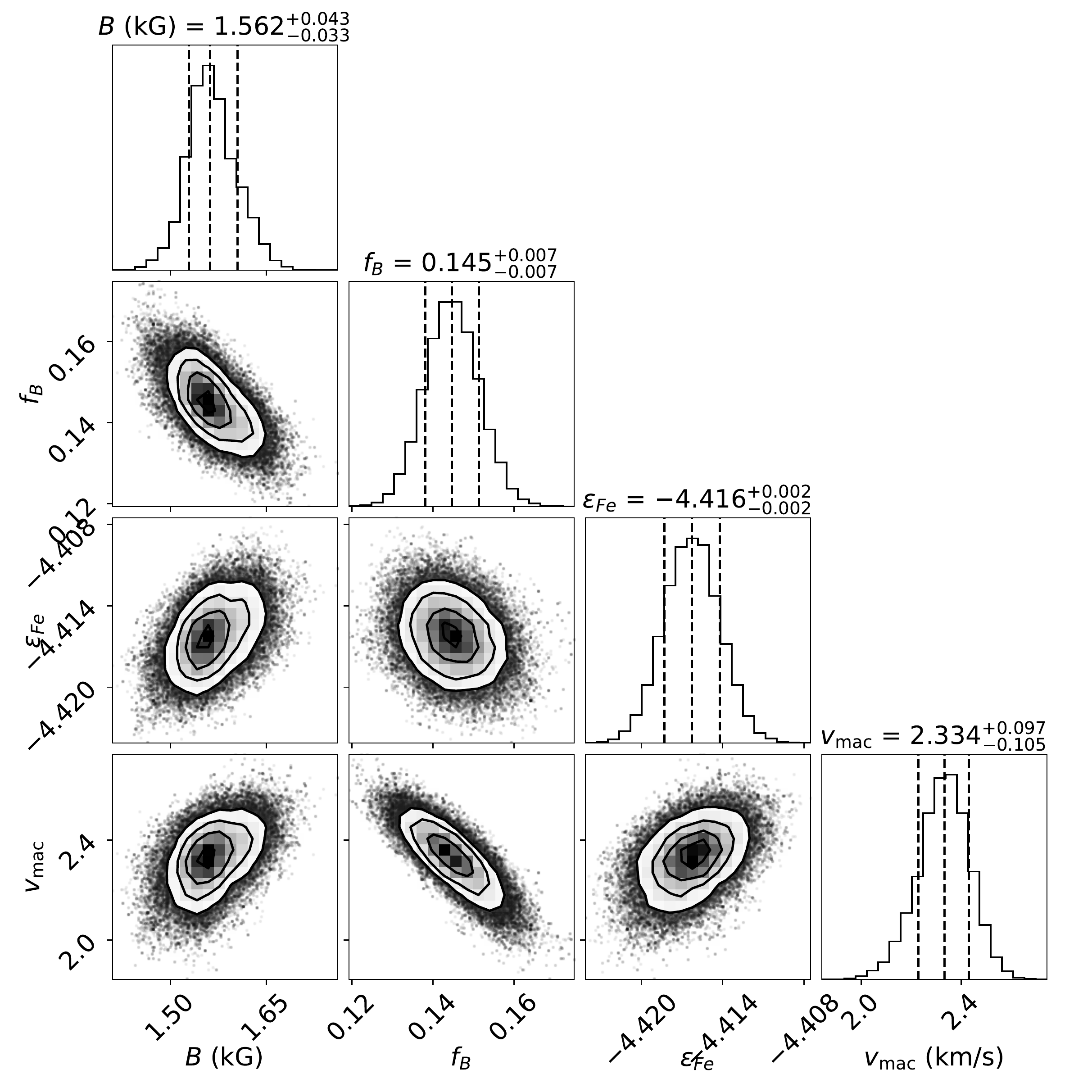}
    \includegraphics[width=0.49\textwidth]{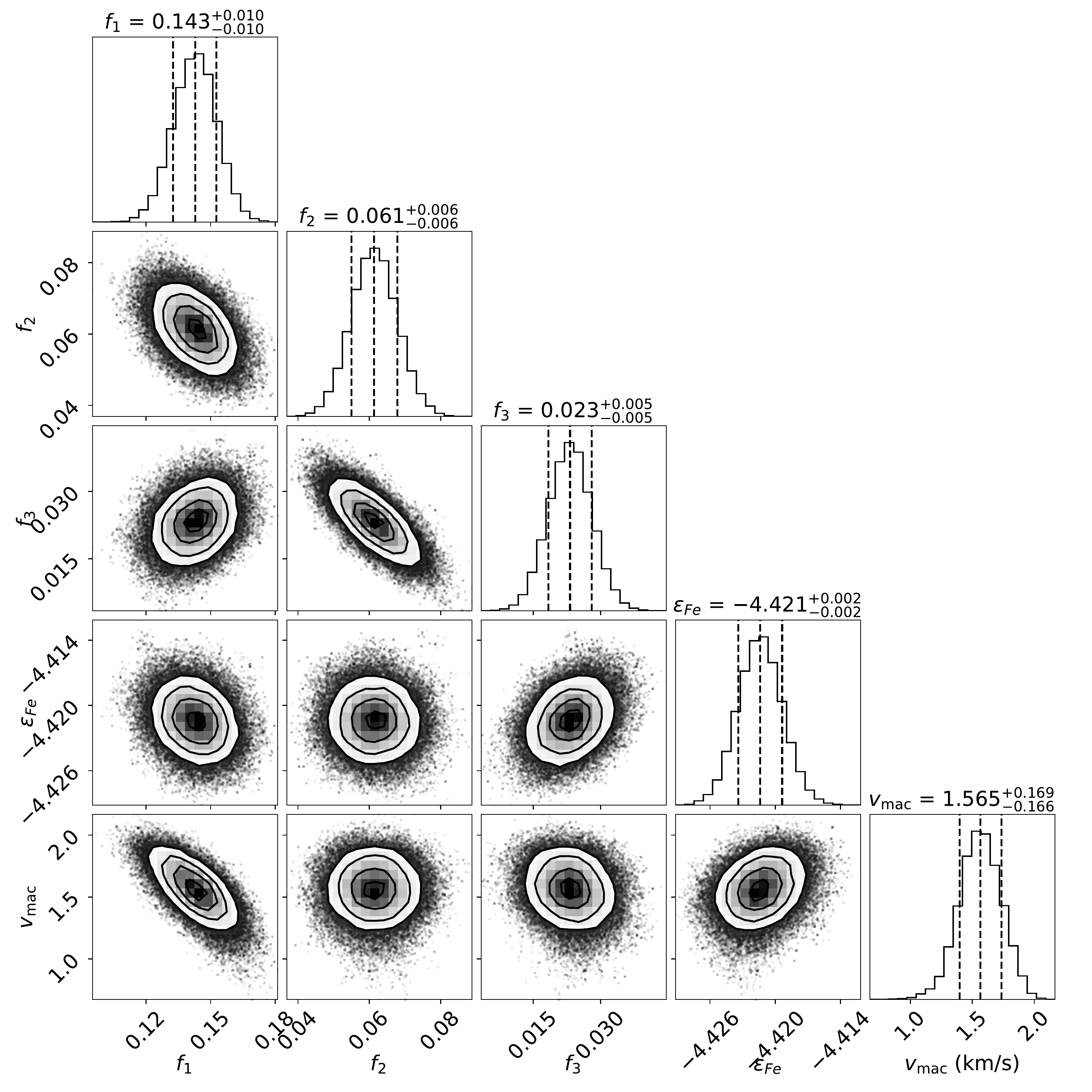}
    \includegraphics[width=0.45\textwidth]{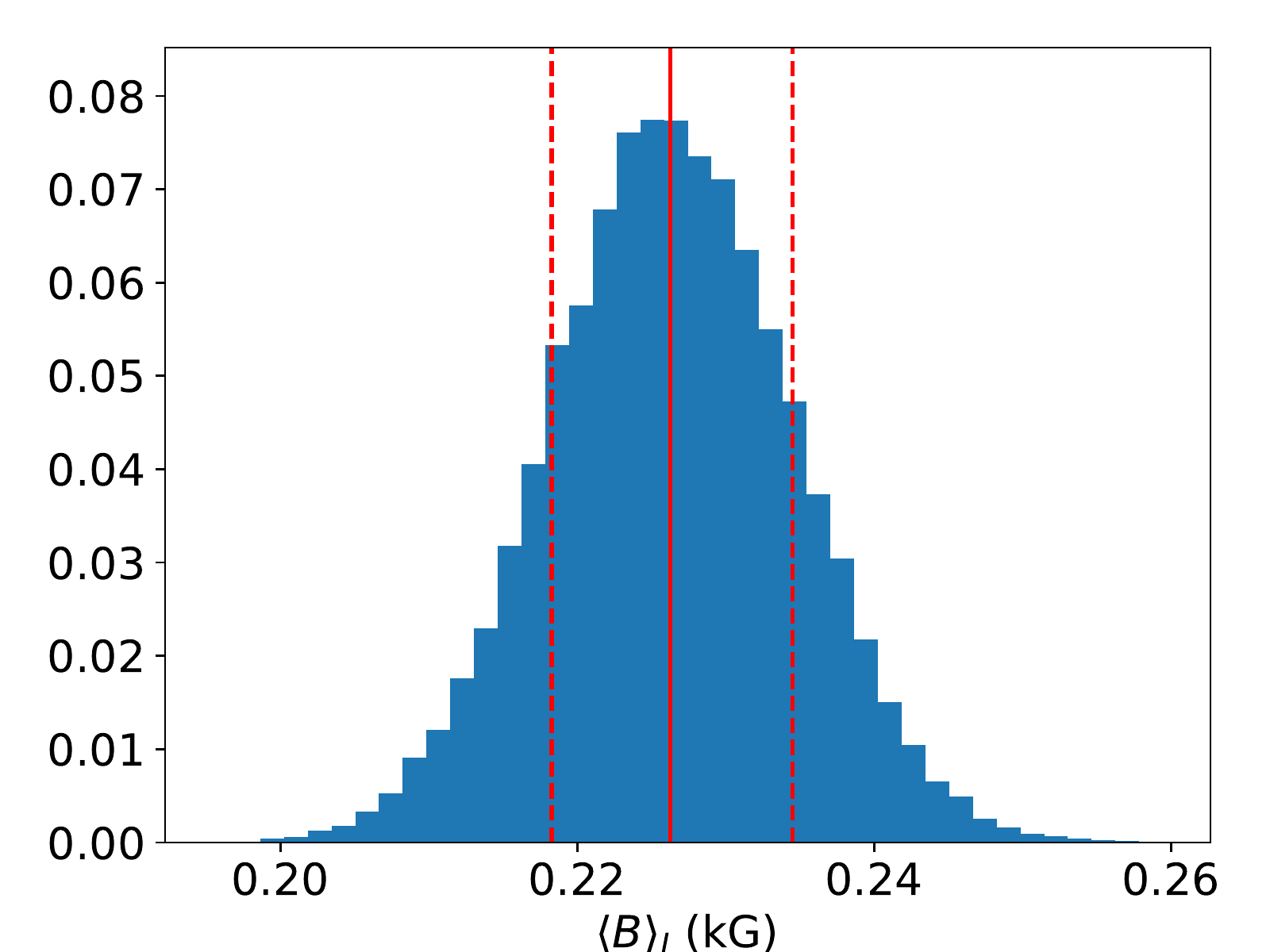}
    \includegraphics[width=0.45\textwidth]{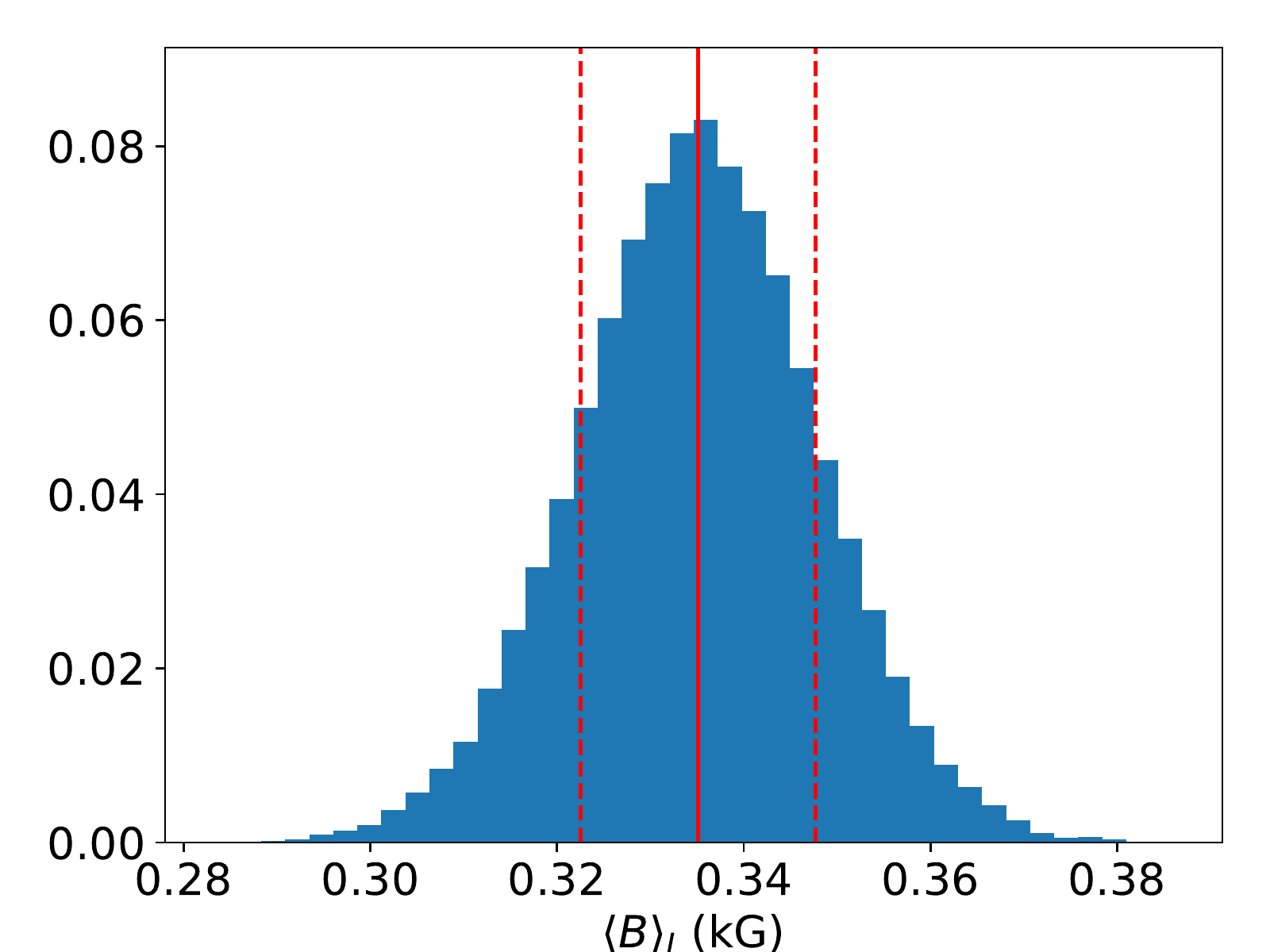}
    \captionof{figure}{Same as Fig. \ref{fig:HD1835_obs} but for HD 102195.}
    \label{fig:HD102195_obs}
\end{minipage}
\vspace{0.3\textheight}
\subsection{HD 130322}
\begin{minipage}{1.0\textwidth}
    \centering
    \includegraphics[width=\textwidth]{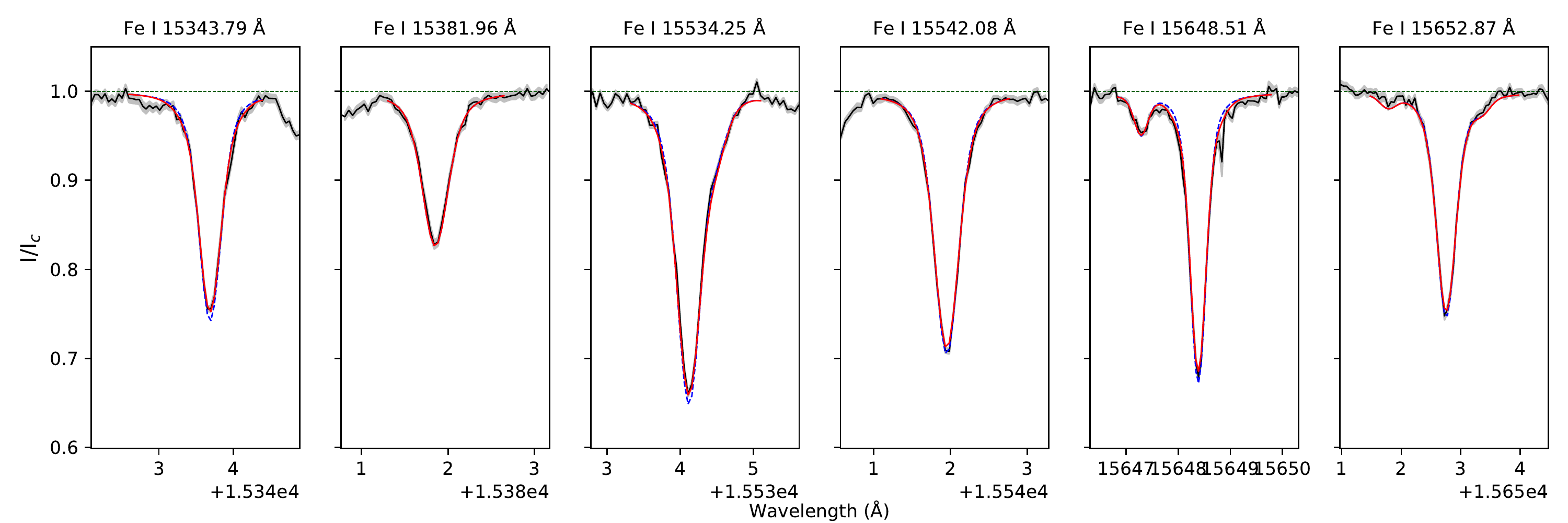}
    \includegraphics[width=0.49\textwidth]{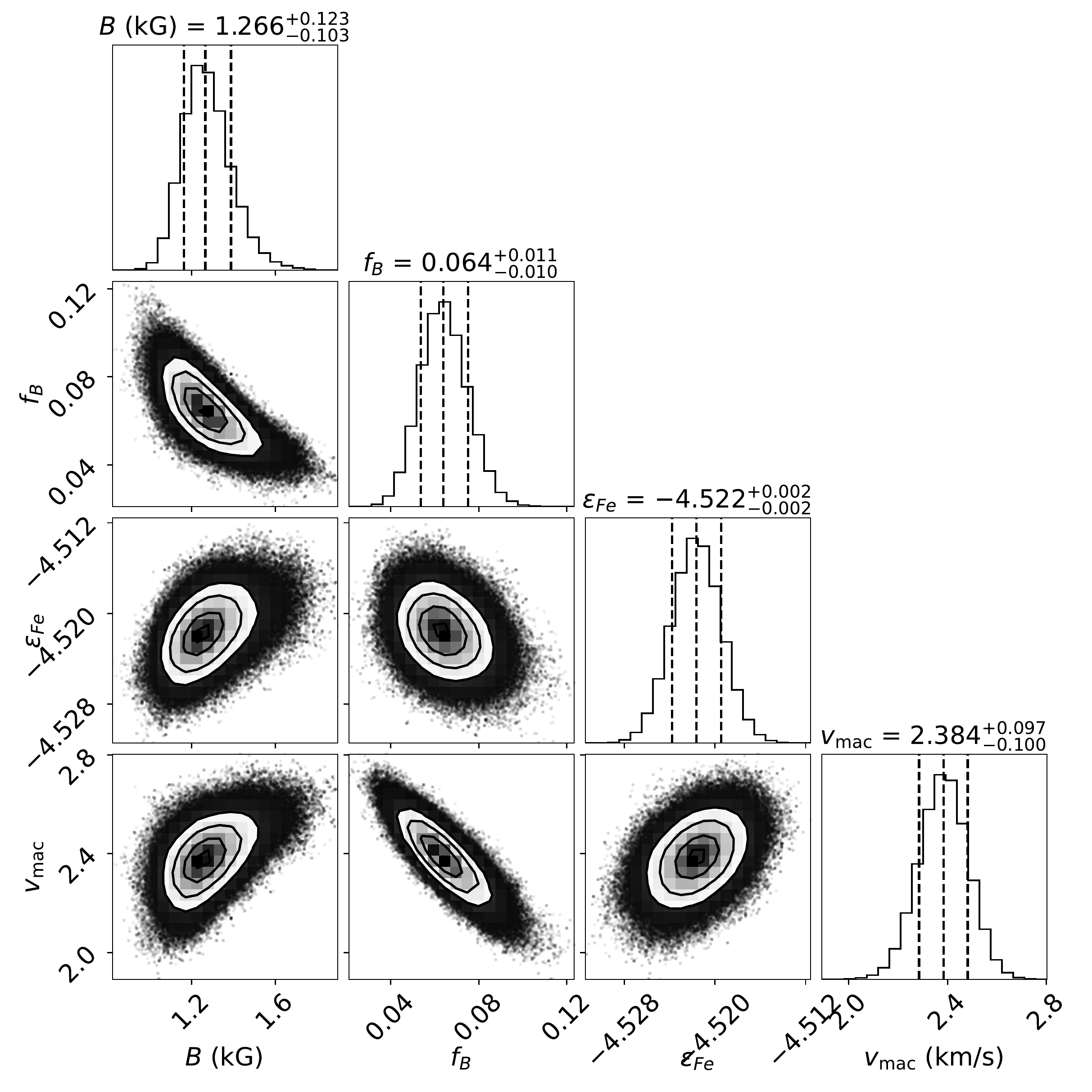}
    \includegraphics[width=0.49\textwidth]{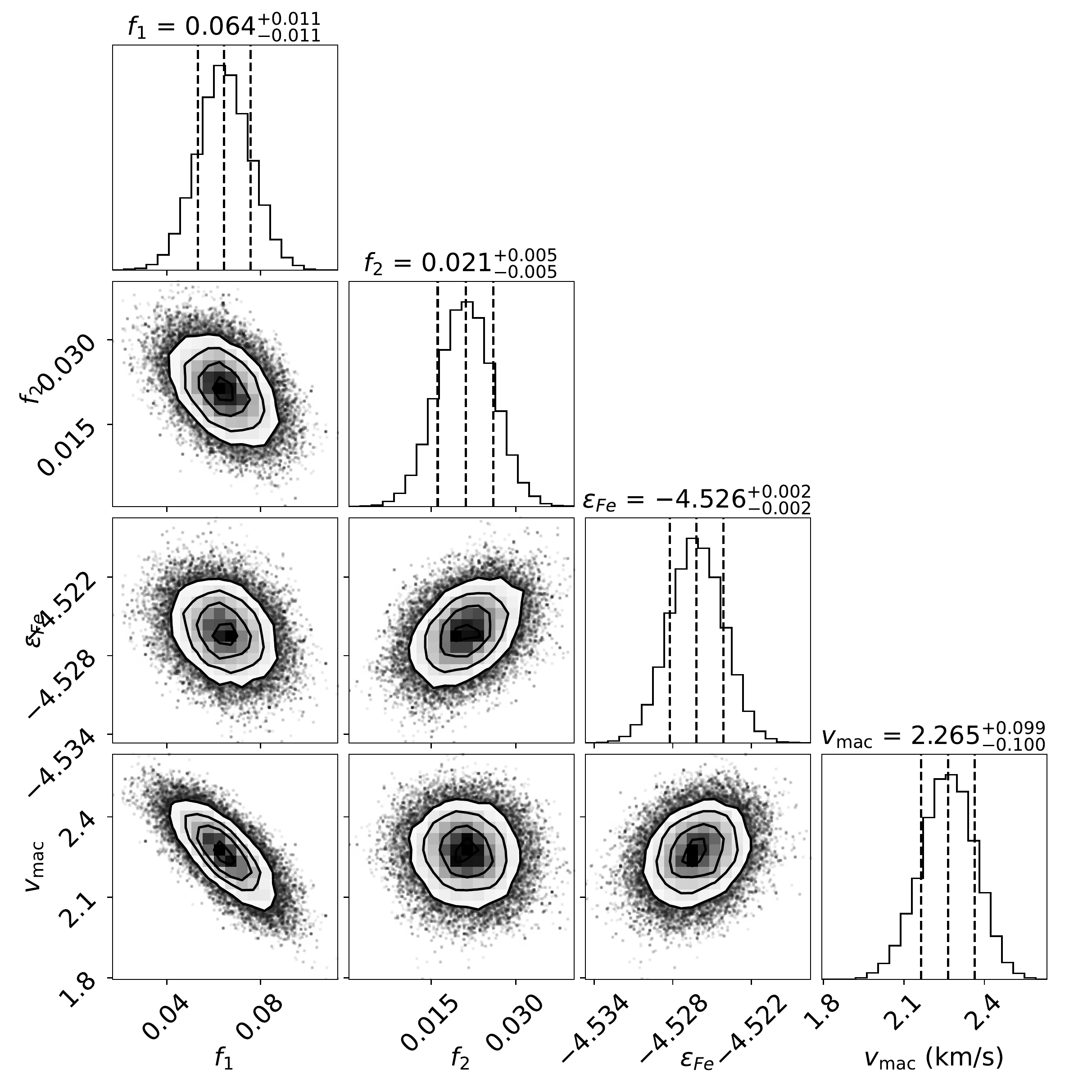}
    \includegraphics[width=0.45\textwidth]{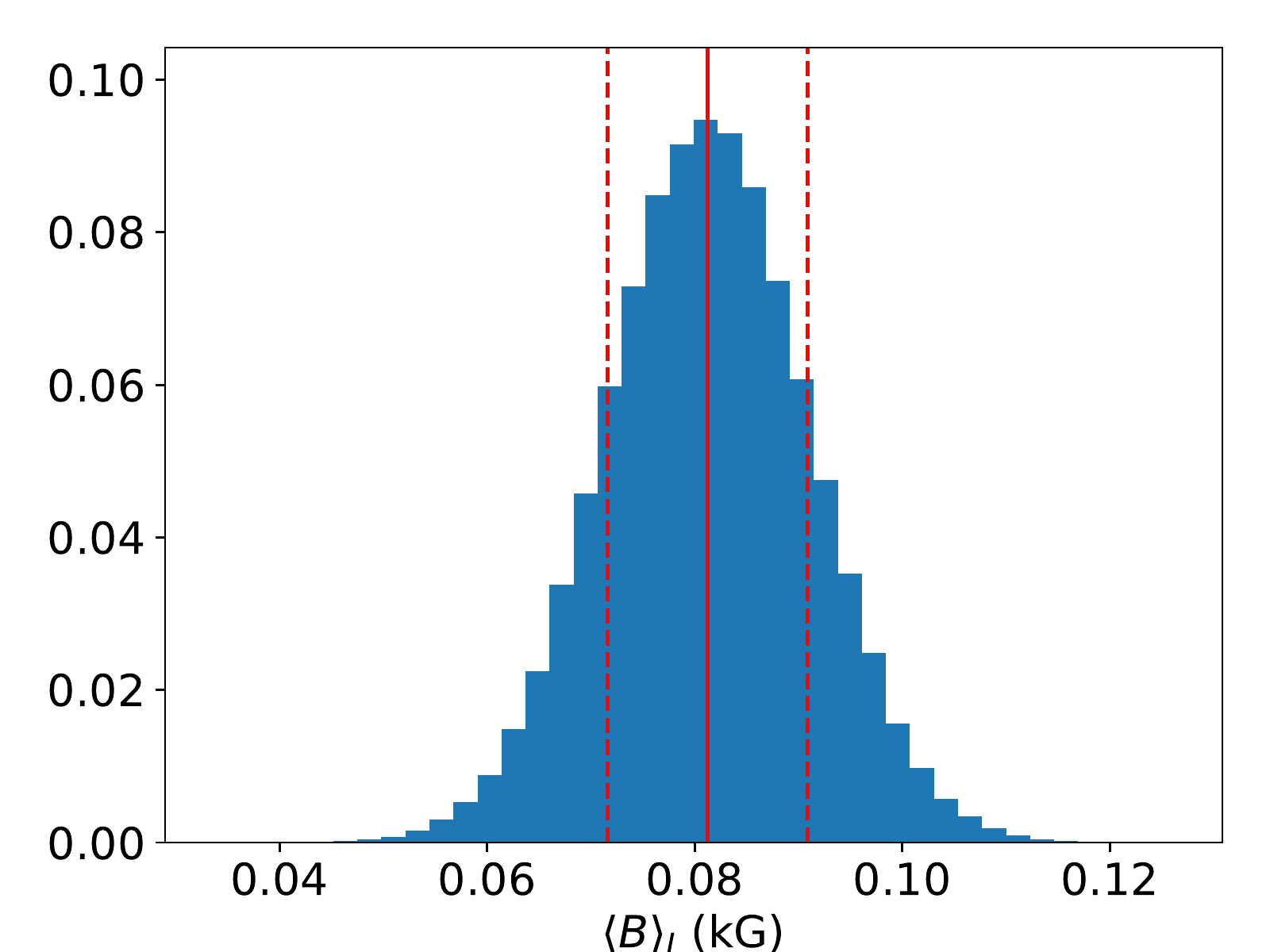}
    \includegraphics[width=0.45\textwidth]{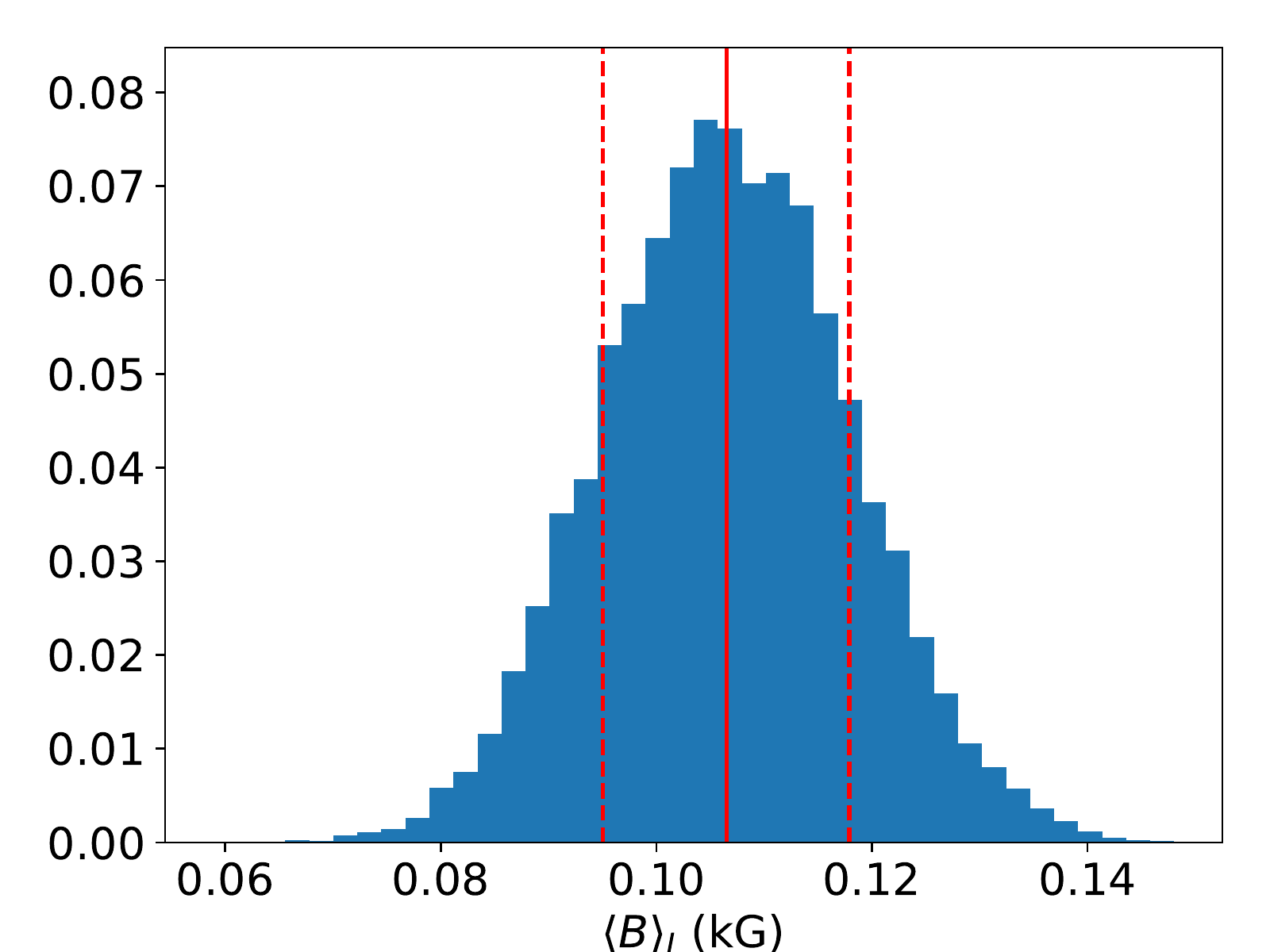}
    \captionof{figure}{Same as Fig. \ref{fig:HD1835_obs} but for HD 130322.}
    \label{fig:HD130322_obs}
\end{minipage}
\vspace{0.3\textheight}
\subsection{HD 131156A}
\begin{minipage}{1.0\textwidth}
    \centering
    \includegraphics[width=\textwidth]{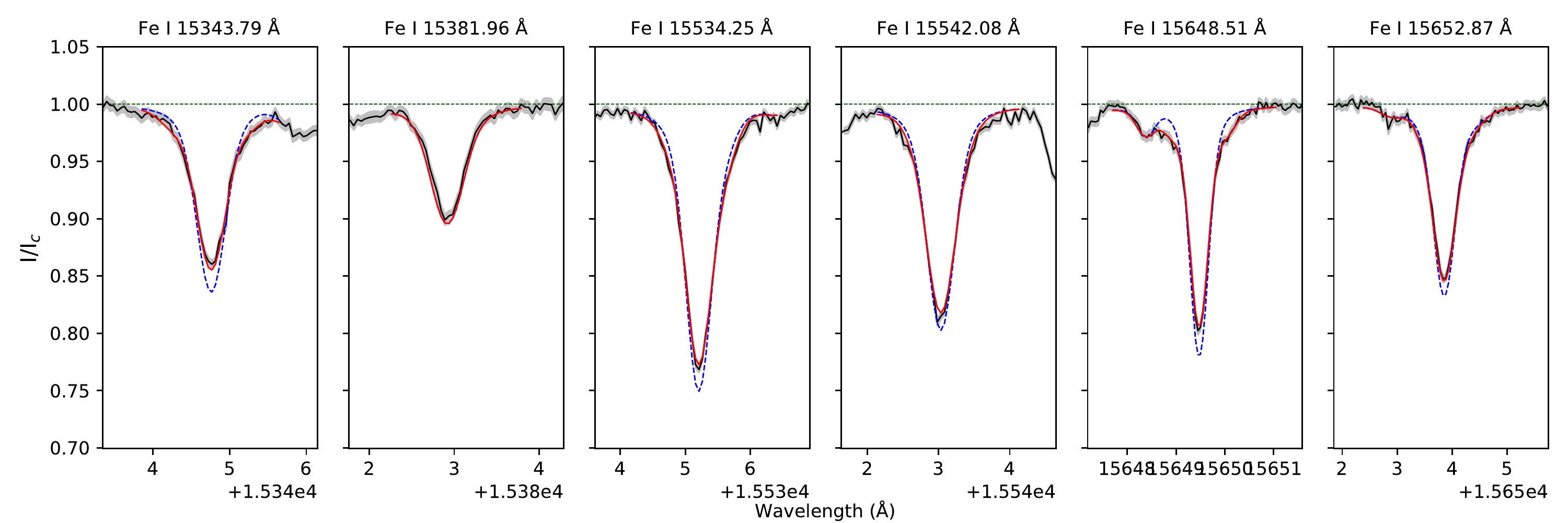}
    \includegraphics[width=0.49\textwidth]{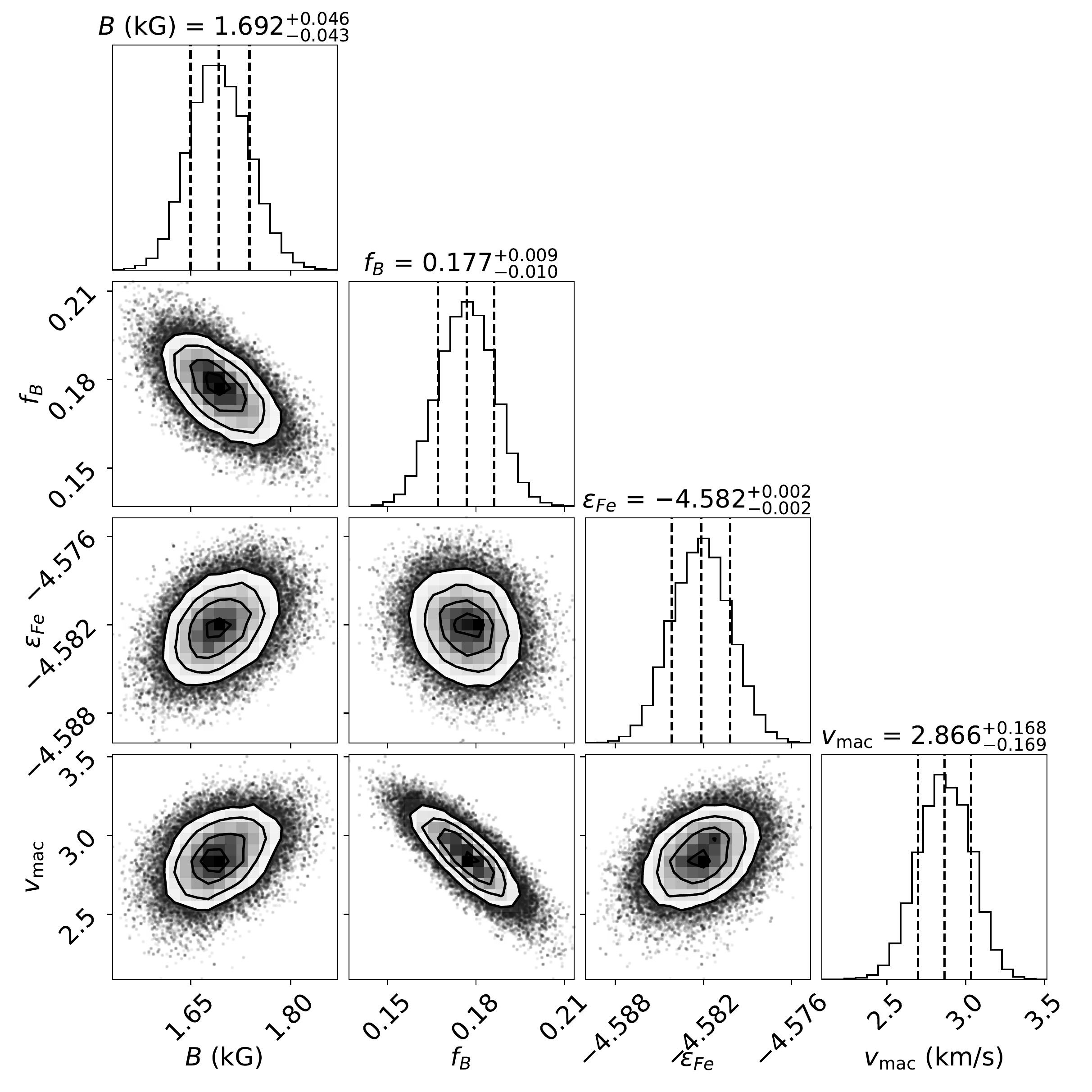}
    \includegraphics[width=0.49\textwidth]{2comp/HD131156A_corner_220323}
    \includegraphics[width=0.45\textwidth]{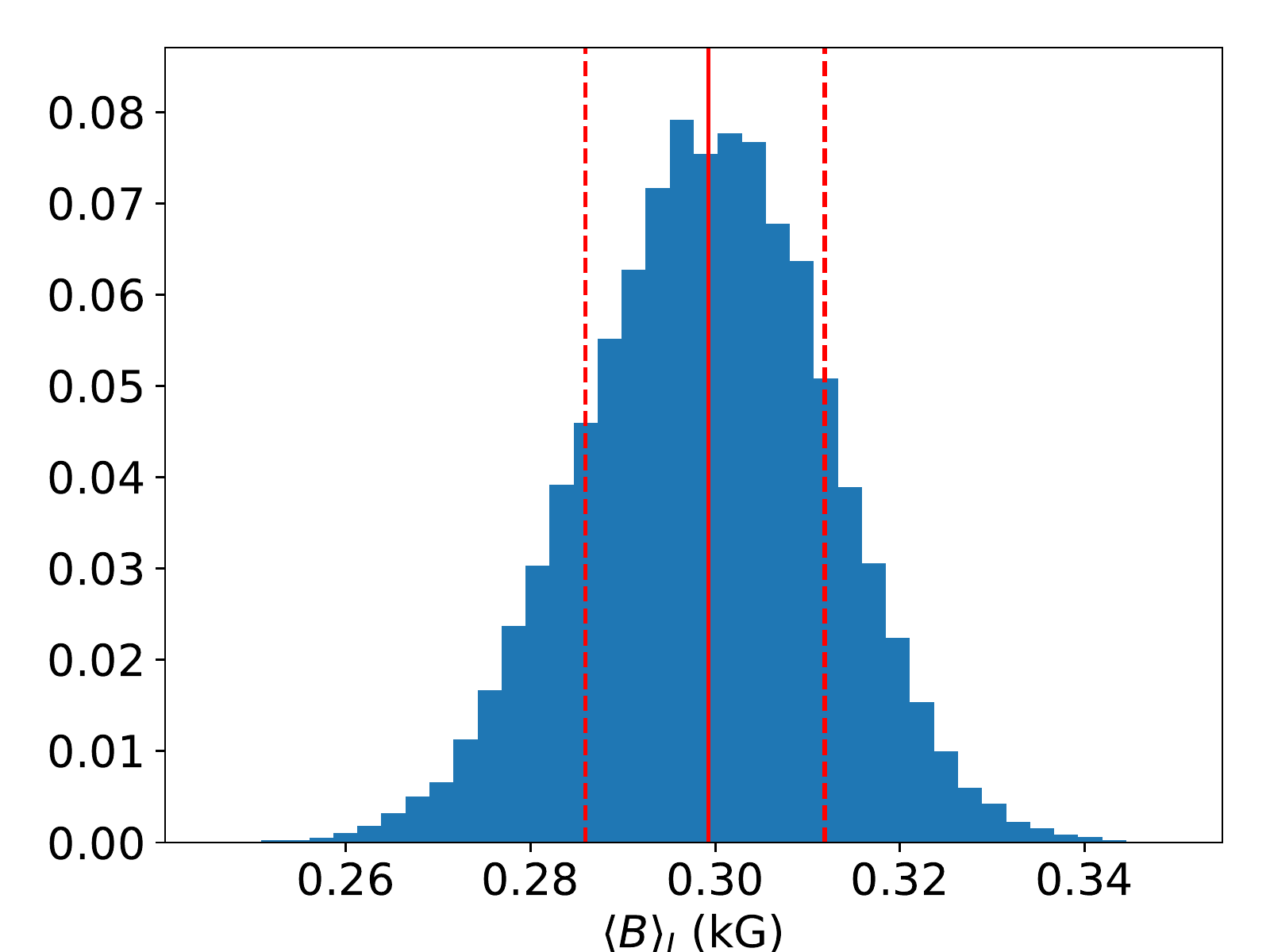}
    \includegraphics[width=0.45\textwidth]{2comp/HD131156A_bavg_220323}
    \captionof{figure}{Same as Fig. \ref{fig:HD1835_obs} but for HD 131156A.}
    \label{fig:HD131156A_obs}
\end{minipage}
\vspace{0.3\textheight}
\subsection{HD 179949}
\begin{minipage}{1.0\textwidth}
    \centering
    \includegraphics[width=\textwidth]{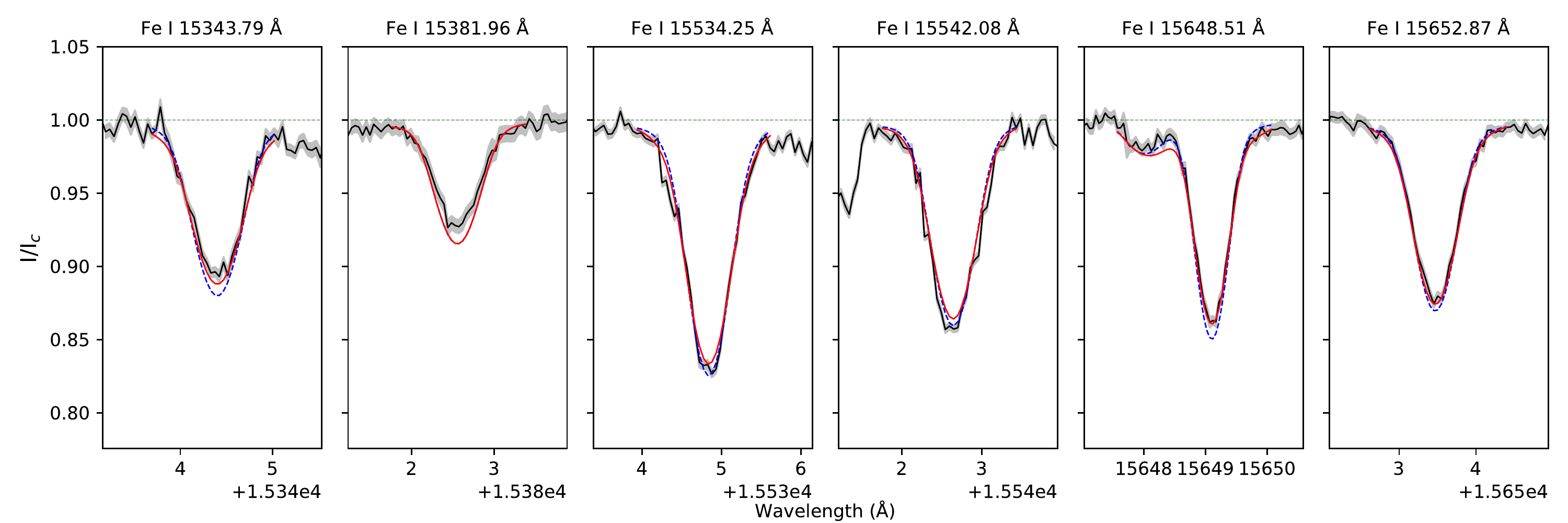}
    \includegraphics[width=0.49\textwidth]{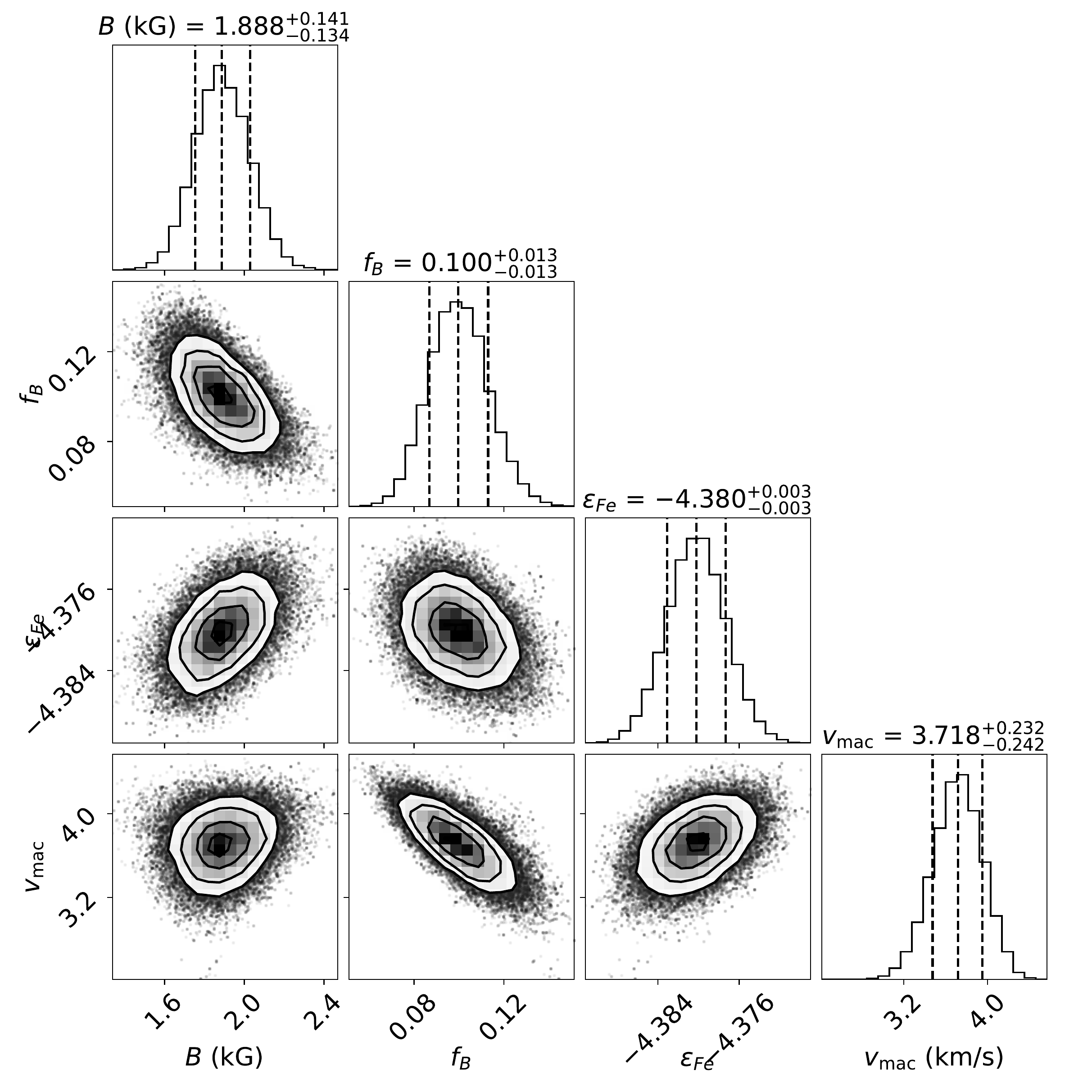}
    \includegraphics[width=0.49\textwidth]{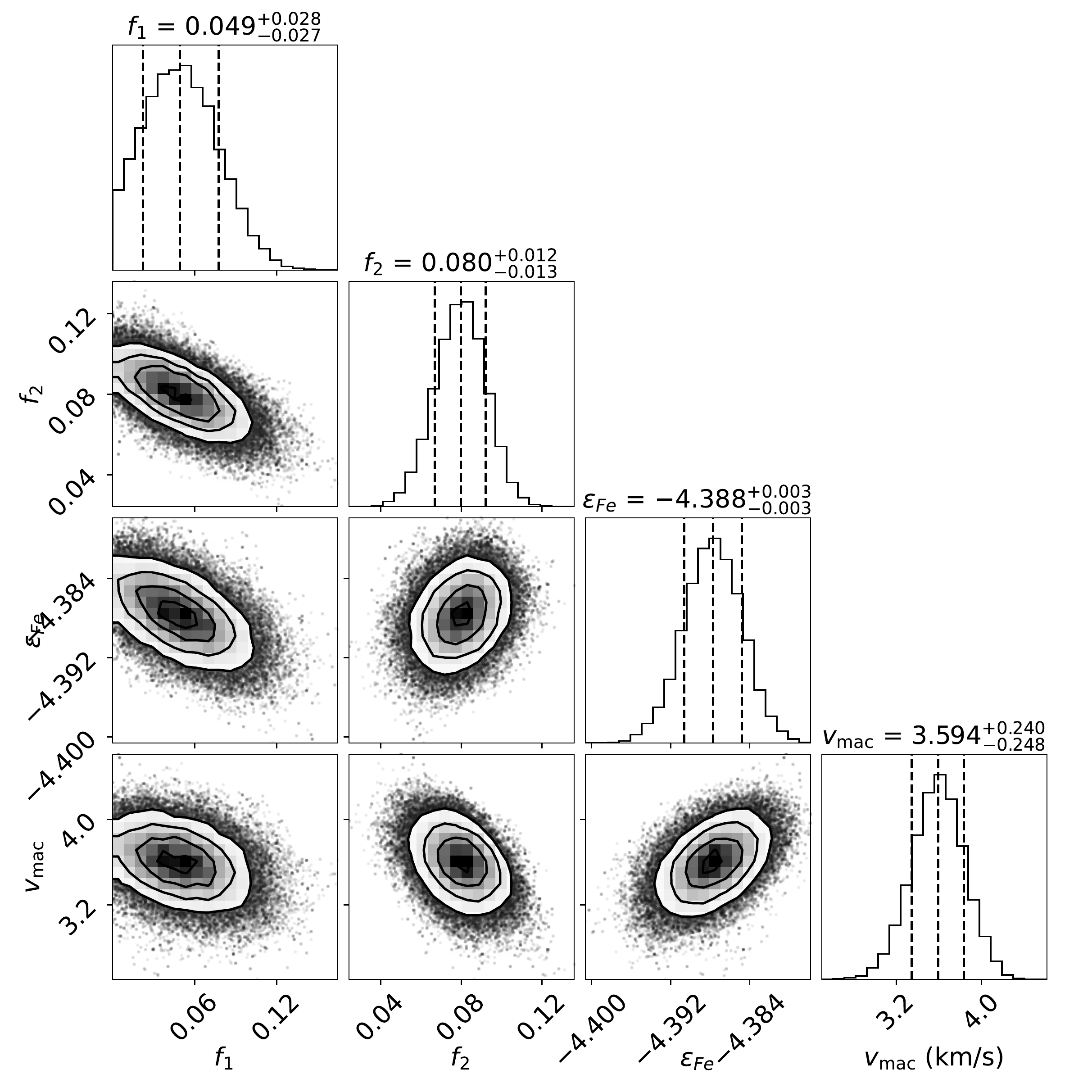}
    \includegraphics[width=0.45\textwidth]{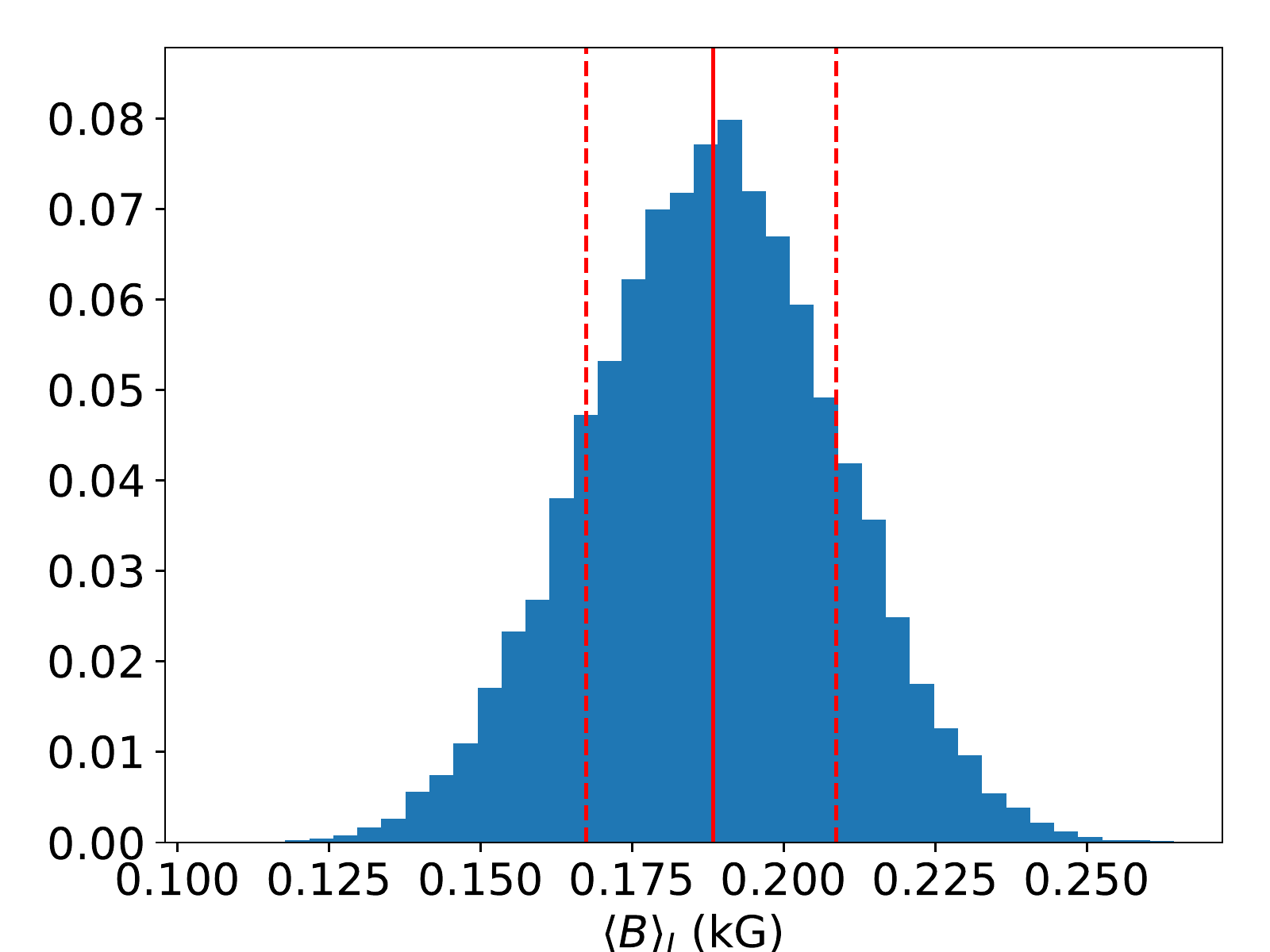}
    \includegraphics[width=0.45\textwidth]{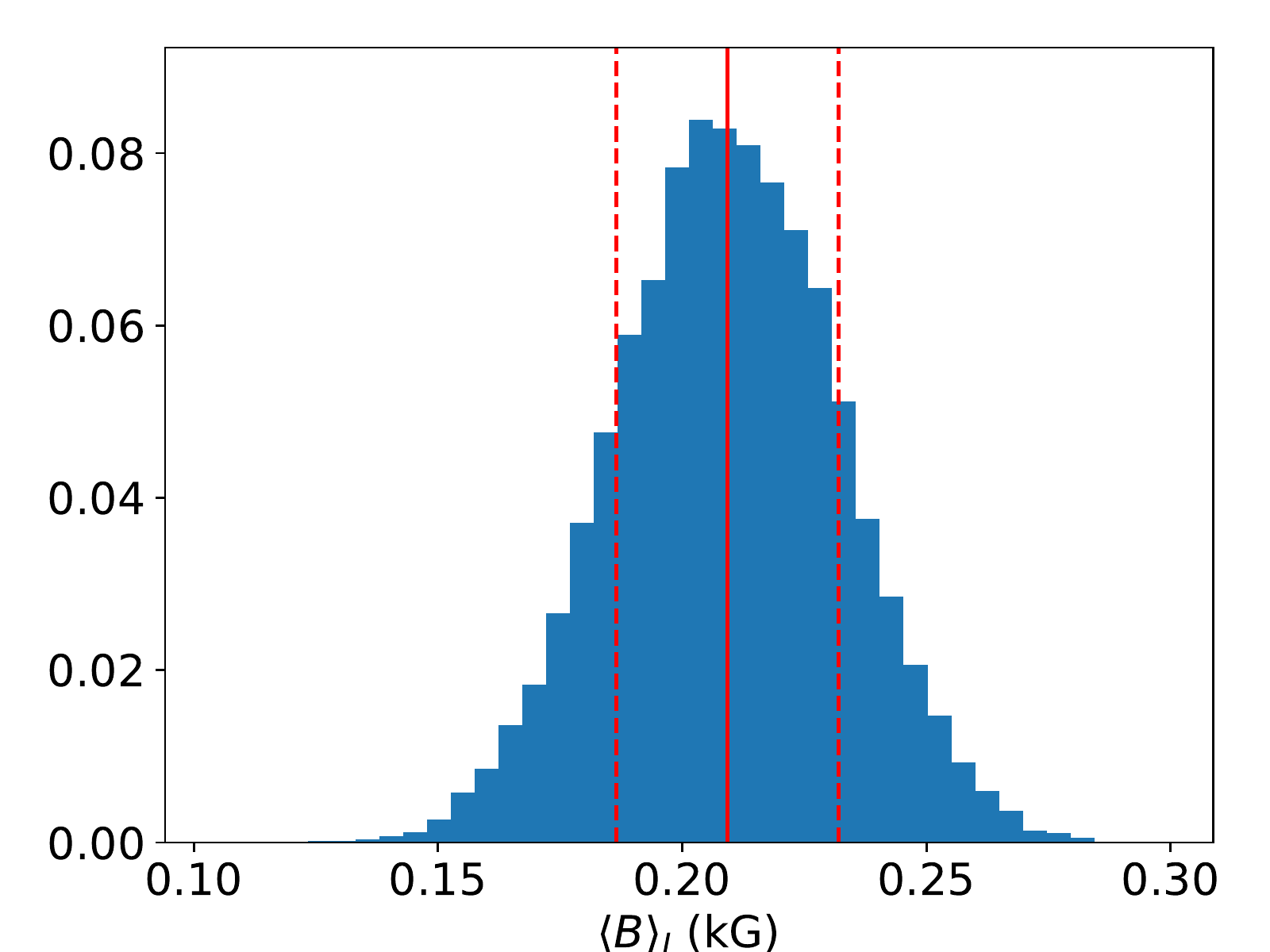}
    \captionof{figure}{Same as Fig. \ref{fig:HD1835_obs} but for HD 179949.}
    \label{fig:HD179949_obs}
\end{minipage}
\vspace{0.3\textheight}
\subsection{HD 206860}
\begin{minipage}{1.0\textwidth}
    \centering
    \includegraphics[width=\textwidth]{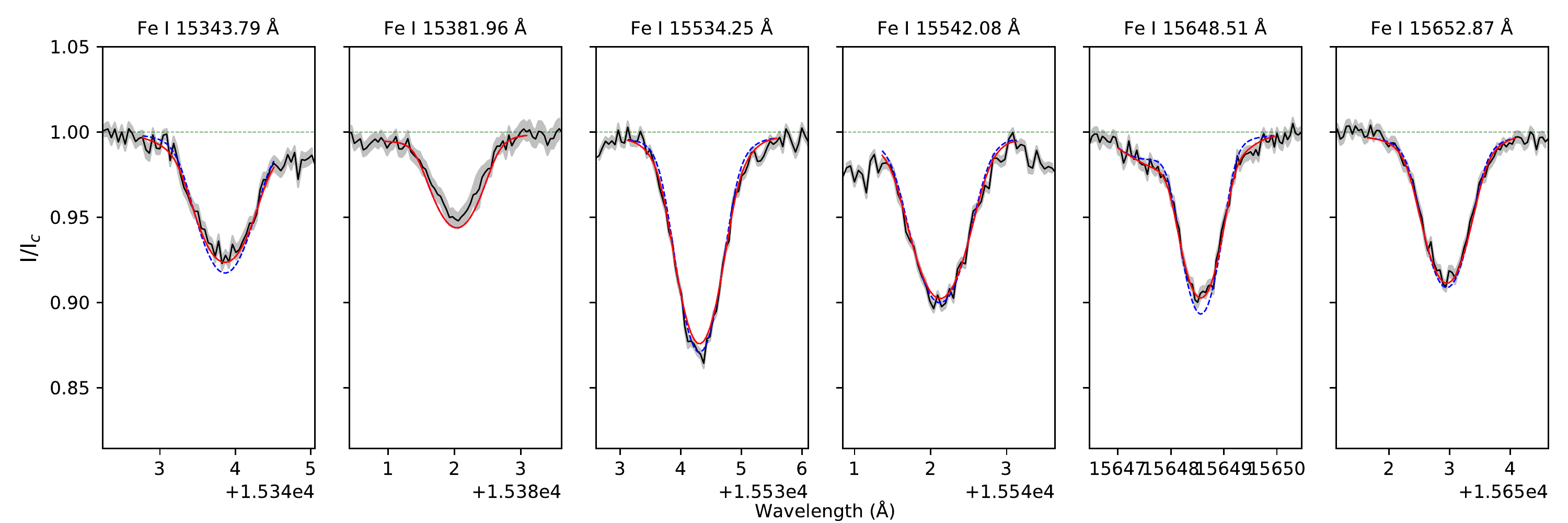}
    \includegraphics[width=0.49\textwidth]{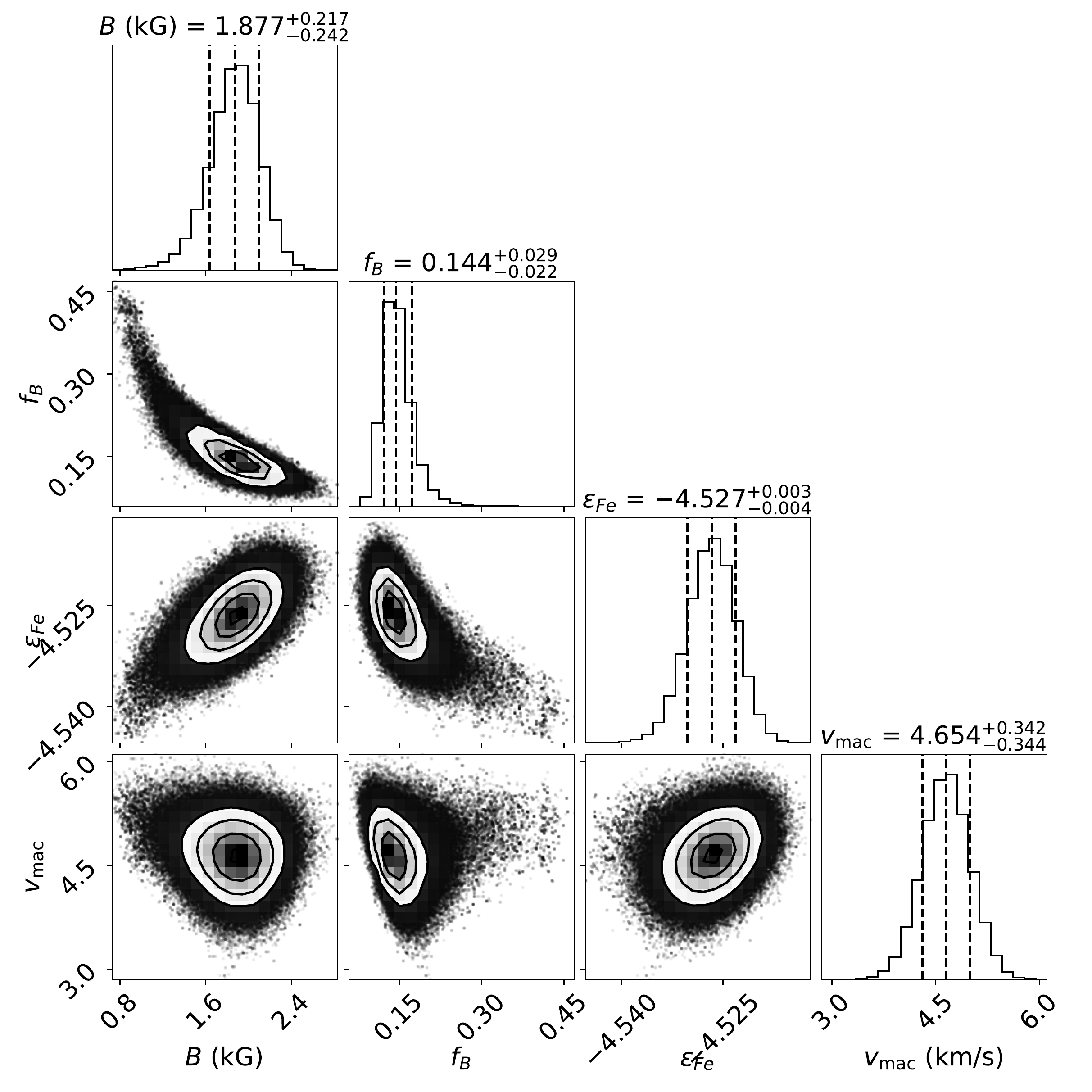}
    \includegraphics[width=0.49\textwidth]{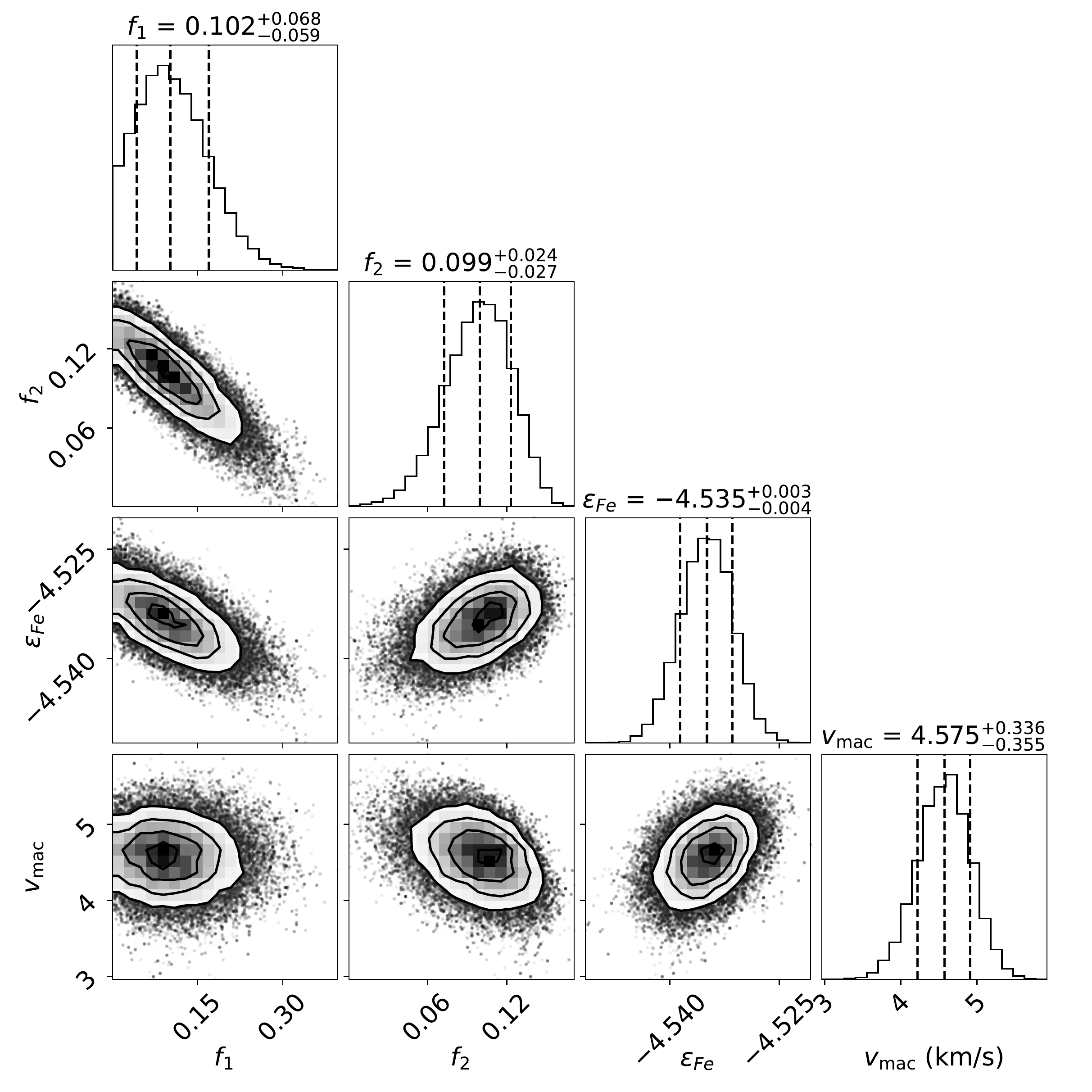}
    \includegraphics[width=0.45\textwidth]{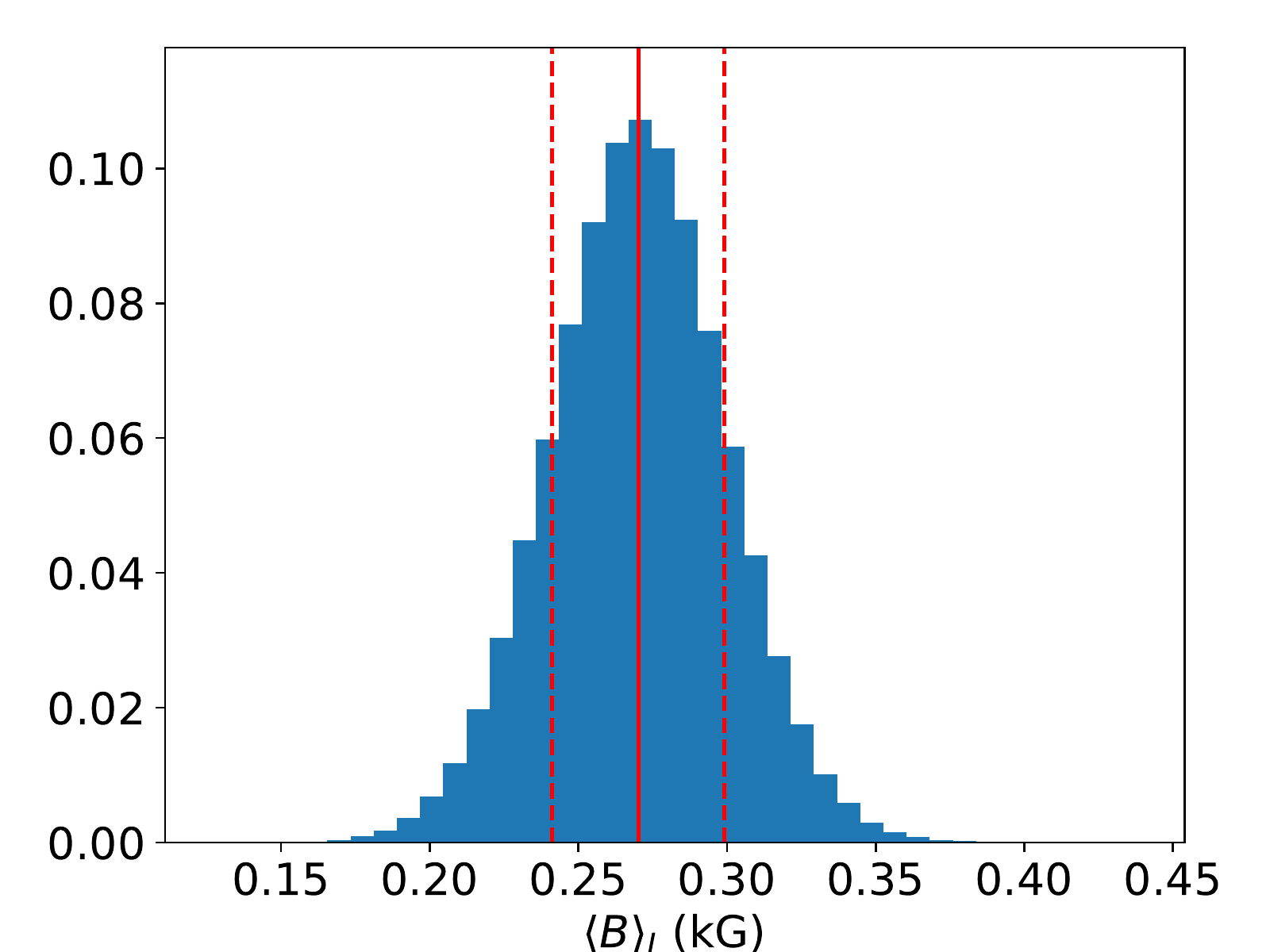}
    \includegraphics[width=0.45\textwidth]{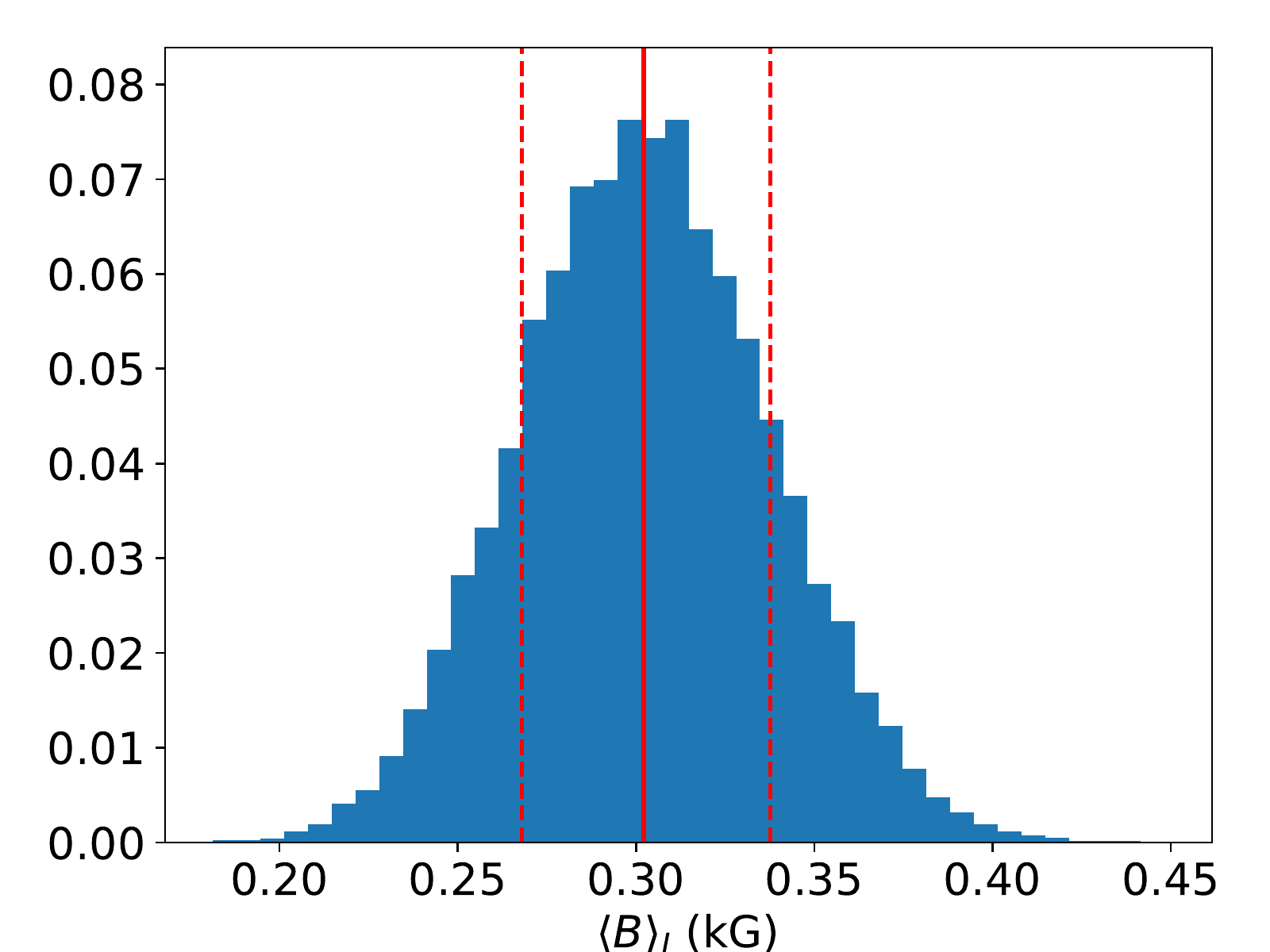}
    \captionof{figure}{Same as Fig. \ref{fig:HD1835_obs} but for HD 206860.}
    \label{fig:HD206860_obs}
\end{minipage}
\vspace{0.3\textheight}
\end{appendix}
\end{document}